\documentclass{aa}  

\usepackage{graphicx}
\usepackage{txfonts}
\usepackage[flushleft]{threeparttable}
\usepackage{tabularx}
\def\arraybackslash{\let\\\tabularnewline} 

\usepackage{xcolor}

\usepackage{gensymb}

\usepackage{soul}

\usepackage[export]{adjustbox}

\usepackage[version=4]{mhchem}
\usepackage{chemformula}

\usepackage{placeins}

\usepackage[normalem]{ulem}

\usepackage{stfloats}

\usepackage[labelformat=simple,justification=centering]{subcaption}
\renewcommand\thesubfigure{(\alph{subfigure})}

\begin{document}

   \title{Chemistry across dust and gas gaps in protoplanetary disks}
   \subtitle{Modelling the co-spatial molecular rings in the HD~100546 disk}

   \author{M. Leemker
          \inst{1}
          \and
          A. S. Booth\inst{1, 2, 3}
          \and
          E. F. van Dishoeck\inst{1, 4}
          \and
          L. W\"{o}lfer\inst{5}     
          \and
          B. Dent\inst{6} 
          }
          
   \institute{Leiden Observatory, Leiden University, P.O. box 9513, 2300 RA Leiden, The Netherlands\\
              \email{leemker@strw.leidenuniv.nl}
         	 \and 
         	 Clay Postdoctoral Fellow
         	 \and 
             Center for Astrophysics \textbar\, Harvard \& Smithsonian, 60 Garden St., Cambridge, MA 02138, USA            
            \and
         	 Max-Planck-Institut f\"ur Extraterrestrische Physik, Giessenbachstrasse 1, 85748 Garching, Germany
         	 \and 
         	 Department of Earth, Atmospheric, and Planetary Sciences, Massachusetts Institute of Technology, Cambridge, MA 02139, USA
         	 \and 
         	 Joint ALMA Observatory, Alonso de Córdova 3107, Vitacura, Santiago, 763 0355, Chile
             }

   \date{Received 22 December 2023; accepted 8 April 2024}

  \abstract
{Nearby extended protoplanetary disks are commonly marked by prominent rings in the dust emission, possibly carved by forming planets. High-resolution observations show that both the dust and the gas are structured. These molecular structures may be related to radial and azimuthal density variations in the disk and/or the disk chemistry. }
{The aim of this work is to identify the expected location and intensity of rings seen in molecular line emission in gapped disks while exploring a range of physical conditions across the gap. In particular, we aim to model the molecular rings that are, in contrast with most other gapped disks, co-spatial with the dust rings at $\sim$20 and $\sim$200~au in the HD~100546 disk using the thermochemical code DALI.  }
{We model observations with the Atacama Large Millimeter/submillimeter Array (ALMA) of CO isotopologues, [C~{\sc I}], HCN, CN, \ce{C2H}, NO, and \ce{HCO+} in the HD~100546 disk. 
An axisymmetric 3D thermochemical model reproducing the radial profiles of the CO isotopologue observations and the double ring seen in continuum emission is used to make predictions for various emission lines. The effect of the amount of gas in the dust gap, the C/O ratio, an attenuated background UV radiation field, and the flaring index on the radial distribution of different molecules are investigated. }
{
The fiducial model of a gapped disk with a gas cavity at $0-15$~au, a dust cavity at $0-20$~au, and a gas and dust gap at $40-175$~au provides a good fit to the continuum and the CO isotopologues in the HD~100546 disk. In particular, the CO isotopologue emission is consistent with a shallow gas gap with no more than a factor of $\sim 10$ drop in gas density at $40-175$~au. Similar to the CO isotopologues, the HCN and \ce{HCO+} model predictions reproduce the data within a factor of a few in most disk regions. However, the predictions for the other atom and molecules, [C~{\sc I}], CN, \ce{C2H}, and NO do not match the intensity nor the morphology of the observations. An exploration of the parameter space shows that in general the molecular emission rings are only co-spatial with the dust rings if the gas gap between the dust rings is depleted by at least four orders of magnitude in gas or if the C/O ratio of the gas is varying as a function of radius. For shallower gaps the decrease in the UV field roughly balances the effect of a higher gas density for UV tracers such as CN, \ce{C2H}, and NO. Therefore, the CN, \ce{C2H}, and NO radicals are not good tracers of the gas gap depth. In the outer regions of the disk around 300~au, these UV tracers are also sensitive to the background UV field incident on the disk. Reducing the background UV field by a factor of 10 removes the extended emission and outer ring seen in CN and \ce{C2H}, respectively, and reduces the ring seen in NO at 300~au. 
The C/O ratio primarily effects the intensity of the lines without changing the morphology much. The [C~{\sc I}], HCN, CN, and \ce{C2H} emission all increase with increasing C/O, whereas the NO emission shows a more complex dependence on the C/O ratio depending on the disk radius.

}
{CO isotopologues and \ce{HCO+} emission trace gas gaps and gas gaps depths in disks. The molecular rings in HCN, CN, \ce{C2H}, and NO predicted by thermochemical models do not naturally coincide with those seen in the dust, contrary to what is observed in the HD~100546 disk. This could be indicative of a radially varying C/O ratio in the HD~100546 disk with a C/O above 1 in a narrow region across the dust rings, together with a shallow gas gap that is depleted by a factor of $\sim 10$ in gas, and a reduced background UV field. The increase in the C/O ratio to $\gtrsim 1$ could point to the destruction of some of the CO, the liberation of carbon from ice and grains, or in case of the outer ring, it could point to second generation gas originating from the icy dust grains.}

   \maketitle

\section{Introduction}

The gas, dust, and ice surrounding a young star are the building blocks of new planets. Many of the nearby, massive, and large protoplanetary disks have been observed to not be smooth but have gaps, rings, and arcs in the continuum emission that are possibly carved by massive forming planets \citep[e.g.,][]{Andrews2018, Huang2018, Long2018, Andrews2020, Francis2020}. Recently, young planets with clear accretion signatures have been directly detected in the PDS~70 disk \citep{Keppler2018, Haffert2019, Benisty2021}. Additionally, a point-like source consistent with a planet in a Keplerian orbit in the dust gap of the HD 169142 disk is seen \citep{Hammond2023} and the presence of a giant accreting planet is debated in the AB~Aur disk \citep{Currie2022, Zhou2023}. Promising signs of planets are actually now seen in disks through indirect methods such as detailed studies of the CO gas kinematics. \citep[e.g.,][]{Pinte2018, Teague2018, Pinte2019, Alarcon2022, Izquierdo2022, Izquierdo2023}. The chemical composition of the solid core and gaseous atmosphere of these planets depends on the chemical composition of the ice and the gas in their protoplanetary disk. Comparing the composition in a protoplanetary disk to that measured in exoplanet atmospheres provides insights into where and when these planets have formed \citep{Oberg2011, Oberg2016, Cridland2016, Mordasini2016, Cridland2017, Cridland2023, Eistrup2023}.

The chemical composition of the gas across a disk is not constant. Signs of chemical substructures have been found in simple gas-phase molecules such as the commonly observed CO, \ce{HCO+}, \ce{N2H+}, CN, HCN, and the recently detected NO \citep[e.g.,][]{Cazzoletti2018, Bergner2019, Qi2019, Aikawa2021, Bergner2021, Law2021, Zhang2021, Leemker2023, Temmink2023}. These structures are seen both at the location of dust structures such as rings, gaps, and arcs, e.g. in the MWC~480, PDS~70, IRS~48, and the HD~100546 disks, as well as in regions uncorrelated to the dust e.g., four out of five disks observed in the MAPS (Molecules with ALMA at Planet-forming scales) program \citep{Facchini2021, Law2021, Booth2021irs48, Booth2021hd, vanderMarel2021irs48, Brunken2022, Jiang2022, Leemker2023}. These ringed structures can thus be due to chemical rings or physical rings in the underlying gas density structure. 

Young planets that are embedded in their native protoplanetary disk can carve a deep gap in the dust, and if they are sufficiently massive, also a gap in the gas as traced by CO emission \citep{Bruderer2014, Zhu2014, Dong2015, Rosotti2016, vanderMarel2016, Andrews2018, Long2018, Binkert2021, Oberg2021m, Zhang2021, Leemker2022}. A gas gap in a planet forming disk not only reduces the amount of gas and dust at that location, it also reduces the attenuation of the UV radiation causing a relative increase in the UV field \citep{Cleeves2011, Facchini2018, Alarcon2020, Rab2020}. This in turn leads to an increase of the dust temperature. The resulting gas temperature in the dust gap is a balance of the heating and cooling processes and can be either warmer or colder than the region just outside the gap \citep[e.g.,][]{Bruderer2013, Facchini2017, Alarcon2020, Broome2023}. This change in temperature and UV field across the gap could lead to rings in other molecules as the chemistry in disks is sensitive to both of these.

A change in the temperature could lead to the freeze-out or thermal desorption of one of the major volatiles: \ce{H2O}, CO, and \ce{N2}.
These snowlines are the midplane locations where 50\% of a molecule is in the gas-phase and 50\% is frozen out onto the dust grains. This is exceptionally clear in the structure of \ce{HCO+} and \ce{N2H+}. These molecules respond very strongly to the desorption of \ce{H2O}, CO, and \ce{N2}, resulting in ring-shaped emission outside (\ce{HCO+} and \ce{N2H+} for the water and CO snowlines, respectively) or inside (\ce{N2H+} for the \ce{N2} snowline) the snowline \citep{Qi2013, vantHoff2017, Leemker2021}. On top of that, the abundance of \ce{HCO+}, \ce{DCO+}, and \ce{N2H+} is set by the ionisation rate in the disk which may radially change due to substructures \citep[e.g.,][]{Aikawa2001, Cleeves2014, Aikawa2021}. 

In addition to snowlines, the stellar UV field is important for setting the chemistry in disks by photodissociating molecules in the surface layers of the disk. This not only destroys molecules, but it also enables formation paths to new ones. For example, the photodissociation of water produces OH which can lead to the formation of NO and the photodissociation of CO leads to the presence of atomic carbon in disks \citep{Thi2010, Bruderer2012, Schwarz2014, Leemker2023}. Moreover, UV radiation can vibrationally excite molecular hydrogen (\ce{H_2^*}) at intermediate disk radii. The extra energy stored in \ce{H_2^*} can then be used to overcome the energy barrier to form CN and NO in the surface layers of disks \citep{Agundez2010, Visser2018, Cazzoletti2018, PanequeCarreno2022, Leemker2023}. Apart from the innermost regions in the midplane, HCN follows the same trend as it is closely related to CN in the disk surface layers. Therefore, ring-shaped CN and HCN emission is expected even if the underlying gas and dust surface density profiles are smooth \citep{Long2021}.

In this paper we aim to investigate the conditions that lead to molecular rings coinciding with dust rings under different physical and chemical conditions. In particular, these results are applied to the HD~100546 disk, one of the gapped disks with molecular rings seen at the location of the continuum rings at 20~au and $\sim200$~au. 
In Sect.~\ref{sec:methods} we present the [C~{\sc I}] data in the HD~100546 disk and describe the setup of our thermochemical models that include a gas and dust cavity as well as a gap between the two rings seen in the HD~100546 continuum and CO isotopologues. Section~\ref{sec:results} presents a fiducial model that fits these data. In addition, the predictions for [C~{\sc I}], HCN, CN, \ce{C2H}, NO, and \ce{HCO+} in this disk under different physical conditions, varying parameters such as a deep gas gap, an elevated C/O ratio, an attenuated background UV field, and disk flaring are presented. The results are discussed in Sect.~\ref{sec:discussion} in the specific context of the HD~100546 observations and our conclusions are summarized in Sect.~\ref{sec:conclusions}.

   \begin{figure*}
   \centering
   \includegraphics[width=\textwidth]{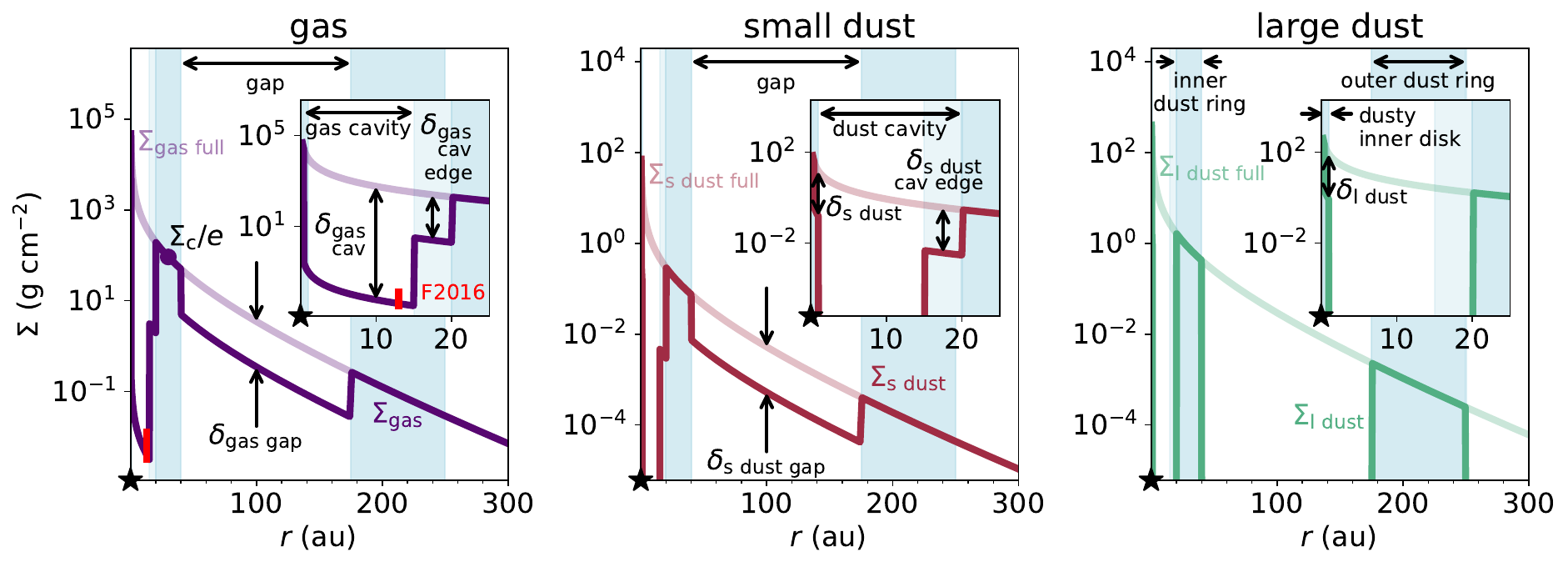}
      \caption{Surface density of the gas (purple), small dust (red), and large dust (green) in the fiducial model for the HD~100546 disk. The light purple, red, and green lines indicate the respective surface densities in a full disk without any gaps, cavities and rings. The blue shaded regions indicate the dusty inner disk and the two dust rings at 20-40~au and 175-250~au respectively. The surface density $\Sigma_{\rm c}/e$ at $r_{\rm c}$ is indicated with a scatter point in the left panel. Additionally, the DALI parameters that control drops and increases in the surface densities are indicated with arrows (see Table~\ref{tab:dali} for their values). The red bar in the inset in the left panel indicates the gas surface density based on CO by \citet{Fedele2016}. The small insets in each panel focus on the inner 25~au of the disk.}
         \label{fig:dali_structure}
   \end{figure*}

\section{Methods} \label{sec:methods}

\subsection{Source: HD~100546}

HD~100546 is a $\sim 2$~M$_{\odot}$, 5~Myr old Herbig Ae/Be star with a luminosity of $\sim$36~L$_{\odot}$ \citep{Kama2016, Arun2019, Pineda2019}. This star is located in a relatively isolated region at a distance of $108.1\pm 0.4$~pc \citep{Gaia2016, GaiaDR3}. The star is surrounded by a massive and gas-rich disk of $0.03-0.1$~M$_{\odot}$ based on CO isotopologue emission, \textit{Herschel} PACS and HIFI observations of the CO rotational ladder up to $J=38$, constraining temperature structure disk layers, and an upper limit on the HD flux \citep{Kama2016, Stapperinprep}. The \textit{Herschel} PACS observations of CO, C, and \ce{C+} indicate that the gas in the surface layers of the disk is warmer than the dust \citep{Bruderer2012, Fedele2016}. Additionally, this instrument revealed gas-phase water that is present outside $\sim40$~au in the HD~100546 disk \citep{vanDishoeck2021, Pirovano2022}. Both the CO gas and the dust in this disk are structured with small scale gas and dust spirals (a few $0\farcs1$), a large scale ($\sim 1"$) dusty spiral and a large scale spiral in the gas kinematics (\citealt{Garufi2016, Follette2017, Sissa2018, Norfolk2022}; W\"{o}lfer et al. subm.). Additionally, the continuum emission in this disk is seen in three locations: an unresolved dusty inner disk, a bright inner ring from $\sim20$~au to 40~au, and a weak outer ring at $<1$\% of the peak continuum intensity from $\sim150$~au to 250~au \citep{Bruderer2012, Walsh2014, Pineda2019, Fedele2021}. This paper primarily focusses on the molecular emission at the location of the second ring at $\sim 200$~au and how this compares to the inner ring at $\sim 20$~au. 

The central cavity and gap between the two dust rings have been hypothesized to be carved by massive planets. The inner planet has been inferred from CO ro-vibrational emission and a feature seen in scattered light \citep[e.g.,][]{Brittain2013, Brittain2014, Currie2015, Brittain2019}. In addition, co-rotating SO emission is seen at the location of the proposed planet \citep{Booth2023}. A second planet has been hypothesized between the two dust rings as a point source is seen at $50-60$~au in scattered light observations \citep{Quanz2013, Currie2014}. However, the nature of these potential planets is debated as no accretion signatures have been found at this location and different results have been found depending on the data quality and imaging techniques \citep{Currie2017, Follette2017, Rameau2017}. Therefore, more research is needed to firmly conclude if these planets are present. 

ALMA observations of molecular lines have revealed a number of different molecules in this disk. The detection of formaldehyde and the complex organic molecule methanol in this warm disk indicates that the molecular inventory is likely inherited from the native dark cloud as the bulk of the disk is too warm for CO freeze-out \citep{Booth2021hd}. The CS and SO emission peak at different radii, with the transition between the two coinciding with the inferred water snowline location at the inner edge of the dust cavity \citep[$\sim$20~au;][]{Keyte2023, Boothinprep}.
Finally, many of the molecular lines except the CO isotopologues are close to co-spatial with the rings seen in the dust and are consistent with a flat disk despite \ce{^12CO} emitting from higher disk layers (\citealt{Law2022, Stapper2023, Boothinprep}; W\"{o}lfer et al. subm.). In this work, we use high resolution ALMA observations at $\sim0\farcs1$ spatial resolution of the dust and CO isotopologues (\citealt{Pineda2019, Perez2020, Wolfer2023}, subm.), intermediate resolution data at $\sim0\farcs2-0\farcs5$ of the dust, CO, \ce{C^17O}, HCN, CN, \ce{C2H}, NO, and \ce{HCO+} \citep{Boothinprep}, and [C~{\sc I}] observations to elucidate the origin of the observed molecular substructures.

\subsection{ALMA data covering [C~{\sc I}]}

The \ce{^12CO} $J=7-6$ and [C~{\sc I}] line emission data used in his paper were taken in Cycle~6 with ALMA in Band 10 (2018.1.00141.S; PI: B. Dent), using 43 antennas with baselines of 15-783\,m. Starting from the archival pipeline-calibrated data we performed further data reduction using CASA \citep{McMullin2007}. Due to a low signal-to-noise ratio, no self-calibration was applied. We first subtracted the continuum using the \texttt{uvcontsub} task, flagging channels containing line emission, and then imaged the lines with \texttt{tCLEAN} for both the continuum-subtracted and non-subtracted data sets. In this process, we imaged the data at the highest spectral resolution possible and adopted a Briggs robust weighting of 0.5, which gave the best trade off between spatial resolution and sensitivity. We also applied a Keplerian mask\footnote{\url{https://github.com/richteague/keplerian_mask}} using an inclination of 42\,\degree, position angle of 140\,\degree, distance of 110\,pc, stellar mass of 2.13\,M$_{\odot}$, and systemic velocity of 5.7\,km\,s$^{-1}$. We further used the multi-scale deconvolver (scales 0,5,10,20) and a slight \texttt{uv-taper} ($0\farcs05\times 0\farcs05,0\,\degree$) to improve the signal-to-noise ratio in the images. This resulted in an image with a spatial resolution of $0\farcs21 \times 0\farcs16\ (28\degree)$ and a maximum recoverable scale of 2" in diameter (108~au in radius). The integrated intensity map and the spectrum are presented in Fig.~\ref{fig:CI_obs}.

   \begin{figure*}[ht!]
   \centering
   \includegraphics[width=\textwidth]{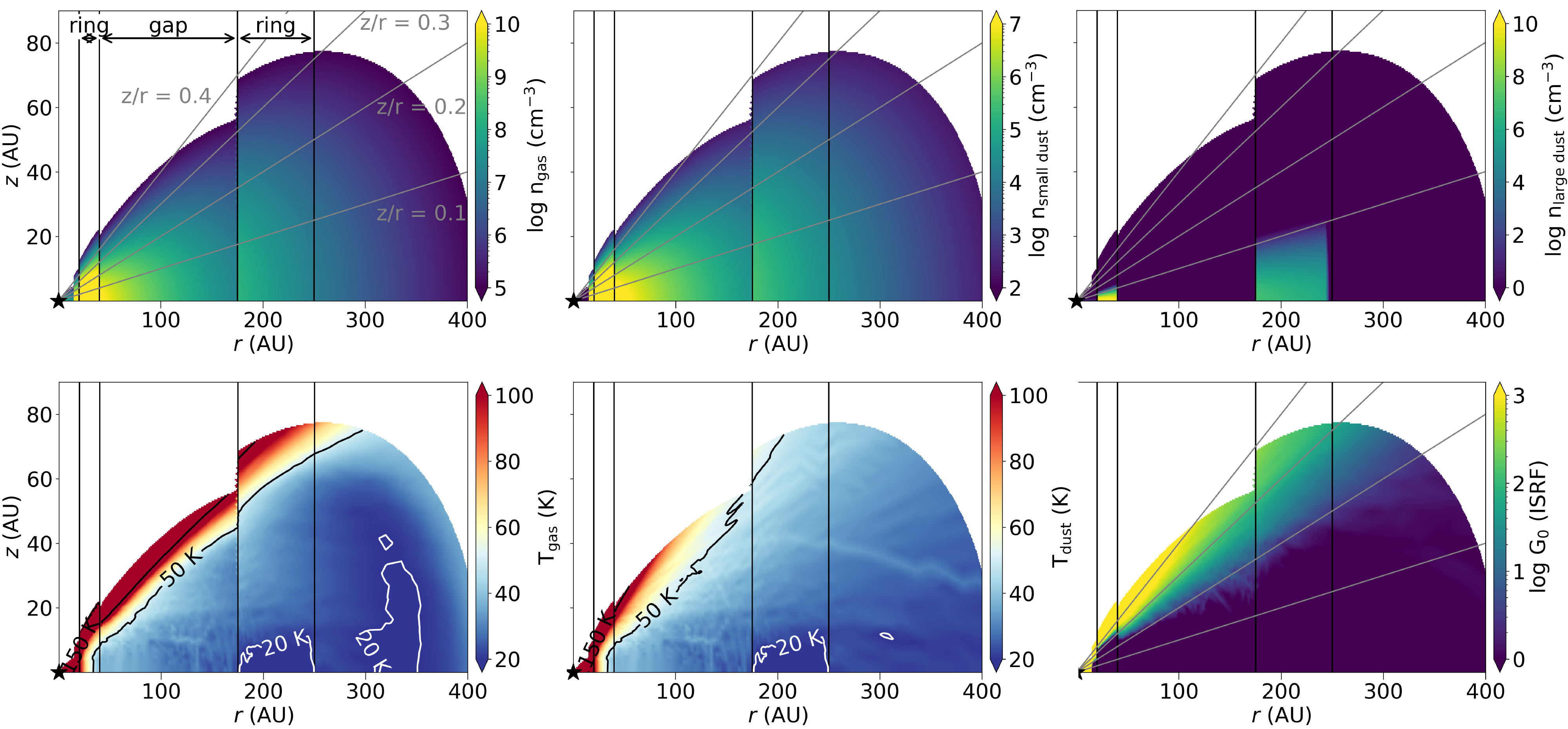}
      \caption{Structure of the fiducial DALI model. From left to right and top to bottom: gas, small dust, and large dust density, gas and dust temperature, and UV field. Only the regions with a gas number density above $10^5$~cm$^{-3}$ are shown. Outside 250~au, the midplane gas density is too low for the gas and dust temperature to be coupled. }
         \label{fig:dali_2D_basic_fid}
   \end{figure*}

\subsection{DALI}
To investigate the relative importance of physical versus chemical rings, we model the dust rings and rings seen in molecular line emission using the thermochemical code Dust And LInes \citep[DALI;][]{Bruderer2009, Bruderer2012, Bruderer2013}. The model presented in this work is based on those presented in \citet{Pirovano2022} and \citet{Keyte2023}. We improve on the dust structure from the latter model by including a dust gap in the outer disk to match the double ringed structure of the continuum observations \citep{Fedele2021, Boothinprep}. Additionally, the gas density structure is improved by including a gas gap between the two dust rings and by including an inner gas cavity just inside the inner dust ring at 20~au to match the observed CO emission in those regions (\citealt{Perez2020, Boothinprep}; W\"{o}lfer et al. subm.). The general setup of the model is described below.

\subsubsection{Gas density structure}

The radial and vertical density structure of the disk follow the standard DALI setup for a flared disk with an exponentially tapered power law in the radial direction and a Gaussian distribution of the gas in the vertical direction \citep[see][and Appendix~\ref{app:dali_structure} for details]{Bruderer2013}. To include the gas cavity and a gas gap in this disk, the gas surface density of the full disk is lowered as follows: 
\begin{equation}
  \Sigma_{\rm gas} =
    \begin{cases}
      \Sigma_{\rm gas\ full}\times 10^{-5} & (0.4~\mathrm{au} < r < 15~\mathrm{au})\\
      \Sigma_{\rm gas\ full}\times 10^{-2} & (15~\mathrm{au} < r < 20~\mathrm{au})\\
      \Sigma_{\rm gas\ full} & (20~\mathrm{au} < r < 40~\mathrm{au})\\      
      \Sigma_{\rm gas\ full}\times \delta_{\rm gas\ gap} & (40~\mathrm{au} < r < 175~\mathrm{au})\\
      \Sigma_{\rm gas\ full} & (175~\mathrm{au} < r < 1000~\mathrm{au}),\\            
    \end{cases}       
\end{equation}
with $\delta_{\rm gas\ gap}$ lowering the gas density in the gap, see the left panel of Fig.~\ref{fig:dali_structure} and the top left panel of Fig.~\ref{fig:dali_2D_basic_fid}.

\subsubsection{Dust density structure}

The small grains follow the gas outside the inner gas cavity as they are expected to be well mixed with the gas. Outside the gas cavity but inside the dust cavity the small grains also follow the gas. Inside the gas cavity down to the dusty inner disk that is observed in continuum emission up to $<2$~au \citep{Benisty2010, Pineda2019, Perez2020}, the small grains are depleted by a factor of $10^{-10}$ to mimic a cavity devoid of any dust. The dusty inner disk inside 1~au is depleted in small (5~nm$-$1~$\mu$m) and large (5~nm$-$1~mm) grains by a factor of $10^{-2}$ (see also the (top) middle panel in Fig.~\ref{fig:dali_structure} and Fig.~\ref{fig:dali_2D_basic_fid}): 
\begin{equation}
  \Sigma_{\rm s\ dust} =
    \begin{cases}
      \Sigma_{\rm s\ dust\ full}\times 10^{-2} & (0.4~\mathrm{au} < r < 1~\mathrm{au})\\
      \Sigma_{\rm s\ dust\ full}\times 10^{-10} & (1~\mathrm{au} < r < 15~\mathrm{au})\\
      (1-f_{\rm ls})\times \Sigma_{\rm gas}/\Delta_{\rm gd} & (15~\mathrm{au} < r < 1000~\mathrm{au}).          
    \end{cases}       
\end{equation}
In the vertical direction the small dust follows the gas.

Observations show that the large dust in the HD~100546 disk is mainly present at three different locations: a dusty inner disk, an inner ring at $20-40$~au, and an outer ring at $175-250$~au \citep{Walsh2014, Pineda2019, Fedele2021}. The cavity in the large dust in our model is 5~au larger than that in the small dust and the gas, similar to what is observed in other transition disks \citep{Perez2015, vanderMarel2016, Leemker2022, Wolfer2023}. This is parametrized as:
\begin{equation}
  \Sigma_{\rm l\ dust} =
    \begin{cases}
      \Sigma_{\rm l\ dust\ full}\times 10^{-2} & (0.4~\mathrm{au} < r < 1~\mathrm{au})\\
      \Sigma_{\rm l\ dust\ full}\times10^{-10} & (1~\mathrm{au} < r < 20~\mathrm{au})\\
      \Sigma_{\rm l\ dust\ full} & (20~\mathrm{au} < r < 40~\mathrm{au})\\
      \Sigma_{\rm l\ dust\ full}\times 10^{-10} & (40~\mathrm{au} < r < 175~\mathrm{au})\\
      \Sigma_{\rm l\ dust\ full} & (175~\mathrm{au} < r < 250~\mathrm{au})\\
      \Sigma_{\rm l\ dust\ full}\times 10^{-10} & (250~\mathrm{au} < r < 1000~\mathrm{au}).    
    \end{cases}       
\end{equation}
The outer ring extends to 250~au, see also the right panel of Fig.~\ref{fig:dali_structure}. Additionally, the large grains are settled to the disk midplane (top right panel in Fig.~\ref{fig:dali_2D_basic_fid}). The dust in the midplane of the outer dust ring is sufficiently cold for some CO freeze-out as the dust temperature drops below 20~K (bottom row, middle panel in Fig.~\ref{fig:dali_2D_basic_fid}).

\subsection{Stellar spectrum}
The stellar spectrum of HD~100546 observed by the Far Ultraviolet Spectroscopic Explorer (FUSE) and the the International Ultraviolet Explorer (IUE) and presented in \citet{Bruderer2012} is used. 
These observations have been dereddend and extended to longer wavelengths using a B9V template \citep{Pickles1998}.
Additionally, an X-ray luminosity of $7.9\times 10^{28}$~erg~s$^{-1}$ \citep{Stelzer2006} with a temperature of $7\times 10^7$~K is included to account for more energetic radiation.

\subsection{Chemistry}
\begin{figure*}[ht!]
   \centering
   \includegraphics[width=\textwidth]{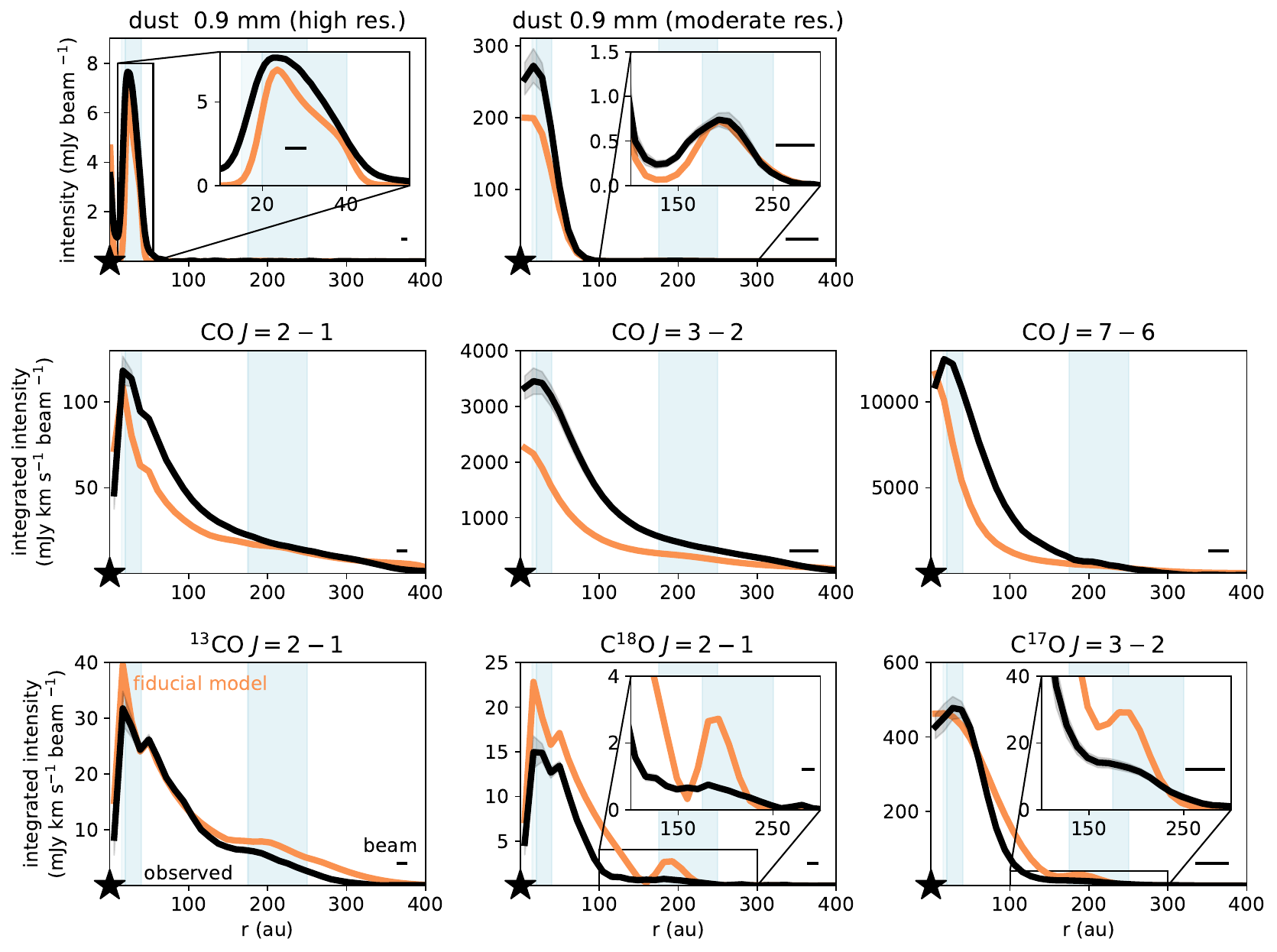}
      \caption{Azimuthally averaged radial profiles of the dust and CO isotopologue emission observed in the HD~100546 disk (black) together with the fiducial model (orange). The beam is indicated with the horizontal bar in the bottom right corner of each panel. The continuum observations are taken from \citet{Pineda2019} (high resolution 0.9~mm continuum), \citet{Boothinprep} (moderate resolution 0.9~mm continuum), the CO, \ce{^13CO}, and \ce{C^18O} $J=2-1$ transitions are first presented in \citet{Perez2020} and W\"{o}lfer et al. subm., the CO $J=7-6$ transition is presented in W\"{o}lfer et al. subm., and the CO $J=3-2$ and \ce{C^17O} $J=3-2$ transitions are presented in \citet{Boothinprep}.  }
         \label{fig:fid_dust_CO}
   \end{figure*}

Fig.~\ref{fig:fid_dust_CO} summarizes the observed distribution of CO isotopologues and dust in the HD~100546 disk in black.
The chemistry is modelled using a chemical network suited for modelling small nitrogen bearing molecules such as CN, HCN, and NO, and small hydrocarbons such as \ce{C2H}. This network was first presented in \citet{Visser2018, Cazzoletti2018}; and \citet{Long2021} and includes 134 species and 1844 reactions. Similar to \citet{Leemker2023}, the binding energy of \ce{C2H3} (ethylenyl) is set to $10^4$~K to mimic the conversion of \ce{C2H3} to larger COMs on the grain surfaces. 
This network includes the main gas-phase formation and destruction routes of the molecules of interest, the photodissociation and photoionization of relevant gas-phase molecules, self-shielding of \ce{H2}, CO, and C, the freeze-out and thermal-, and non-desorption of the major volatiles. The non-thermal desorption rates of \ce{H2O} and CO are updated from $1.3\times 10^{-3}$ and $2.7\times 10^{-3}$ molecules desorbed per grain per incident UV photon to $5\times 10^{-4}$ and $1.4\times 10^{-3}$ molecules desorbed per grain per incident UV photon respectively \citep{Oberg2007, Oberg2009, Paardekooper2016, GonzalezDiaz2019, Fillion2022}. Additionally, some grain surface reactions are included such as the formation of \ce{H2} on grains and PAHs. Moreover, the hydrogenation of O, C, N, and CN on grains to form \ce{H2O}, \ce{CH4}, \ce{NH3}, and HCN ices are included. Nevertheless, no detailed ice chemistry such as the conversion of CO to \ce{H2CO} or \ce{CH3OH} is modelled as the HD~10546 disk is warm and no large-scale CO freeze-out is expected.

The abundance of the CO isotopologues is predicted with the CO isotopologue network presented in \citet{Miotello2016} consisting of 185 species and 5755 reactions. This network includes the \ce{^12C}, \ce{^13C}, \ce{^16O}, \ce{^17O}, and \ce{^18O} isotopes, with their ratios taken as \ce{^12C}/\ce{^13C} = 77, \ce{^16O}/\ce{^18O} = 560, and \ce{^16O}/\ce{^17O} = 1792 \citep{Wilson1994}. Similar to the nitrogen network, this network includes the freeze-out and thermal desorption of simple species and their isotopologues, some grain surface reactions to model the hydrogenation of C, N, and O to \ce{CH4}, \ce{NH3}, \ce{H2O}, and their isotopologues, and isotope selective photodissociation. In addition, self-shielding of \ce{H2}, CO, and their isotopologues, and of C are included as well as fractionation reactions. Both chemical networks are started with molecular initial conditions and a fiducial total (gas $+$ ice) C/O ratio of 0.4 (see Table~\ref{tab:dali} and \ref{app:dali_chem_init}).

For the \ce{HCO+} abundance, the chemical network presented in \citet{Leemker2021} is used. This is a small chemical network with ten reactions and the same number of species centred around \ce{HCO+}. This network does include the freeze-out and thermal desorption of \ce{H2O}, but not that of CO. The \ce{HCO+} is formed through an ion-molecule reaction of CO with \ce{H_3^+} and it is destroyed by dissociative recombination with an electron and by the reaction with gas-phase water. The formation of \ce{H_3^+} is driven by ionization of \ce{H2} by cosmic rays creating \ce{H_2^+} that subsequently reacts with \ce{H2} to form \ce{H_3^+}. \ce{H_3^+} is destroyed by reactions with electrons, gas-phase water and CO.  As the only carbon-bearing molecules in this network are CO and \ce{HCO+}, this network is only run for the models with C/O $<1$. This network is evolved using the density and temperature structure of the DALI model together with initial CO and \ce{H2O} abundances to match the C/O ratio. All chemical networks are evaluated at 5~Myr, the approximate age of the HD~100546 star \citep{Arun2019, Pineda2019}.

\subsection{Raytracing}
The synthetic ALMA observations are produced using the raytracer in DALI. The level populations are calculated without assuming local thermodynamical equilibrium (LTE) using the collisional rate coefficients for \ce{^12CO}, \ce{^13CO}, \ce{C^18O}, \ce{C^17O}, C, HCN, CN, \ce{C2H}, NO, and \ce{HCO+} in the LAMDA database\footnote{\url{https://home.strw.leidenuniv.nl/~moldata/}} \citep{Launay1977, Johnson1987, Roueff1990, Schroder1991, Staemmler1991, Schoier2005, Dumouchel2010, Lique2010, Yang2010, Spielfiedel2012, Denis2020, Ben2021}. An overview of the investigated transitions is presented in Table~\ref{tab:trans}. The observed \ce{C2H} and NO data cover two respectively three lines that are blended with themselves. Therefore, these two and three transitions were raytraced and added assuming the emission is optically thin. Finally, all synthetic image cubes are continuum subtracted and convolved to a beam matching that of the observations of that particular species (see Table~\ref{tab:trans}).

\section{Results} \label{sec:results}

Thermo-chemical models are not expected to reproduce the observed line emission within the uncertainty of the data as the predicted molecular line emission depends both on the abundance of a certain molecule as well as the temperature of the emitting layer. The parameter studies of the DALI thermo-chemical code show a factor of $\sim$2 difference in the CO line flux due to e.g. uncertainties in reaction rates in the chemical networks \citep{Bruderer2012, Bruderer2013} and a somewhat larger factor is expected for molecules other than CO.
Therefore, a model is expected to reproduce the data within a factor of $\sim 2-3$ in general \citep{Kama2016, Sturm2023}. In addition, a thermochemical model reproducing the observed line intensities within this factor is not unique and degeneracies in the model exist due to e.g., differences in the physical and thus thermal and chemical structure of the model. 
Instead of matching the observed line intensities within the uncertainty of the data, we look for trends in the abundances and line intensities for a range of key parameters such as gap depth and chemical composition. As the intensity of the continuum and CO isotopologues are less uncertain in the model than the intensity of [C~{\sc I}], HCN, CN, \ce{C2H}, NO, and \ce{HCO+}, the fiducial model aims to reproduce the dust and the CO isotopologue emission.
Section~\ref{sec:methods_fid} presents fiducial model reproducing the CO isotopologue and continuum observations and the subsequent sections present the trends seen in HCN, CN, \ce{C2H}, NO, \ce{HCO+}, and [C~{\sc I}] in the fiducial model and variations on the fiducial models.

\subsection{The fiducial model} \label{sec:methods_fid}

The key feature of the emission in the HD~100546 disk is that all molecular rings are co-spatial with the dust rings, whereas the CO isotopologue emission appears smooth. The only structure seen in the CO isotopologue emission is a central gas cavity and a weak shelf of \ce{C^18O} and \ce{C^17O} emission extending from 150~au to 250~au (see insert in the bottom middle and right corner of Fig.~\ref{fig:fid_dust_CO}). Therefore, the aim of the fiducial model is not to fit the CO isotopologues perfectly. Instead, the aim of the fiducial model is to provide a basis that roughly fits the continuum and the CO isotopologues and explore under what conditions the molecular rings coincide with those seen in the continuum emission. The observations of the outer molecular rings are all consistent with emission originating from close to the midplane \citep{Boothinprep}. Therefore, the fiducial model has a low flaring index of $\psi = 0$ and a low scale height of $h_c = 0.1$ at the characteristic radius. This flat disk model roughly reproduces the observed emitting heights in the HD~100546 disk as presented in Appendix~\ref{app:COemission_and_heights} and in \citet{Law2022, Stapper2023}; W\"{o}lfer et al. subm. A comparison of our fiducial model with those presented in \citet{Pirovano2022} and \citet{Keyte2023} is shown in Fig.~\ref{fig:dali_structure_lit}, where the main differences are that the model presented in this work is significantly less flared than those presented in \citet{Pirovano2022} and \citet{Keyte2023}, and has a gas cavity at $< 15$~au that is not present in \citet{Keyte2023}. An overview of the parameters of the fiducial model is presented in Table~\ref{tab:dali}.

The gas and dust density, temperature, and UV field in this model are presented in Fig.~\ref{fig:dali_2D_basic_fid}. 
The midplane gas density drops down to a low value of $10^7$~cm$^{-3}$ at 250~au, which is too low for the gas and dust to be thermally coupled in the midplane. Therefore, the gas temperature at these locations is colder than that of the dust. The small dust follows the gas outside the gas cavity at 15~au. The large dust is only seen in the two dust rings at $20-40$~au and $175-200$~au. A steep drop in the UV field is seen at $z/r = 0.2$. This layer also roughly separates the warm surface layer of the disk from the layers below 50~K closer to the midplane. This flat model intercepts less UV radiation from the central star than more flared models, which causes a small amount of CO freeze-out in the outer dust ring.

The resulting model emission of the mm-dust and the CO isotopologues is presented in Fig.~\ref{fig:fid_dust_CO} (orange lines). The top row presents the fits of the fiducial model to the dusty inner disk and the dust ring from $20-40$~au seen in the high resolution continuum (left panel) as well as the much weaker dust ring at $175-250$~au seen at moderate spatial resolution (middle panel; \citealt{Pineda2019, Boothinprep}). The weak outer continuum ring is not seen in the left panel due to the high spatial resolution and small maximum recoverable scale (26~au in radius) of that dataset.

The gas density structure is improved by including a 15~au gas cavity that is seen in the high spatial resolution $J=2-1$ transition of \ce{^12CO}, \ce{^13CO}, and \ce{C^18O} presented in the middle and bottom left panels in Fig.~\ref{fig:fid_dust_CO} (\citealt{Perez2020}; W\"{o}lfer et al. subm.). To match these data, the gas density inside the gas cavity is lowered by a factor of $10^{-5}$ and between the gas (15~au) and dust cavity (20~au) by a factor of $10^{-2}$. Outside the gas cavity, the \ce{^13CO}, \ce{C^18O}, and \ce{C^17O} emission matches that of the observations (\citealt{Perez2020, Boothinprep}; W\"{o}lfer et al. subm.), with an especially close match being achieved for the \ce{^13CO} and \ce{C^17O} emission. The ring at 200~au in the modelled \ce{C^18O} and \ce{C^17O} emission is due to a very low \ce{C^18O} and \ce{C^17O} column density just inside this ring as the column density becomes too low for self-shielding to be efficient (see Fig.~\ref{fig:N_COisos}). The \ce{^12CO} $J=3-2$ and $J=7-6$ emission presented in \citet{Boothinprep} and W\"{o}lfer et al. subm. is underproduced by at most a factor of $\sim 2$ likely due to the gas temperature that is slightly too low in this flat model. However, the main focus of the fiducial model is on the rarest CO isotopologues (\ce{C^18O} and \ce{C^17O}), the dust rings and the molecular line emission.

The lack of mm-dust between the two dust rings in the HD~100546 disk may be due to a massive planet carving a gap in the disk \citep{Pinilla2015, Fedele2021, Pyerin2021}. Therefore, the gas density in this region may be lower than in a full disk model. This is explored in Fig.~\ref{fig:dali_dgas_gap} and we find that CO isotopologue emission is consistent with a shallow gas gap where the gas is depleted by at most one order of magnitude, assuming that the gas and dust gap have the same widths.

  \begin{figure*}[ht!]
   \centering
   \includegraphics[width=\textwidth]{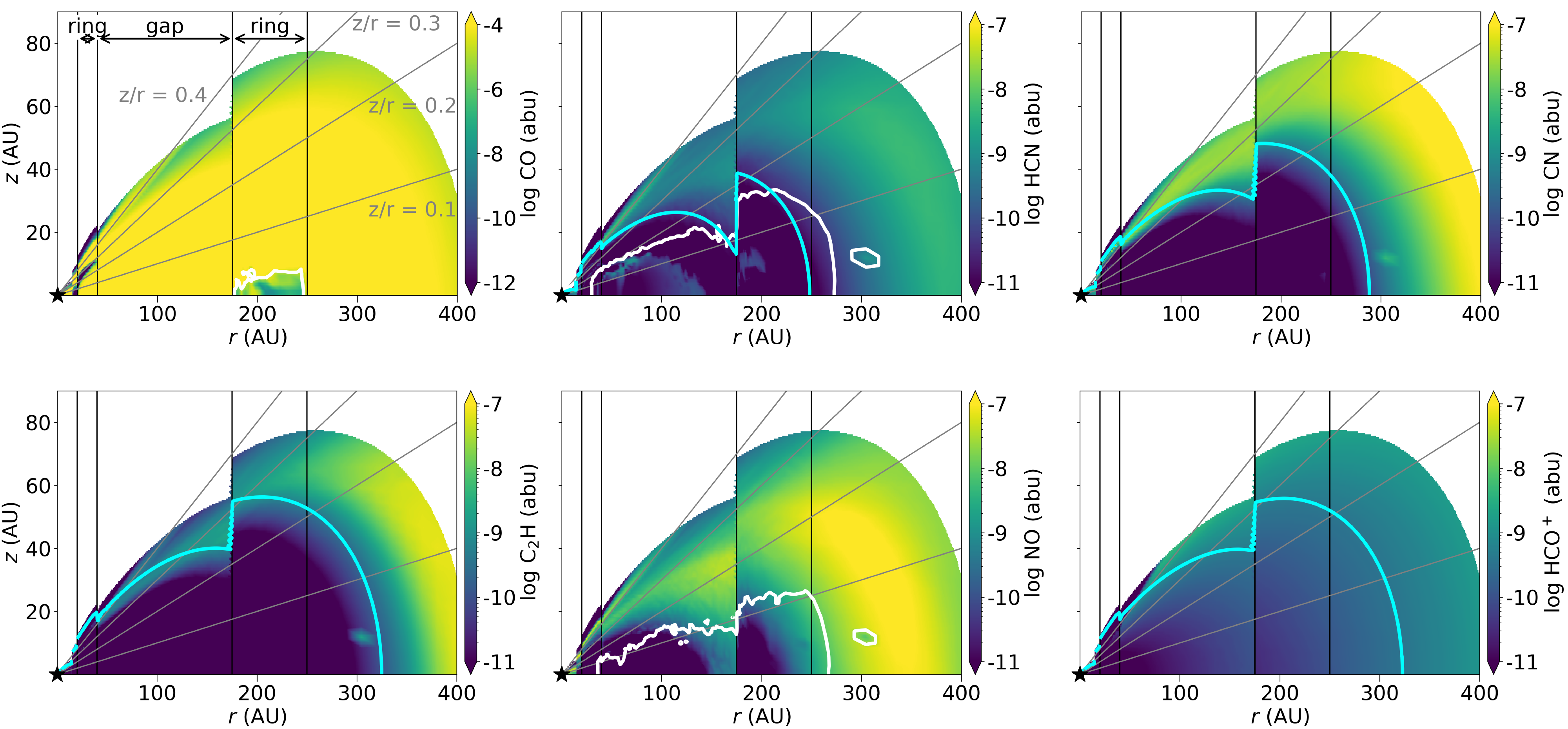}
      \caption{2D abundance maps of CO, HCN, CN, \ce{C2H}, NO, and \ce{HCO+} in the fiducial model. Note that the CO abundance is shown on a different colorscale. The white contour indicates the snowline of CO, HCN, and NO in their respective channels. The blue contour shows the critical density assuming a temperature of 50~K. Only the regions with a gas number density above $10^5$~cm$^{-3}$ are shown. As the CO and NO critical densities are lower than this, those contours are not shown in the respective panels.}
         \label{fig:dali_2D_mol_fid}
   \end{figure*}

\subsection{2D abundance maps}

The 2D abundance structures of the molecules of interest, HCN, CN, \ce{C2H}, NO, and \ce{HCO+}, are presented in Fig.~\ref{fig:dali_2D_mol_fid} together with the CO abundance for reference. The abundances are calculated with respect to the total hydrogen density: $n_{\rm h} = n(\mathrm{H}) + 2\times n(\mathrm{H_2})$. All of these molecules except CO follow the same global morphology: a surface layer with a high abundance that extends from the inner dust ring or the gap out to the outer disk at 400~au where the layer moves down to the midplane. Additionally, HCN and CO are abundant at lower disk layers and the NO and \ce{HCO+} layers extend to the disk midplane from $\sim150$~au outward. The CO is abundant throughout the entire disk except in the surface layers where CO is photodissociated and in the midplane in the outer dust ring where CO is frozen out onto the dust grains. 

HCN, CN, \ce{C2H}, and NO are sensitive to the UV field. The UV radiation in the surface layer forms vibrationally excited molecular hydrogen, \ce{H2^*}, whose abundance is calculated using a two-level approximation in the chemical network with far-UV photons pumping \ce{H2} to \ce{H2^*} and collisional decay and spontaneous radiative decay transforming \ce{H2^*} to \ce{H2} \citep[for details see][]{Bruderer2012}. \ce{H_2^*} then reacts with N to form NH which then reacts with \ce{C+} to form \ce{CN+} that is converted to CN and HCN. Additionally, a pathway through CH and N to form CN is also seen in this region. As the UV field only penetrates the surface layers of the disk, the HCN, CN, and \ce{C2H} are also constrained to this layer and thus reside in a layer where the number density of molecular hydrogen is below the respective critical densities (cyan contours in Fig.~\ref{fig:dali_2D_mol_fid}). The HCN molecule is not only abundant in a surface layer at $z/r > 0.2$; a secondary layer formed through different chemical reactions exists around a height of $z/r=0.1$. Here, the HCN is mainly formed through a reaction of atomic carbon with \ce{H2} to form \ce{CH2} through radiative association which then reacts with O and N to form CO and HCN. As this layer is below the HCN snow surface (white contour in Fig.~\ref{fig:dali_2D_basic_fid}), this HCN quickly freezes-out.

Somewhat deeper in the disk at $z/r=0.2$ NO is abundant. The NO is mainly formed through the reaction of NH with O with a minor contribution of the reactions of OH and N ($<10\%$). The NO layer lies somewhat deeper in the disk than that of HCN, CN, and \ce{C2H} due to the destruction of NO by atomic carbon and nitrogen that (re)forms CO, CN, and \ce{N2}. At lower layers below $z/r = 0.1$, NO quickly freezes-out onto the dust grains as the NO snowline is located at a scale height of 0.1 in the dust gap and the outer dust ring. The thin NO layer at $z/r =0.4$ is due to the photodissociation of water forming OH (see also \citealt{Leemker2023}).

   \begin{figure*}[ht!]
   \centering
   \includegraphics[width=\hsize]{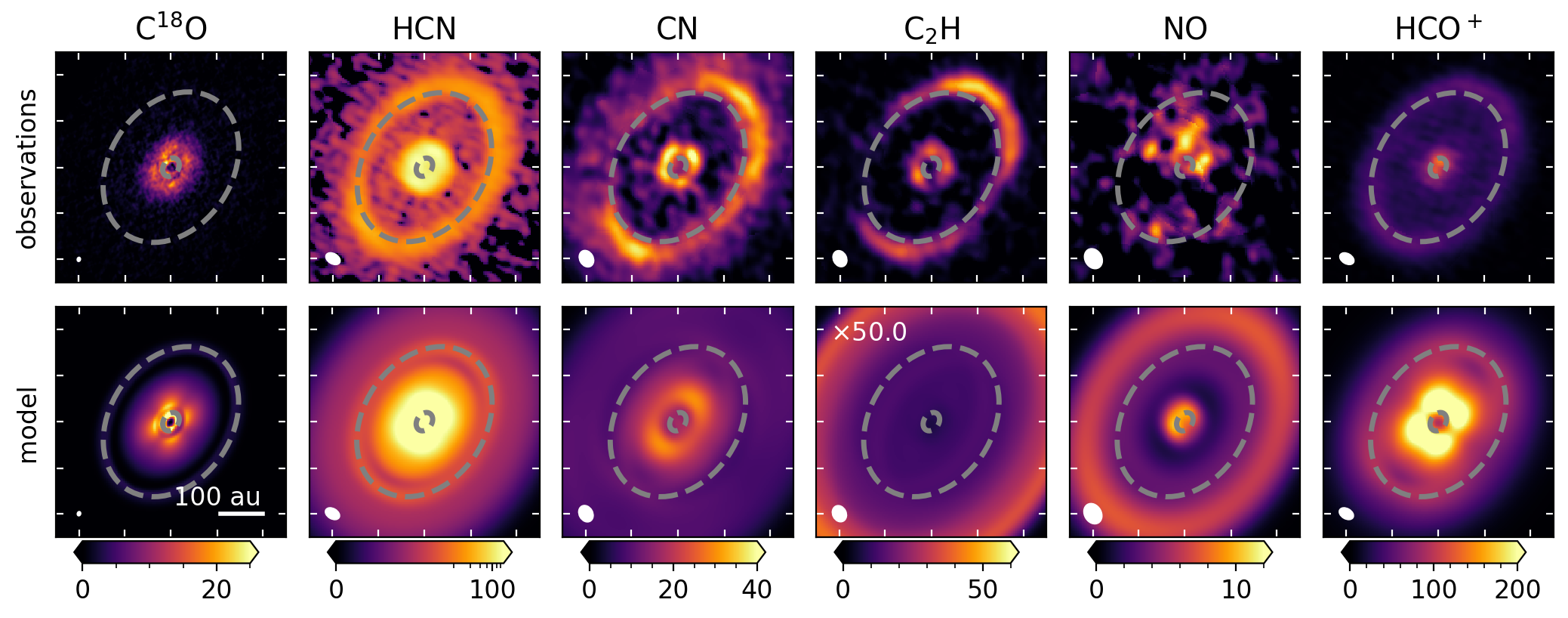}
      \caption{Integrated intensity (moment 0) maps of \ce{C^18O}, HCN, CN, \ce{C2H}, NO, and \ce{HCO+} emission in the HD~100546 disk. The top row presents the observations (\citealt{Perez2020, Boothinprep}; W\"{o}lfer et al. subm.) and the bottom row the predictions of the fiducial model. All intensities are in mJy~km~s$^{-1}$~beam$^{-1}$. The dotted grey ellipses indicate the dust rings seen in the high spatial resolution (inner ring) and in the moderate spatial resolution (outer ring) 0.9~mm continuum. The tickmarks on the panels are spaced by $1"$, the horizontal bar in the bottom left panel indicates a 100~au scale. The HCN is shown on a logarithmic scale to highlight the weak outer ring and the modelled \ce{C2H} emission has been multiplied by a factor of 50. The beam of each of the observations is indicated with the white ellipse in the bottom left corner. }
         \label{fig:mom0_fid}
   \end{figure*}

Finally, the \ce{HCO+} abundance outside 15~au is driven by the cosmic ray ionization rate enhancing the \ce{H3+} abundance. In addition to ionisation, the \ce{HCO+} ion traces the water snowline: a high \ce{HCO+} abundance is only seen outside the water snowline where water cannot destroy \ce{HCO+} \citep{Phillips1992, Bergin1998}. The water snowline in this model is located at the outer edge of the gas cavity at 15~au. Therefore, the \ce{HCO+} abundance at larger radii scales as $\sqrt{\zeta_{\rm c.r.}/n_{\rm H_2}}$ with a weak dependence on temperature \citep{Leemker2021}. This may overpredict the \ce{HCO+} abundance outside the inner dust ring in this disk as cold, gas-phase water is seen outside $\sim40$~au in this disk and water photodesorption is not included in the small chemical network \citep{vanDishoeck2021, Pirovano2022}. 

Altogether, the 2D abundance structures of HCN, CN, \ce{C2H}, NO, and \ce{HCO+} are dominated by a layer with a high abundance centred at $z/r \sim 0.2-0.3$. This layer is located at or above the critical density of HCN, CN, and \ce{C2H}, therefore, these molecules may not be in LTE and the high excitation lines may be less populated whereas the ground level may be more populated than what is expected in LTE. This layer with a high abundance of $10^{-7}-10^{-9}$ is similar to the observed emitting layer of CO and higher than the layer of $z/r \leq 0.1$ that the HCN, CN, \ce{C2H}, and \ce{HCO+} observations suggest \citep{Boothinprep}. The high abundance layer moves down slightly inside the gas gap due to the lower gas density, but apart from this, no clear jump in the abundance at the location of the outer dust ring is seen.

\subsection{Integrated intensity maps} \label{sec:rad_prof}

The integrated intensity (moment 0) maps of the data and the fiducial model are presented in Fig.~\ref{fig:mom0_fid}. The modelled \ce{C^18O} emission reproduces the observations within a factor of two for most disk regions with the outer ring in \ce{C^18O} being more prominent in the model than in the data (see also Fig.~\ref{fig:fid_dust_CO}, bottom row). For the other molecules on the other hand, the observed intensities or the double ringed profile are not reproduced. Only the modelled HCN and \ce{HCO+} emission do show a hint of an outer ring at the location of the second dust ring whereas the \ce{C2H} and NO show a (weak) ring outside the outer dust ring at larger distances than observed. 
The lack of molecular rings at the location of the outer dust ring in the models is not due to optically thick dust hiding the line emission at that location. This is because the modelled continuum emission in the outer disk is optically thin and therefore does not absorb line emission that is emitted from the disk midplane at that radial location. In addition, the HCN, CN, \ce{C2H}, NO, and \ce{HCO+} are most abundant in an elevated layer, well above the heights where the large dust is abundant (see Fig.~\ref{fig:dali_2D_basic_fid} and Fig.~\ref{fig:dali_2D_mol_fid}). The intensity of weak ring at $\sim 375$~au seen in the \ce{C2H} model predictions  has an intensity that is similar to the 2$\sigma$ level of the observations. Therefore, this ring may be hidden in the noise in the observations.  

The inner ring that is observed in HCN and CN is reproduced by the model within a factor of $\sim 2$ but the observed \ce{C2H} ring at 40~au is not seen in the model. The modelled \ce{HCO+} emission is ring-shaped in the inner disk with a ring at 60~au, whereas the data are centrally peaked at the spatial resolution of the observations. The intensity in the modelled ring is two times higher than what is observed at the central location. 

The model predicts outer rings in HCN and \ce{HCO+} that are only a factor of $\sim 3$ weaker and a factor of two brighter than what is observed, respectively. However, the model does not reproduce the outer ring in \ce{C2H} as the modelled ring is two orders of magnitude weaker and 170~au further out than seen in the data. In addition, the predicted \ce{C2H} intensity falls below the sensitivity of the data out to 300~au. The observed outer ring in CN is not recovered by the model and the modelled NO emission is mostly seen at the inner dust ring at $\sim 20$~au and outside the outer dust ring at $280$~au whereas the observations show some emission between the two dust rings and possibly at the location of the outer dust ring. In summary, Fig.~\ref{fig:mom0_fid} shows that despite a good fit to the continuum and CO isotopologue emission, the model does not reproduce the observations of molecules other than CO isotopologues, especially for CN, \ce{C2H}, and NO. The chemical network used to calculate these abundances reproduces the CN, HCN, HNC, and \ce{C2H} emission in other protoplanetary disks within a factor of a few in general \citep[e.g.,][]{Cazzoletti2018, Visser2018, Miotello2019, Long2021}. Therefore, the difference between the model predictions and the observations in the HD~100546 disk are likely due to the physical or chemical structure of the disk rather than uncertainties in the chemical network.

   \begin{figure*}
   \centering
   \includegraphics[width=\hsize]{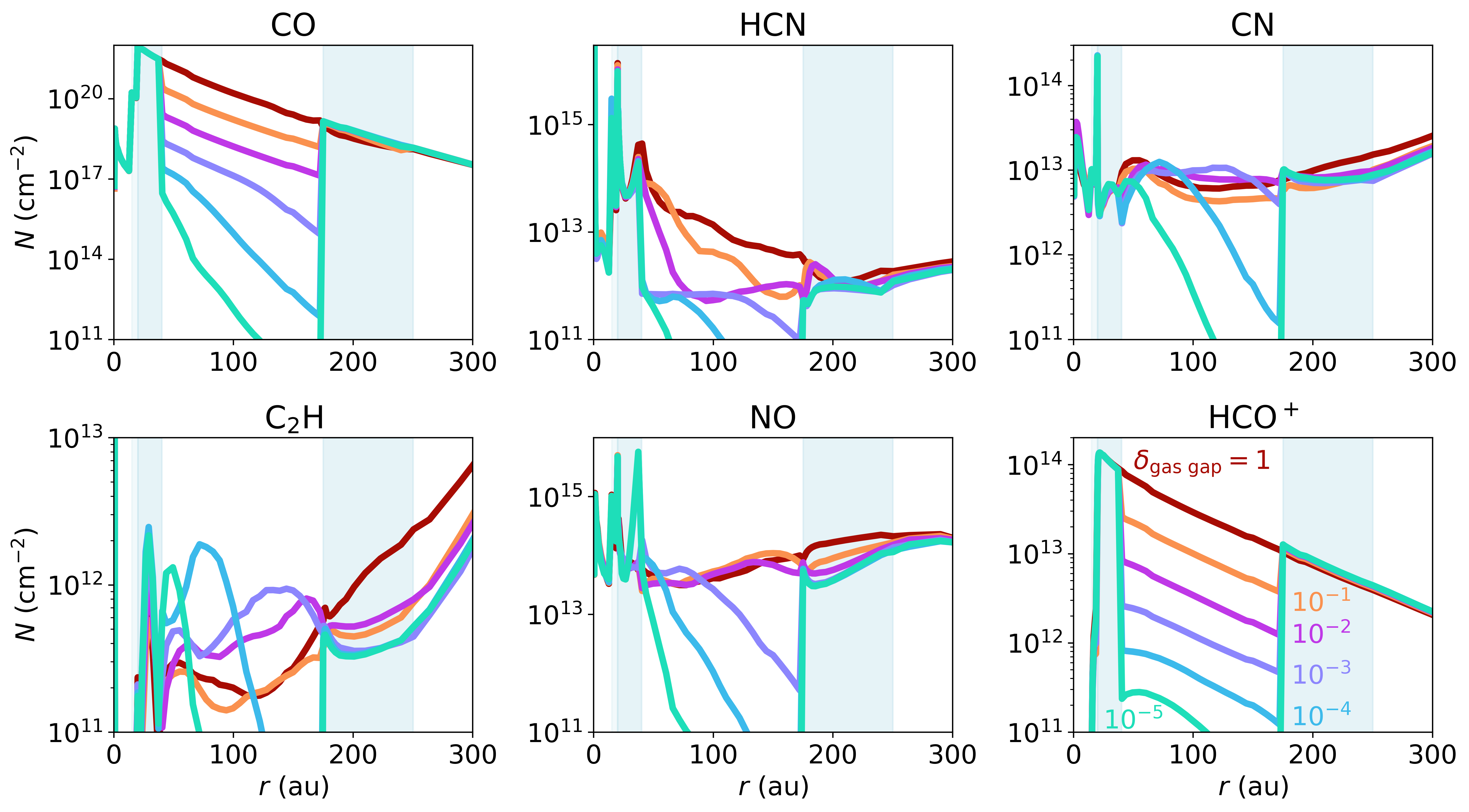}
      \caption{Column densities of CO, HCN, CN, \ce{C2H}, NO, and \ce{HCO+} in the fiducial model with a C/O ratio of 0.4 and a gas gap depth of 1 (no gap; dark red), $10^{-1}$ (fiducial model; orange), $10^{-2}$ (pink), $10^{-3}$ (purple), $10^{-4}$ (blue), and $10^{-5}$ (very deep gap; light blue). The small dust density is dropped by the same factor in the gap as in the gas.}
         \label{fig:dali_N_dgap}
   \end{figure*}

\subsection{The effect of a gas gap of at least one order of magnitude}
A possible solution to create double rings in the molecular line emission is to induce a deep gas gap between the two dust rings. The \ce{^12CO} and \ce{^13CO} emission do not show any evidence of a gas gap deeper than the fiducial depth of a factor of 10, neither does the \ce{C^17O} where only a shoulder of emission extending out to $200-300$~au is observed (see Fig.~\ref{fig:dali_dgas_gap}). Still, a depletion of gas could affect the molecular line emission. Therefore, the effect of a deeper gas gap is investigated in this subsection. The small dust is assumed to follow the drop in the gas density.

\subsubsection{Column densities}

The effect of the depth of the gap between 40 and 175~au on the column density is presented in Fig.~\ref{fig:dali_N_dgap} where the column densities inside 250~au are smoothed in log space with a savgol filter in scipy over 10 points and a 3rd order polynomial to remove any artificial spikes due to low photon statistics in the model. The CO column density directly traces the drop in the gas density for gaps up to three orders of magnitude in gas density. For deeper gaps, CO photodissociation becomes important and the CO column density drops even faster than the gas density. 

The \ce{HCO+} column density directly follows that of the gas where a drop of two orders of magnitude in gas column density results in a one order of magnitude drop in the \ce{HCO+} column density. The reason for this is that the \ce{HCO+} number density outside the water snowline follows $n({\rm HCO^+}) \propto \sqrt{\zeta_{\rm c.r.}\times n({\rm H_2})}$, with $\zeta_{\rm c.r.}$ the cosmic ray ionisation rate and $n({\rm H_2})$ the density of molecular hydrogen \citep[see eq. B.6 in][]{Leemker2021}. Therefore, the \ce{HCO+} column density in the gap is directly related to that of the gas. 

   \begin{figure*}[ht!]
   \centering
   \includegraphics[width=\hsize]{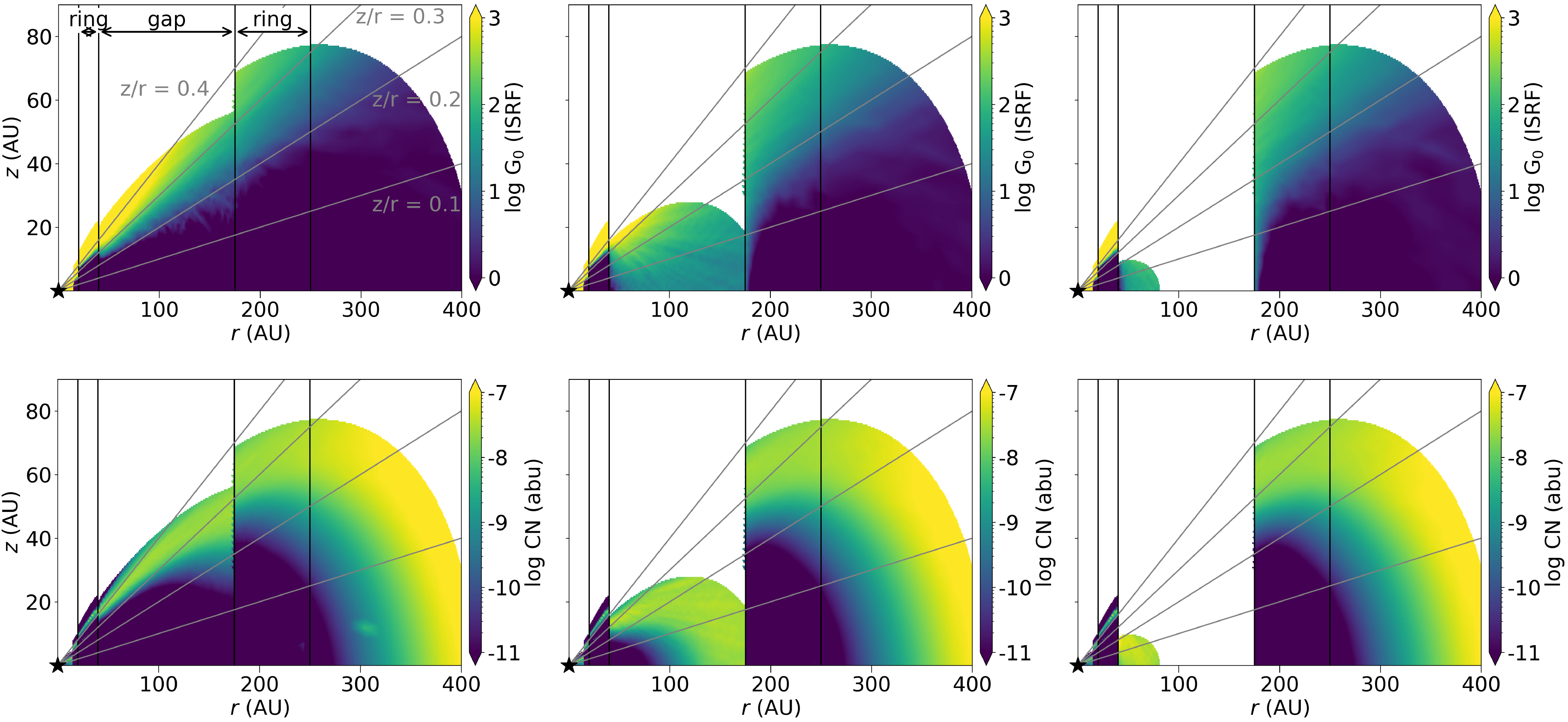}
      \caption{The UV field in units of G$_{\rm 0}$ (top) and CN (bottom) abundance for different gas gap depth. From left to right: $\delta_{\rm gas\ gap} = 10^{-1}$, $10^{-3}$, and $10^{-5}$. 
      Only the regions with a gas number density above $10^5$~cm$^{-3}$ are shown. }
         \label{fig:dali_2D_dgap}
   \end{figure*}

The CN, \ce{C2H}, and NO show a different pattern in their column densities than CO, but a similar pattern to each other, when the gap depth is increased. For shallow to somewhat deep gas gaps ($\delta_{\rm gas\ gap} \sim 1-10^{-2}$), their column densities are independent of the gap depth within a factor of a few. For very deep gas gaps ($\delta_{\rm gas\ gap} \sim 10^{-4}-10^{-5}$), a steep decrease in the column density is seen in the outer parts of the gap. The sharp peaks in the HCN, CN, \ce{C2H}, and NO column densities at the inner ring and inner region of the gap (15, 20, and 40~au) may be sensitive to the precise conditions in this region.

The behaviour in the gap can be understood from the 2D UV field and CN abundance which are presented in Fig.~\ref{fig:dali_2D_dgap}. Shown are the results for a shallow gap ($10^{-1}$) and two deep gaps ($10^{-3}$ and $10^{-5}$), in the left column, a shallow gas gap is present, but the CN is only abundant in a layer high above the midplane at $z/r =0.3$ in the gap. Changing the amount of gas in this region only shifts this layer up or down, but it does not affect the column density greatly as the midplane abundance of CN is low. For a deep gas gap (middle and right columns), the UV field is more intense in the gap than in the model with a shallow gas gap with an intensity up to 100~G$_{\rm 0}$ in the midplane in the gap. As CN is sensitive to the UV field, its abundance increases in the gap. The resulting CN column density is an interplay of an increase in the abundance that is most prominent at small radii, and a decrease in the gas column density that is most effective at large radii. This causes a peak in the CN column density that moves to smaller radii of 75 and 50~au with deeper gaps of $10^{-4}$ and $10^{-5}$, respectively. As \ce{C2H} and NO are also sensitive to the UV field, their abundances and column densities show the same qualitative behaviour. Note the increasing \ce{C2H} column density at radii larger than 200~au. Especially for \ce{C2H}, the inward travelling peak in the column density is clearly visible at 160~au, 140~au, 70~au, and 50~au for $\delta_{\rm gas\ gap} = 10^{-2}, 10^{-3}, 10^{-4}, 10^{-5}$, respectively.

The HCN column density always decreases or stays constant with deeper gaps due to the moderate HCN abundance in the midplane together with the high abundance in the surface layer. The component in the midplane slowly decreases up to a gap depth of $10^{-2}$, lowering the HCN column density simultaneously. For deeper gaps, the midplane component of HCN vanishes and the HCN column density steeply drops with gas density, similar to the pattern seen for CN, \ce{C2H}, and NO.

   \begin{figure*}[p]
   \centering
   \includegraphics[width=\hsize]{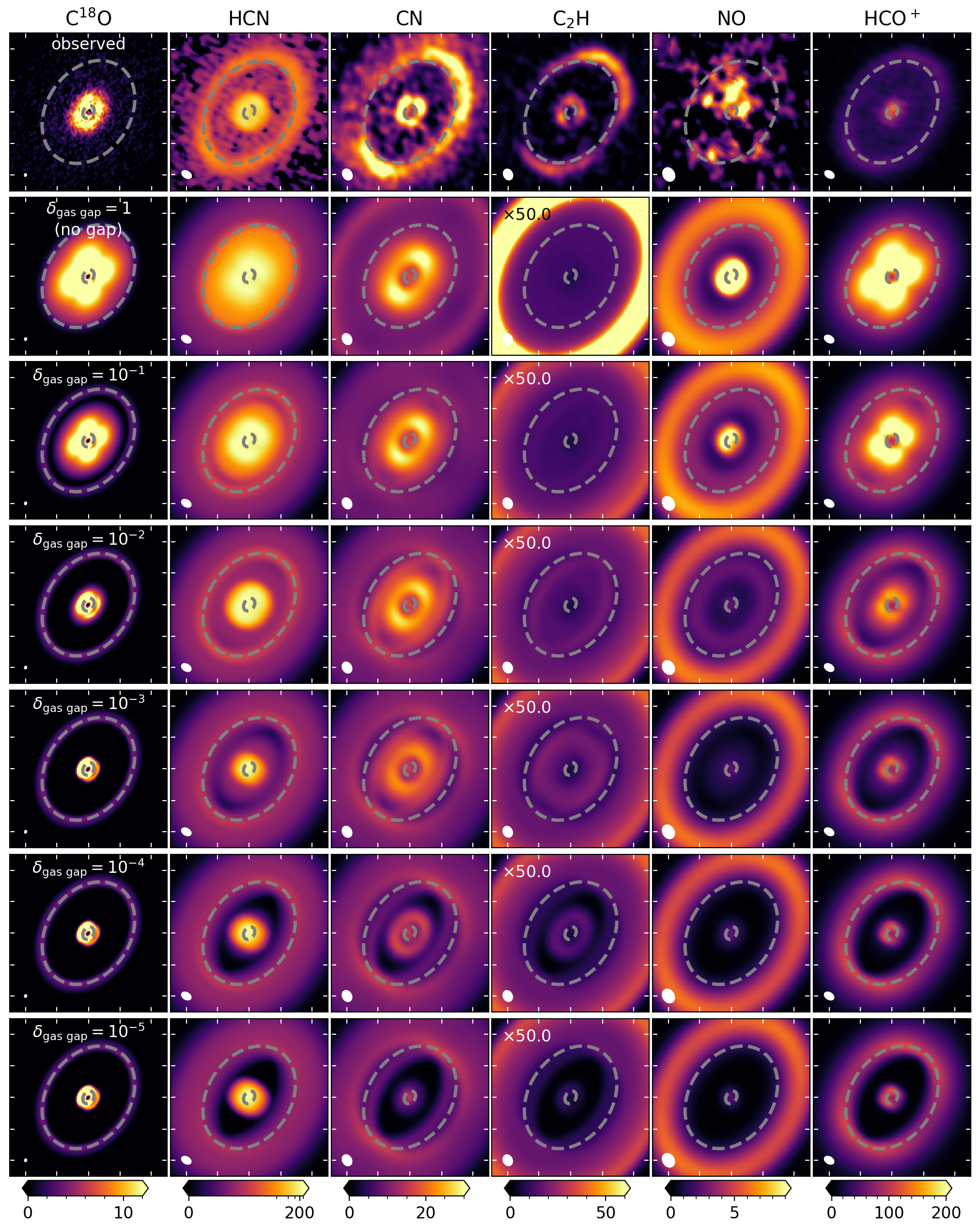}
      \caption{Integrated intensity (moment 0) maps in mJy~km~s$^{-1}$~beam$^{-1}$ for \ce{C^18O}, HCN, CN, \ce{C2H}, NO, and \ce{HCO+} from left to right. The different rows present the results for different depth of the gas gap with the gap depth ($\delta_{\rm gas\ gap}$) indicated in the top of the left column. The HCN emission is shown on a logarithmic scale to highlight the weak outer ring. The beam is indicated in the bottom left corner of each panel and the tickmarks on the axes are spaced 1" apart. 
      }
         \label{fig:dali_mom0_dgap}
   \end{figure*}

\subsubsection{Emission maps}

The integrated intensity (moment 0) maps for different gas gap depths are presented in Fig.~\ref{fig:dali_mom0_dgap} and the azimuthally averaged radial profiles are presented in Fig.~\ref{fig:dali_azi_avg_dgap}. The fiducial model is presented on the third row and the moment maps are identical to those presented in Fig.~\ref{fig:mom0_fid} but on a different colorscale. In general, the molecular line emission shows ringed emission for all models that have some degree of gas depletion between the dust rings ($\delta_{\rm gas\ gap} = 10^{-1}-10^{-5}$). However, double rings are only seen for the models with a very deep gas gap ($\delta_{\rm gas\ gap} = 10^{-4}-10^{-5}$). These rings are typically located outside the dust rings instead of at or just beyond the dust as seen in the observations. For example, the NO emission shows a very clear double ring for these models, but the outer ring is located at $\sim$280~au where no NO is seen in the data. The \ce{C2H} emission shows a ring of emission inside the gas gap that moves inwards with increasing gap depths following the corresponding \ce{C2H} column density. This effect is also seen in the CN and NO emission, although it is less pronounced. The NO emission just outside the inner ring is sensitive to a sharp peak in the column density and may be very sensitive to the conditions in this region. The remaining three molecules, \ce{C^18O}, HCN and \ce{HCO+} all show double ringed emission profiles for all models with some gas depletion. 

The model with a deep gas gap of $10^{-4}$ times less gas best reproduces the double ringed nature of the HCN, CN, and \ce{C2H} observations, whereas the model with a factor of ten more gas ($\delta_{\rm gas\ gap} = 10^{-3}$) reproduces the diffuse \ce{HCO+} emission between the dust rings better. The first model reproduces the morphology of the molecular line emission just outside the inner dust ring for CN and \ce{C2H} while also predicting a reasonable intensity for HCN. Additionally, this model reproduces the presence of an outer ring at $200-250$~au seen in HCN, CN, and \ce{HCO+}. The CO isotopologue emission on the other hand is best reproduced by a disk with a shallow gap of only $10^{-1}$ times less gas as the \ce{C^18O} emission in the gap is directly related to the drop in gas density.
In summary, there is not a single gas gap depth that can explain the observed emission of all molecules simultaneously.

\subsubsection{Molecular column densities and line ratios}

Ratios of column densities can give additional insights  into the underlying processes as they are less sensitive to the absolute model results. The column density ratios of \ce{HCO+}/CO, \ce{C2H}/CO, CN/HCN, and CN/NO are presented in Fig.~\ref{fig:Nratios_diff_gaps} for a model without a gas gap and models with gap depths up to $10^{-5}$. The first three ratios increase with radius outside the outer gas cavity in the full gas disk model, whereas the CN/NO ratio first increases inside the dust gap, then slowly decreases up to the outer dust ring and then somewhat increases again. This is caused by the different radii where the CN and NO column densities peak in the disk. The \ce{HCO+}/CO and \ce{C2H}/CO ratios increase due to the increasing column density in \ce{HCO+} and \ce{C2H} as that of CO smoothly decreases with radius. Finally, the increase in the CN/HCN ratio is driven by the decreasing column density of HCN. Comparing these ratios to ratios of \ce{HCO+}/CO and \ce{C2H}/CO in the inner and outer dust ring presented in \citep{Boothinprep} shows that the fiducial model is consistent with the \ce{HCO+}/CO ratio but that the \ce{C2H}/CO ratio likely requires a deep gas gap of multiple orders of magnitude or an overall increase in the \ce{C2H} column density.

When a gas gap is introduced, the ratios presented in Fig.~\ref{fig:Nratios_diff_gaps} all increase compared to the model without a gas gap except for a disk with a gap of $10^{-1}$ and $10^{-2}$ times less gas where the CN/NO ratio stays roughly constant in the gap. The same trend is seen in the ratios for models with $\delta_{\rm gas\ gap} = 1-10^{-2}$ of the corresponding molecular lines presented in Fig.~\ref{fig:Iratios_diff_gaps}. For deeper gas gaps, the line ratios become insensitive to the gas gap depths due to the finite resolution of the observations. This is because part of the bright emission in the rings leaks into the gap regions, causing these regions to be dominated by the rings for deep gas gaps. Altogether, the column density ratios \ce{HCO+}/CO, \ce{C2H}/CO, and CN/HCN of show a strong dependence on the gas gap depth that is much weaker in the ratios of the emission lines. 

\subsection{Different C/O ratios}

   \begin{figure*}[ht!]
   \centering
   \includegraphics[width=\hsize]{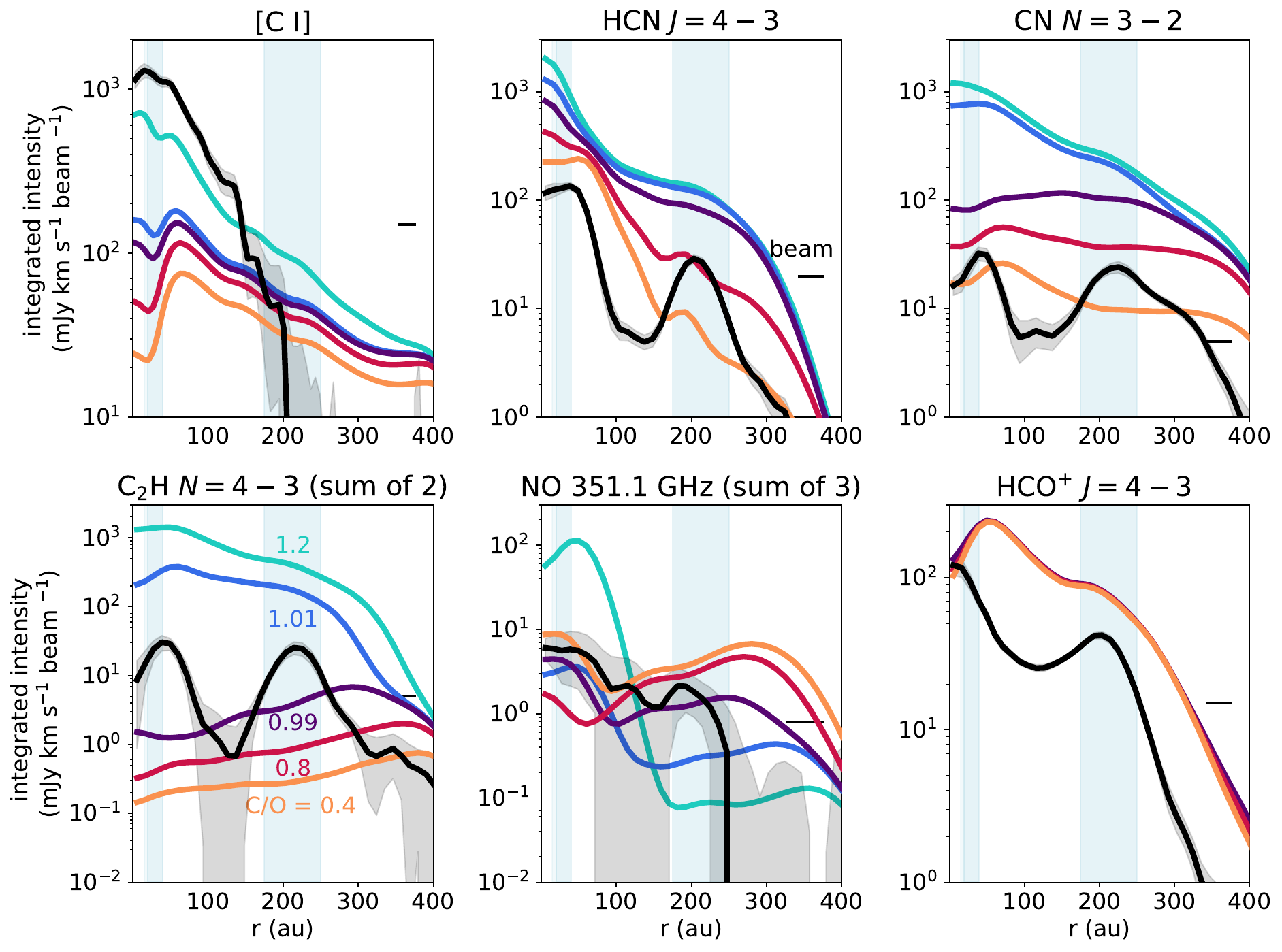}
      \caption{Azimuthally averaged integrated intensity of the molecular line emission for different C/O ratios in a model with a gas gap depth of $10^{-1}$. The black line indicated the observed intensities.  }
         \label{fig:I_dC_O}
   \end{figure*}

\subsubsection{Column densities and emission profiles}

The observed \ce{C2H} emission is two orders of magnitude brighter than predicted by the fiducial model. Observations in other disks show that the \ce{C2H} emission becomes significantly brighter when the C/O ratio of the gas exceeds 1 \citep[e.g.,][]{Bergin2016, Miotello2019, Bosman2021C2H, Cridland2023}. In this section, the effect of an increased C/O ratio is explored in the fiducial model with a gas gap of $\delta_{\rm gas\ gap} = 10^{-1}$. The 2D abundance maps for \ce{C2H} and NO are presented in Fig.~\ref{fig:dali_2D_dC_O}. The 2D abundance maps in the left column are the maps for the fiducial model and are the same as those presented in Fig.~\ref{fig:dali_2D_mol_fid}. The middle and right columns show these maps for a C/O ratio just below one at 0.99 and just above one at 1.01, while the C/H ratio is the same as in the fiducial model. As the C/O ratio increases, the layer where \ce{C2H} is abundant moves down from $z/r = 0.3$ for C/O = 0.4 to $z/r = 0.2$ for C/O = 0.99 and then it jumps down to $z/r = 0.15$ for a C/O ratio that is only 0.02 higher at 1.01. The peak \ce{C2H} abundance in this layer also jumps from $\sim 2\times 10^{-9}$ to $4\times 10^{-9}$ and to $7\times 10^{-8}$ for C/O = 0.4, 0.99, and 1.01, respectively. A similar trend is seen for HCN, CN, and C. 

The NO abundance shows the opposite behaviour as NO is expected to become less abundant when the C/O ratio is high. This is also seen in the 2D abundance maps as the NO abundance quickly decreases if the C/O is increased from 0.4 to 0.99 by removing water. As the C/O ratio further increases, the NO layer splits into two regions around $z/r = 0.05$ below its snow surface and one at $z/r = 0.2-0.3$. \ce{HCO+} and CO both carry one oxygen atom and one carbon atom and are therefore not as sensitive to the C/O ratio. The column densities for these molecules for C/O ratios between 0.4 and 1.2 are presented in Fig.~\ref{fig:N_dC_O} in Appendix~\ref{app:ratios_C_O}. 

The step-like behaviour of the CN and \ce{C2H} abundance is also seen in the azimuthally averaged integrated intensity profiles presented in Fig.~\ref{fig:I_dC_O}. If the C/O ratio increases from 0.99 to 1.01, the CN and \ce{C2H} integrated intensities increase by a factor of a few to one order of magnitude and by two orders of magnitude in the gap, respectively. The C/O ratio does change the integrated intensity across the disk, but the locations of the CN and \ce{C2H} peaks remain around those seen in the data except for a large peak at $300-400$~au seen in \ce{C2H}. Interestingly, the observed \ce{C2H} emission falls between that predicted by models with C/O = 0.99 and 1.01, possibly indicating a C/O ratio in the rings that is very close to 1.00. 

The HCN emission only shows a double peak profile for the models with C/O = 0.4 and 0.8 as the column density inside the gap increases faster with increasing C/O than that just outside the gap. The model with a C/O of 0.8 matches the intensity in the outer dust ring particularly well whereas in the model with C/O = 0.4 matches the inner ring within a factor of $\sim2$. 

The top left panel in Fig.~\ref{fig:I_dC_O} presents the azimuthally averaged radial profile of the observed [C~{\sc I}] emission in the HD~100546 disk. Unlike the HCN, CN, \ce{C2H}, and \ce{HCO+} emission, the [C~{\sc I}] does not show a double peaked profile at the dust rings. This could partially be due to the small maximum recoverable scale of 2" ($\sim 108$~au in radius), filtering the emission at larger distances. Therefore, the model may overpredict the observed line intensity outside this radius. Inside $\sim108$~au, the fiducial model underpredicts the emission of atomic carbon by an order of magnitude. The only model that recovers the observed [C~{\sc I}] intensity up to 150~au is a model with a larger C/O of 1.2.

The effect of the C/O ratio on the NO emission is that the NO peak moves inwards from 300~au to $\sim 40$~au. The high intensity at $\sim 40$~au is due to some additional NO at $z/r = 0.1-0.2$ in the models with C/O $>1$. The intensity in the outer disk decreases with increasing C/O ratio due to the lower abundance. 

In summary, increasing the C/O ratio increases the predicted intensity for [C~{\sc I}], HCN, CN, \ce{C2H}, and NO inside 100~au and for NO it decreases the intensity outside 100~au. The observed emission lines are reproduced by different C/O ratios for each molecule. For example, the region inside 150~au is best reproduced by C/O = 1.2 for [C~{\sc I}], but this overpredicts the CN and \ce{C2H} emission by more than an order of magnitude. At the location of the outer dust ring, the cyanides (HCN and CN) and NO are compatible with a C/O ratio of 0.8, but the \ce{C2H} requires a C/O ratio very close to 1 (0.99 < C/O < 1.01).

\subsubsection{Column density and emission line ratios}

To eliminate some of the uncertainties of comparing absolute modelled intensities to the observed ones, the column density and line ratios of \ce{HCO+}/\ce{C^17O}, \ce{C2H}/\ce{C^17O}, CN/HCN, and CN/NO are presented in Appendix.~\ref{app:ratios_C_O}. The latter three column density ratios all increase in most disk regions with increasing C/O due to the enhancement of \ce{C2H} and CN, and the decrease of NO. In particular, the CN/NO ratio outside 100~au and the \ce{C2H}/\ce{C^17O} in the entire disk are sensitive to small changes in the C/O. Comparing the observed emission line ratios to the modelled ones shows that the C/O needs to be very close to 1.00 as the \ce{C2H}/\ce{C^17O} ratio for C/O = 0.99 underpredicts the observed value and that of C/O = 1.01 overpredicts the observations. The CN/NO line ratio only traces the C/O ratio for $r>175$~au, and is therefore not a good tracer of the C/O in the disk.

\subsection{Background UV}

One striking feature that most of the models presented here show is that the CN, \ce{C2H}, and NO emission are too bright outside the outermost dust ring. This raises the question if the modelled external UV field is too strong as all of these molecules at least partially trace the UV field. The models assume that the external UV field is 1~$G_0$, the mean UV field in the interstellar medium. However, HD~100546 is an isolated star with a tenuous envelope that may result in a lower external UV field \citep{Grady2001}. 

\begin{figure*}[ht!]
   \centering
   \includegraphics[width=\hsize]{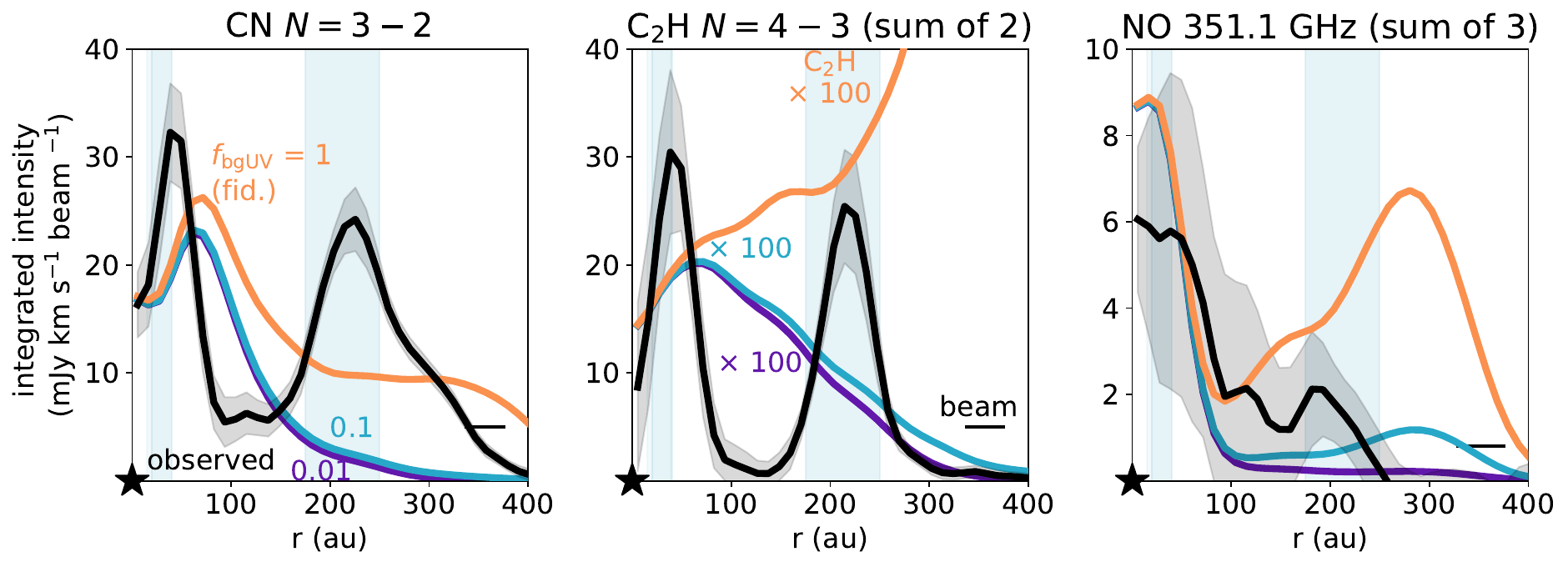}
      \caption{Azimuthally averaged integrated intensity of the CN, \ce{C2H}, and NO  for a C/O = 0.4 and for different background UV radiation fields of 1~G$_{\rm 0}$ (orange), 0.1~G$_{\rm 0}$ (blue), and 0.01~G$_{\rm 0}$ (purple). The modeled \ce{C2H} emission is increased by a factor of 100 to show it on the same scale as the observations shown in black. }
         \label{fig:I_dbgUV}
   \end{figure*}

The effect of a lower UV field on the CN, \ce{C2H}, and NO column density is presented in Fig.~\ref{fig:N_dbgUV}. Lowering the background UV field by a factor of ten decreases the CN and \ce{C2H} column densities by that same factor at $300-400$~au. The NO decreases by a factor of 5 from $200-350$~au. This is also reflected in the azimuthally averaged radial intensity profiles presented in Fig.~\ref{fig:I_dbgUV}. The shelf of CN and the peak of \ce{C2H} at 375~au are not visible for the lowered background radiation fields, creating a ring of emission at the location of the outer dust ring. The effect of a lower background UV field on the [C~{\sc I}], HCN, and \ce{HCO+} emission is small. 

In particular, a lower background UV radiation field more closely reproduces the double peak profile of the \ce{C2H} emission, with both peaks having a similar intensity. The \ce{C2H} intensities in these models are a factor of 200 lower than what is observed for C/O = 0.4, thus a lower background UV field better reproduces the morphology of the \ce{C2H} emission but it does not solve the difference between the intensity seen in the observations and in the models. 
Additionally, the NO emission in the outer disk is not overpredicted if the background radiation field is lowered. Overall, a lower background UV field seems to improve the comparison with observations outside the outer dust ring.

   \begin{figure*}[ht!]
   \centering
   \includegraphics[width=\hsize]{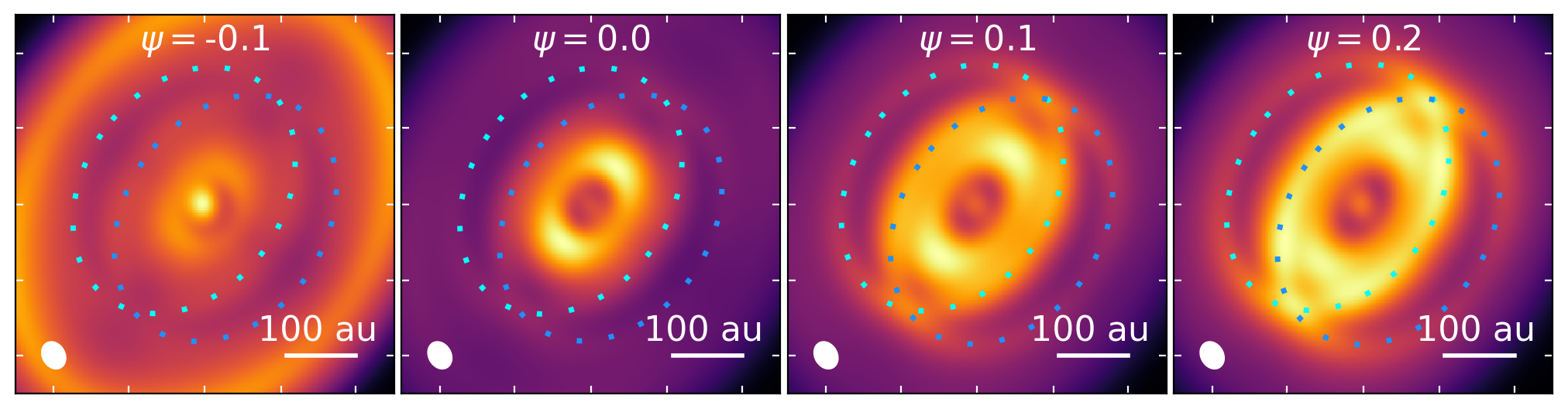}
      \caption{Moment 0 maps of the CN emission for models with different flaring indices $\psi = -0.1$ to $+0.2$. The light and dark blue ellipses indicate the projected location at which the modelled CN number density peaks at the inner rim of the outer dust ring, which corresponds to $z = 53-68$~au. The panels are shown on an individual color scale for clarity.  }
         \label{fig:mom0_psi}
   \end{figure*}

\subsection{Flaring}

The observations of HCN, CN, \ce{C2H}, and \ce{HCO+} show two emission rings that are concentric around the position of the star. This indicates that these molecules emit from a layer close to the disk midplane ($z/r\leq 0.1$, \citealt{Boothinprep}) as an elevated layer would shift these rings w.r.t. the position of the star. Such an offset is indeed seen in the models for CN. Figure~\ref{fig:mom0_psi} presents this with the light and dark blue ellipses for different flaring indices of the disk. The ellipses indicate the projected location of the peak CN number density at the inner rim of the outer dust ring. For the two flattest models ($\psi = -0.1$ and 0) the CN at $r=175$~au peaks at $z=55-53$~au whereas for the two more flared models ($\psi = 0.1$ and 0.2), the CN peaks at a height of $66-68$~au at $r=176$~au. These ellipses trace two rings in the CN emission in the moment 0 maps. These two rings increase and move slightly further apart with increasing flaring index, tracing the height where the CN is abundant. As this height only changes by $\sim10$~au between a very flared and a close to flat disk, flaring alone cannot explain the observed emitting heights. However, the emission in the moment~0 map not only depends on the vertical structure but also on the density and temperature. The main difference between the models with $\psi = 0-0.2$ is the expected CN intensity at the location of the blue ellipses at the outer ring. Even though the flared disk model produces an outer ring that has a similar intensity to the inner ring at the major axis, we emphasize that the elevated emitting surface of CN creates projection effects that are not seen in the HD~100546 observations.  Observations of \ce{C^18O} in the Elias~2-27 disk show a ring that is close to being centered at the position of the star. Yet the \ce{C^18O} channel maps show that the emitting surface is flared, with $z/r \sim 0.2$ \citep{PanequeCarreno2021}. Therefore, deriving the emitting height from integrated intensity maps may hide part of the true vertical structure of the disk. The column density and radial emission profiles for other molecules for different flaring indices are presented in Appendix~\ref{app:flaring}.

\section{Discussion} \label{sec:discussion}

Correlations between the radial structures in the dust and gas may point to a common origin such as massive planets carving deep gas gaps in disks. Additionally, such correlations between gas and dust structures can point to the sublimation of ices as seen in the IRS~48 disk \citep{vanderMarel2021irs48, Booth2021irs48, Brunken2022, Leemker2023}. The observations in the HD~100546 disk show a clear correlation between the dust rings and the HCN, CN, \ce{C2H}, and \ce{HCO+} emission suggesting a direct relation between the structures in the gas and the dust. However, the rare CO isotopologue emission, tracing the CO column density, does not show the same double ringed structure. In this work, we use thermochemical models to investigate under what conditions molecular rings are co-spatial with the dust rings and which physical conditions are important for the ring locations.

\subsection{Radial structures }

\subsubsection{CO isotopologues, \ce{HCO+}, and HCN}

The thermochemical models spanning a range of gas gap depth ($\delta_{\rm gas\ gap} = 1-10^{-5}$) show that two molecules directly respond to a the depletion of gas in the gap: CO isotopologues and \ce{HCO+}.
The CO isotopologue emission does not show a deep gap in the observations that is seen in models with a deep gas gap. Therefore, the HD~100546 disk is consistent with a shallow gas gap ($\delta_{\rm gas\ gap} = 10^{-1}$).
A clear double ring is seen in \ce{HCO+} model emission if the modelled gas gap is at least two orders of magnitude deep (or one order of magnitude if the disk is viewed face-on due to projection effects, see Fig.~\ref{fig:dali_azi_avg_dgap_face-on} in Appendix~\ref{app:molecules_radial}). A weak outer ring is seen in the HCN model predictions. Interestingly, despite the correlations seen in the models, no general correlation is observed in the MAPS sources between the HCN and \ce{HCO+} rings and the dust, and only a weak ($\sim 3\sigma$) correlation is seen between the CO isotopologue emission and the dust \citep{Law2021, Jiang2022}. 

The trend in the \ce{HCO+} and CO isotopologues across the gas gap is also seen in other modelling works. 
Modelling of \ce{HCO+} across a gas gap outside the CO snowline predicts that the ratio of the \ce{HCO+} to the total gas column density increases when a gas gap is introduced \citep{Alarcon2020, Smirnov-Pinchukov2020}. A difference in the various model predictions is seen for the model with only a dust gap and no gas gap ($\delta_{\rm gas\ gap} =1$). This model presented in \citet{Smirnov-Pinchukov2020} shows an increased \ce{HCO+}/CO column density ratio in the dust gap whereas this is not seen in the models presented in this work (Fig.~\ref{fig:Nratios_diff_gaps}). A possible cause for this is that the gap in \citet{Smirnov-Pinchukov2020} is located outside the CO snowline. Part of the CO ice desorbs and reacts to form \ce{HCO+}, boosting the \ce{HCO+} column density in their model. This is not seen in our models as CO is abundant throughout the disk, with only some CO freeze-out in the outer dust ring below a height of $\sim 8$~au, instead of gas-phase CO being abundant only in the dust gap as in the models presented in \citet{Smirnov-Pinchukov2020}.

\subsubsection{CN, \ce{C2H}, and NO}

The other modelled molecules follow a more intricate relation with the gas gap depth. 
For gaps up to approximately three orders of magnitude deep, the CN, \ce{C2H}, and NO rings peak at various radii both inside (CN, \ce{C2H}) and outside (NO) the dust rings and do in general not coincide with the dust rings or each other. Additionally, the inner and outer rings of a particular molecule are predicted to have very different intensities. This is similar to the results of \citet{Cazzoletti2018} and of the MAPS program which observed CN and \ce{C2H} and found no general and significant correlation between structures in the dust and in the molecular line emission in four out of five disks, despite some overlapping features \citep{Law2021, Jiang2022}. The CN/HCN peaks in some of the dust gaps in the disks in the MAPS sample, similar to the model predictions for the HD~100546 disk \citep{Bergner2021}. For the models with a gap of $10^{-1}-10^{-3}$, the \ce{C2H} is predicted to peak inside the dust gap, similar to what is seen in the MWC~480 disk, the only MAPS disk where gas structures other than CO isotopologue emission do correlate with the dust, and in the AS~209 disk \citep{Alarcon2021}. The location of the inner CN ring in the models presented in our work (see Fig.~\ref{fig:dali_N_dgap}) is only sensitive to gas gap depth if the gap is depleted by five orders of magnitude in gas. Therefore, chemistry is the main driver of the ring locations for shallow gas gaps. 

In summary, the radial locations of the rings seen in the HD~100546 disk cannot be explained by a single gas gap depth as the CO isotopologue emission, HCN, and \ce{HCO+} are best reproduced by a shallow gap ($\delta_{\rm gas\ gap} = 10^{-1}-10^{-2}$), whereas the double rings in CN and \ce{C2H} are only reproduced for a deep gas gap of $\delta_{\rm gas\ gap} = 10^{-4}-10^{-5}$, partially due to the vertical structure of the disk. 
Unlike HD~100546, the PDS~70 and IRS~48 disks both have a deep gas cavity just inside the molecular rings, forcing the line emission to be co-spatial with that of the dust. Therefore, only if a deep physical gap is present in a disk, for example due to a massive planet, the molecular rings are likely also physical rather than chemical rings.

\subsection{Chemistry in the rings and gap}
In addition to physical rings in the disk density structure, chemical rings can also affect the observed emission. The chemistry across the gap may vary due to e.g., thermal and non-thermal sublimation of the ice mantles on the dust and UV shielding by small dust grains. 

\subsubsection{\ce{HCO+} tracing a low gas-phase water abundance}
The abundance of \ce{HCO+} and gas-phase water anti-correlate in protoplanetary disks as the main destruction pathway of \ce{HCO+} is through gas-phase water \citep{Phillips1992, Bergin1998, Leemker2021}.
Water ice is likely present on the dust grains in the gap between the two dust rings in the HD~100546 disk \citep{Honda2016}. Photodesorption of this water ice could explain the gas-phase water seen from just outside the inner dust ring (starting from $35-40$~au) to $\sim300-400$~au \citep{Du2017, vanDishoeck2021}. The outer radius is not well constrained as the \textit{Herschel} HIFI line profile that is used to constrain the emitting location is most sensitive to the inner radius rather than the outer radius (priv. comm. with M. R. Hogerheijde). Additionally, the line profile does not show evidence for a jump in the \ce{H2O} abundance at the location of the outer dust ring. Modelling of the cold water lines shows that gas-phase water has an abundance of $3\times 10^{-9}$ in the outer disk ($\sim 40-300$~au, \citealt{Pirovano2022}). Photodesorption of water ice is not included in the network used to predict the \ce{HCO+} abundance in this work. As \ce{HCO+} is destroyed by gas-phase water, the predicted \ce{HCO+} abundance and emission are thus overestimated in the gas gap and also at larger radii out to 300~au. The difference between the modelled \ce{HCO+} emission in the gap and the observed intensity could thus be arising from this effect. As the gas gap does not contain large grains, water ice must be present on small grains.

Thermal desorption of water is predicted at the water snowline which is located at 15~au, at the outer edge of the gas cavity in the models. The modelled \ce{HCO+} abundance anti-correlates with that of gas-phase water, with a small increase of less than a factor of ten in the \ce{HCO+} abundance in the gap (see Fig.~\ref{fig:dali_2D_mol_fid}). The modelled peak location of the \ce{HCO+} emission is 40~au whereas the data shows centrally peaked emission at a spatial resolution of 31~au, indicating that the water snowline is located inside the central beam of the observations. Additionally, \citet{Pirovano2022} found that the abundance of gas-phase water inside the HD~100546 dust cavity is low at $< 10^{-9}$. Therefore, the water snowline in the HD~100546 disk is likely unresolved at the 31~au resolution of the \ce{HCO+} observations and thus located inside 15~au.

\subsubsection{Grain surface chemistry}
Another mechanism that could lead to the rings seen in HCN, CN, \ce{C2H}, and \ce{HCO+} and the detection of NO is grain surface chemistry. The temperature structure of the fiducial model suggests that the outer dust ring is sufficiently cold to have some HCN, CN, \ce{C2H}, NO, and CO freeze-out (see Fig.~\ref{fig:dali_2D_mol_fid}). 
Non-thermal desorption could somewhat enhance the gas-phase abundance of HCN and NO by $10^{-3}-10^{-4}$ times their ice abundance as both of these molecules have significant ice abundance ($>10^{-7}$) in that region. The high NO abundance in the IRS~48 disk may have an origin in the ice with \ce{N2O} or HNCO ices as possible parents \citep{Leemker2023}. \ce{C2H} and CN on the other hand do not have a high abundance in the ice and these emission rings thus require a different origin such as gas-phase formation or sublimation of larger molecules that form CN and \ce{C2H} after sublimation.

   \begin{figure*}[ht!]
   \centering
   \includegraphics[width=\hsize]{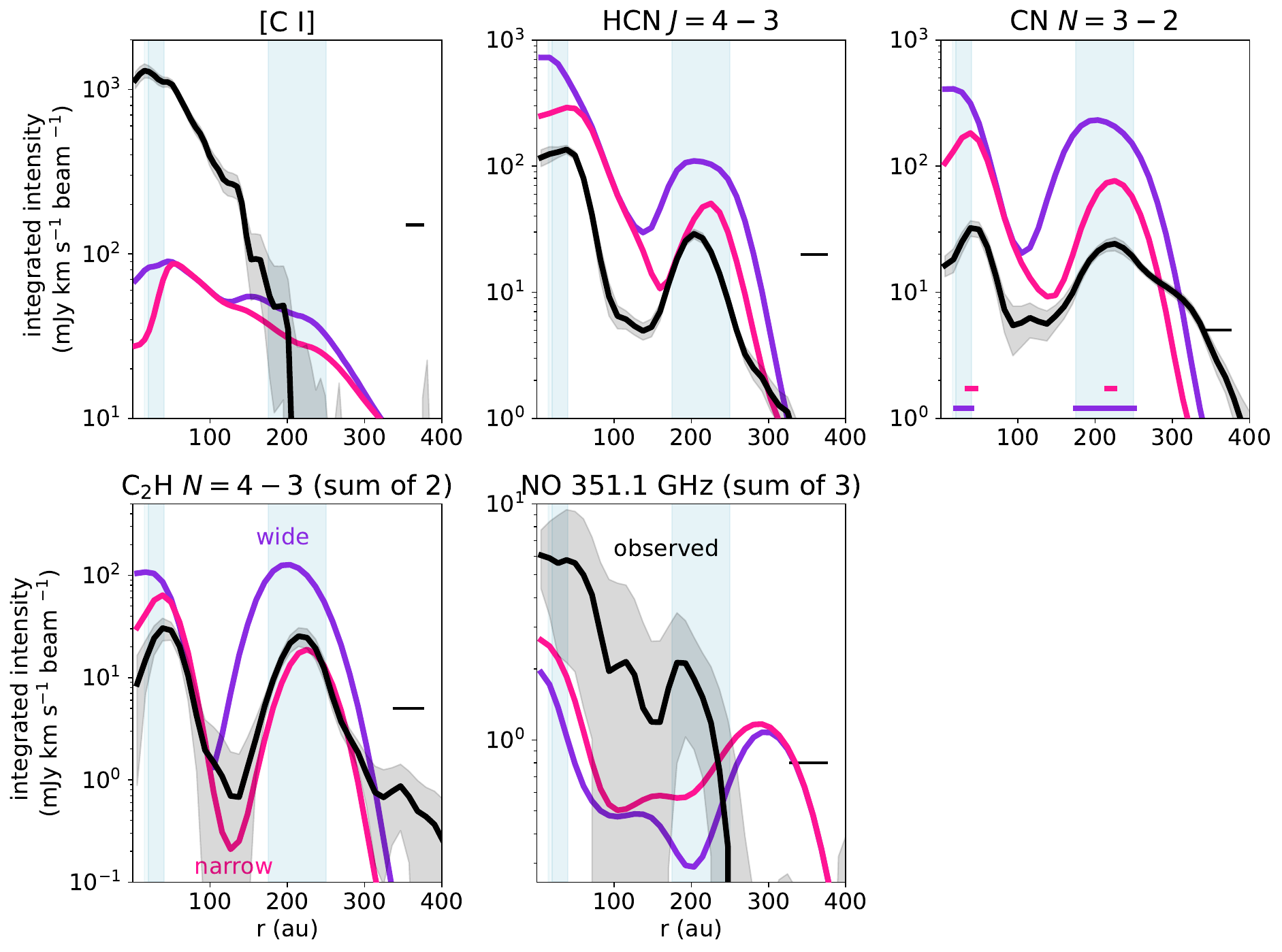}
      \caption{Azimuthally averaged radial profiles of [C~{\sc I}], HCN, CN, \ce{C2H}, and NO for models with high C/O ratio of 1.01 in the regions indicated by the pink and purple horizontal bars in the top right panel and the fiducial C/O ratio of 0.4 in the other disk regions. The background UV field is reduced by a factor of 10 in these models. The pink model has a high C/O in two 10~au wide rings at the observed ring location and the purple model has an increased C/O across the entire width of both dust rings. }
         \label{fig:radially_varying_C_O}
   \end{figure*}

\subsubsection{C/O ratio across the HD~100546 disk}
The sublimation of ices could also be reflected in the gas-phase C/O ratio as most ices are oxygen rich.
Of the ratios investigated in this work, the \ce{C2H}/\ce{C^17O} ratio is the most sensitive to the C/O ratio. The CN/NO ratio has been suggested as a tracer of the gas-phase C/O ratio \citep{HilyBlant2010, Daranlot2012, LeGal2014}. However, the models presented in this work show that the relation between CN/NO and C/O is more complicated.

Our models suggest that the C/O ratio in the dust rings needs to be elevated from 0.4 to $\sim $0.8-1 to reproduce the HCN, CN, and \ce{C2H} emission in the dust rings, while the C/O in the gas gap is roughly consistent with the fiducial value of 0.4. The C/O in the rings is a factor of $\sim 2$ higher than the C/O of 0.5 derived from CS and SO by \citet{Keyte2023}. In particular, the \ce{C2H} emission is very sensitive to the C/O ratio, consistent with what is found in earlier works \citep[e.g.,][]{Bergin2016, Miotello2019, Bosman2021C2H}. Additionally, the bright \ce{C2H} emission without having a deep gas gap points to an elevated C/O as also suggested in the AS~209 disk \citep{Alarcon2021}. The \ce{C2H} emission at the outer dust ring in HD~100546 could be due to freeze-out of oxygen bearing molecules in this region leading to a high C/O ratio \citep[][and Fig.~\ref{fig:dali_2D_mol_fid}]{Oberg2011, vanderMarel2021C_O}. Interestingly, \citet{Jiang2023} find that the modelled \ce{C2H} intensity peaks at or inside the ring of a gap carved by an accreting planet due to the sublimation of carbon-rich organic molecules. The rings in the HD~100546 disk are located just outside the dust rings but possibly a migrating planet could still heat the gas in this regions to create a ring in \ce{C2H}.

Two other mechanisms that could lead to an elevated C/O ratio are the liberation of carbon from ice and grains and destruction of CO by \ce{He+} \citep[e.g.,][]{Bergin2014, Anderson2017, Klarmann2018, Wei2019, Bosman2021C2H}. The latter process is included in the chemical network. The dim \ce{C2H} emission could indicate that the cosmic ray ionisation rate or the X-ray flux are higher than the value of the X-ray flux adopted in this model. A more recent value of the X-ray flux measured by \citet{Skinner2020} is a factor of $\sim 4-$6 higher than the value reported in \citet{Stelzer2006} that is used in this work, however, \citet{Bruderer2012} found that the X-ray flux only affected the CO, [C~{\sc I}], [C~{\sc II}], and [O~{\sc I}] line fluxes by less than $\sim 5$\%. For the HCN, CN, \ce{C2H}, and NO emission, only the inner 100~au is affected by the increased X-ray luminosity and the increased X-ray luminosity does not affect the emission of [C~{\sc I}] by much (see Appendix~\ref{app:Lx}). The largest effect on the emission morphology is seen in the \ce{C2H} profile, where the increases X-ray flux creates a \ce{C2H} ring at 30~au, similar to the morphology of the observations. However, the increased X-ray luminosity also increases the emission of CN inside the ring at 40~au that is not seen in the observations.

The spatial overlap between the dust rings and the \ce{C2H} emission suggests that destruction of carbon grains could contribute to the elevated C/O ratio in the disk surface layers. This could enhance the emission of carbon bearing molecules that tend to be abundant at higher disk layers than NO. Therefore, a vertical gradient in the C/O ratio could help to reproduce the \ce{C2H}, [C~{\sc I}], and NO at the cost of CN and HCN. Instead of an enhancement in the C/H ratio, the O/H ratio could also be suppressed by, for example the freeze-out of oxygen bearing molecules. As the C/H and O/H ratios in the HD~100546 photosphere are consistent with the Solar values, the increase in the C/O ratio in the disk is likely local \citep{Kama2016}. Another explanation for the spatial overlap between molecular rings and the outer dust ring could be that the gas in the outer dust ring is second generation gas due to collisions of larger bodies, possibly indicating that the outer dust ring is a debris disk belt in a protoplanetary disk.

\subsubsection{\ce{HCO+}, CN, and HCN tracing radiation and gas gap depth}

\ce{HCO+} not only traces the desorption of water, it is also sensitive to the ionization in the disk \citep[e.g.,][]{Aikawa2001,  Cleeves2015twhya}. The observed column density ratios of \ce{HCO+}/CO in the MAPS disks increase with radius or stay roughly constant at $\sim 10^{-6}-10^{-4}$ \citep{Aikawa2021}. The general trend of an increasing ratio with radius is recovered by our models, but the observed values in MAPS require a gas gap depth of $\sim 10^{-2}-10^{-4}$. Comparing this to the HD~100546 disk shows that either the gas gap is shallow ($\delta_{\rm gas\ gap} = 1-10^{-1}$) for the outer half of the gap, or that the ionisation rate in the modelled gap is too high. 

The ratio of CN/HCN, tracing the UV indicates that the gap in HD~100546 needs to be $10^{-1}-10^{-2}$ deep (see Fig.~\ref{fig:Iratios_diff_gaps}). Radial variations in the CN/HCN column density ratio likely trace radial variations in the UV field due to changes in the dust density structure due to enhanced HCN photodissociation compared to CN photodissociation \citep{Bergner2021}. The models presented in this work show a similar trend where the CN/HCN column density ratio peaks inside the gas gap and decreases in the outer dust ring (see Fig.~\ref{fig:Nratios_diff_gaps}). Similarly, this ratio has a low value of a few times $10^{-2}- 10^{-1}$ in the inner dust ring where the small dust surface density is two orders of magnitude higher than that in the outer dust ring where the ratio is close to ten (see Fig.~\ref{fig:dali_structure}). This is similar to the observed value of 0.4 in the inner ring and $\sim 2$ in the outer dust ring \citep{Boothinprep}. The model without a gap in the gas and small dust grains but with a gap in the large dust grains does not show a peak in the CN/HCN column density ratio in the gap, indicating that this ratio is primarily sensitive to the small dust grains that have a larger opacity in the UV than the large dust grains.

In summary, the rings seen in the HD~100546 disk likely have a chemical origin as the physical rings in the gas density structure alone cannot explain the observed molecular rings. The C/O ratio in the rings needs to be elevated above the fiducial value of 0.4 to reproduce the observed intensity in the \ce{C2H}, CN, and [C~{\sc I}]. In addition, a decreased background UV radiation field is necessary to suppress the outer ring in \ce{C2H} and NO. Figure~\ref{fig:radially_varying_C_O} presents two models with a background UV field attenuated by a factor of 10 and where the C/O ratio is increased to 1.01 in two rings. In the purple model, the C/O ratio is increased across the entire width of the two dust rings ($20-40$~au and $175-250$~au). The molecular rings predicted by this model are wider than what is seen in the HD~100546 disk observations. Therefore, a second model is presented in pink. In this model, the C/O ratio is only increased in two 10~au wide rings from 35 to 45~au and from 215 to 225~au to match the location of the observed emission ring. This model reproduces the double ring seen in HCN, CN, and \ce{C2H} and the increase in the NO emission towards the position of the star. Additionally, this model does reproduce the intensity of these molecules within a factor of $\sim 5$ and only the [C~{\sc I}] is underpredicted by approximately one order of magnitude inside 150~au. The width of the rings with a high C/O ratio may be larger than 10~au as the spatial resolution of the observations is $\sim 0\farcs2 - 0\farcs5$, still the purple model shows that these rings cannot be as wide as the dust rings.

\section{Conclusions} \label{sec:conclusions}
In this paper, we use an exploratory grid of thermochemical models to explore trends in gapped disks, and to reproduce the observed co-spatial rings of mm-dust and molecular emissions in the HD100546 disk \citep{Boothinprep}.
We used the thermochemical code DALI to explore the radial location and intensity of molecular rings seen in HCN, CN, \ce{C2H}, NO, and \ce{HCO+}. In particular, we investigated under which conditions these molecular rings overlap with those seen in the continuum. The fiducial model provides a good fit to the continuum emission and the emission of CO isotopologues with a deep dust and gas cavity inside 20~au and 15~au, respectively, and a deep dust gap but a shallow gas gap from $40-175$~au. The abundance of the other molecules of interest, HCN, CN, \ce{C2H}, and NO are predicted with a dedicated nitrogen network and that of \ce{HCO+} is predicted with the network presented in \citet{Leemker2021}. We find that the following trends are predicted for disks in general:

\begin{itemize}

\item In general, molecular rings and dust rings do not coincide unless gaps are very deep.

\item The modelled emission of CO isotopologues and \ce{HCO+} directly trace the gaps and gap depths in the model gas density structure whereas the other molecules do not.

\item The thermochemical model does predict molecular rings in e.g., HCN, CN, \ce{C2H}, and NO but these generally do not follow the dust rings.

\item For gas gaps of one to approximately three orders of magnitude deep, the column density of UV tracers, e.g., CN, \ce{C2H}, and NO, show only a minor sensitivity to the gas gap depth and are thus not good tracers of this. For very deep gas gaps ($\delta_{\rm gas\ gap} \sim 10^{-4}-10^{-5}$) these molecular rings emit at a similar location as the dust rings.

\item A globally elevated C/O ratio mainly affects the total modelled intensity of C- or O-bearing molecules such as [C~{\sc I}], HCN, CN, \ce{C2H}, and NO but not the radial location of the emission rings. However, the CN/NO ratio is not a good tracer of the disk C/O ratio inside 175~au. The modelled CN and \ce{C2H} column density and emission increase by roughly one to two orders of magnitude, respectively, when the bulk model C/O ratio is increased from 0.99 to 1.01.

\item A lower background UV radiation field reduces the predicted intensity and/ or moves the ring seen in UV tracers to smaller radii.

\item The flaring index of the disk only has a minor effect on the height of the layer where a high CN abundance is seen, despite models with lower flaring indices being more flat.

\end{itemize}

In addition, we draw the following conclusions specifically for the HD~100546 disk:

\begin{itemize}

\item The dust and CO isotopologue gas emission in the HD~100546 disk are well reproduced by a model with a deep central gas ($r<15$~au) and dust cavity ($r<20$~au) and a deep dust gap but a shallow gas gap ($\delta_{\rm gas\ gap} = 10^{-1}$) from $40-175$~au. A drop in gas density by more than one order of magnitude is excluded by the CO isotopologue data.

\item The best-fit C/O for [C~{\sc I}], CN, and is an elevated C/O of  $\sim 0.8-1.2$. In particular, the \ce{C2H} requires a C/O ratio very close to 1. The NO emission inside 250~au is reproduced by models with a C/O ratio less than 1.2. 

\item The models predict emission rings in CN, \ce{C2H}, and NO at $300-400$~au that are not seen in the observations. A lower background UV radiation field in the models reduces the column density and emission of CN, \ce{C2H}, and NO and it increases that of HCN in the outer disk ($r> 100$~au) to values close to those observed.

\item A radially varying C/O ratio with a C/O ratio of 1.01 in two narrow regions at 40 and 220~au where the observed molecular rings peak, and the fiducial C/O ratio of 0.4 in the other disk regions reproduces the double ringed morphology seen in the HCN, CN, and \ce{C2H} emission and it roughly reproduces the morphology of the NO. The [C~{\sc I}] emission is still underproduced by an order of magnitude.

\item To reconcile the observed ring locations and intensities with those predicted by models, a combination of a gas gap with a factor of 10 less gas, a low background UV field and an increased C/O ratio of 1.01 in a region narrower than the dust rings (at 40 and 220~au) are needed.

\end{itemize}

This work shows that thermochemical models do not predict molecular rings to be co-spatial with those seen in the dust in general. Moreover, these models indicate that if observations show multiple molecular rings that are co-spatial with the dust, the chemical composition in that region is likely different than in the other disk regions. The constraints from modelling multiple molecules across a gas gap provides a first step towards a full picture of the chemical makeup of the planet-forming material. Multi-line observations of multiple molecules at high spatial and spectral resolution and sensitivity are needed to better constrain the layer from which these molecules emit. These observations can then be used to test thermochemical models and derive the chemical composition of the planet-forming material. In particular, observations of [C~{\sc I}] in more disks are needed to determine whether the [C~{\sc I}] emission is indeed brighter than expected based on thermochemical models or that the difference is specific to the HD~100546 disk.

\begin{acknowledgements}
We thank the referee for the detailed comments. 
The authors would like to thank Luke Keyte and Dylan T. Natoewal for the useful discussions on the DALI modelling. In addition, Michiel Hogerheijde is thanked for the discussion on water in the HD~100546 disk.
This work is supported by grant 618.000.001 from the Dutch Research Council (NWO). 
Astrochemistry in Leiden is supported by funding from the European Research Council (ERC) under the European Union’s Horizon 2020 research and innovation programme (grant agreement No. 101019751 MOLDISK), by the Netherlands Research School for Astronomy (NOVA).

This paper makes use of the following ALMA projects: 2011.0.00863.S, 2015.1.00806.S, 2016.1.00344.S, 2018.1.00141.S, 2021.1.00738.S. ALMA is a partnership of ESO (representing its member states), NSF (USA) and NINS (Japan), together with NRC (Canada), MOST and ASIAA (Taiwan), and KASI (Republic of Korea), in cooperation with the Republic of Chile. The Joint ALMA Observatory is operated by ESO, AUI/NRAO and NAOJ. The National Radio Astronomy Observatory is a facility of the National Science Foundation operated under cooperative agreement by Associated Universities, Inc.

\end{acknowledgements}

\bibliographystyle{aa}
\bibliography{refs.bib}

\begin{appendix}

\section{Observed lines}
\begin{figure*}[b!]
   \centering
  \begin{subfigure}{0.67\columnwidth}
  \centering
  \includegraphics[width=1\linewidth]{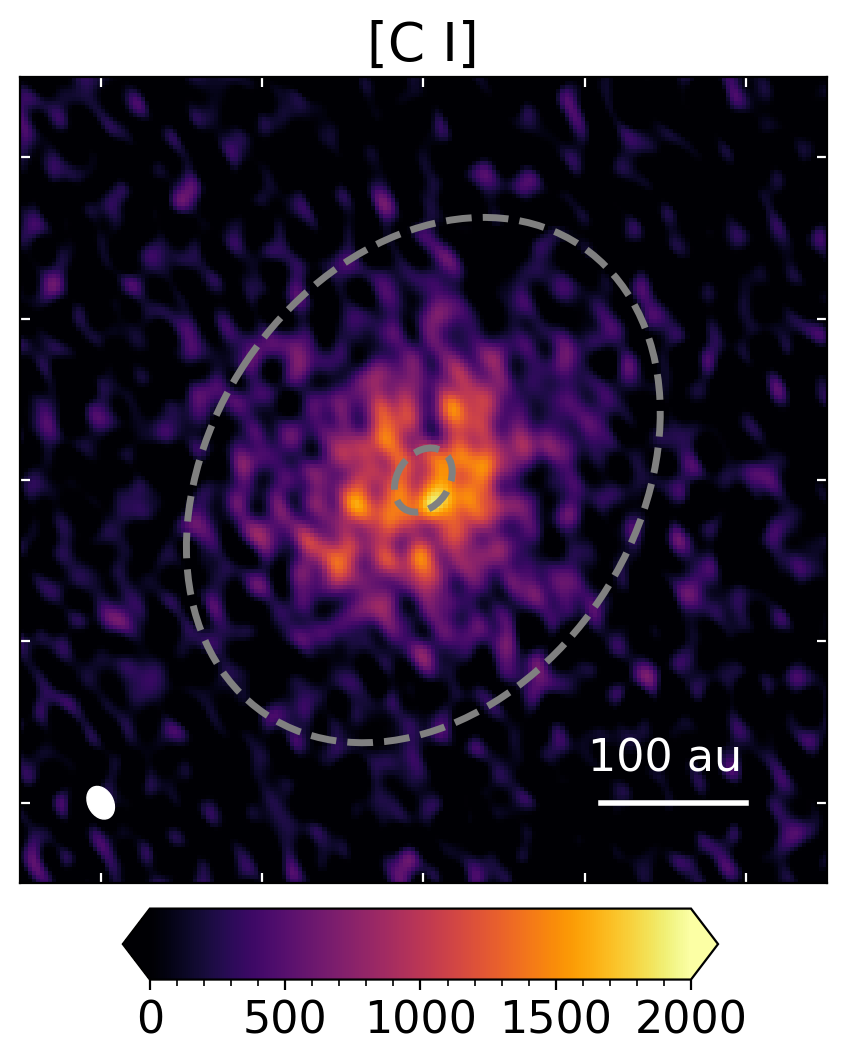}
\end{subfigure}%
\begin{subfigure}{1\columnwidth}
  \centering
    \includegraphics[width=1\linewidth]{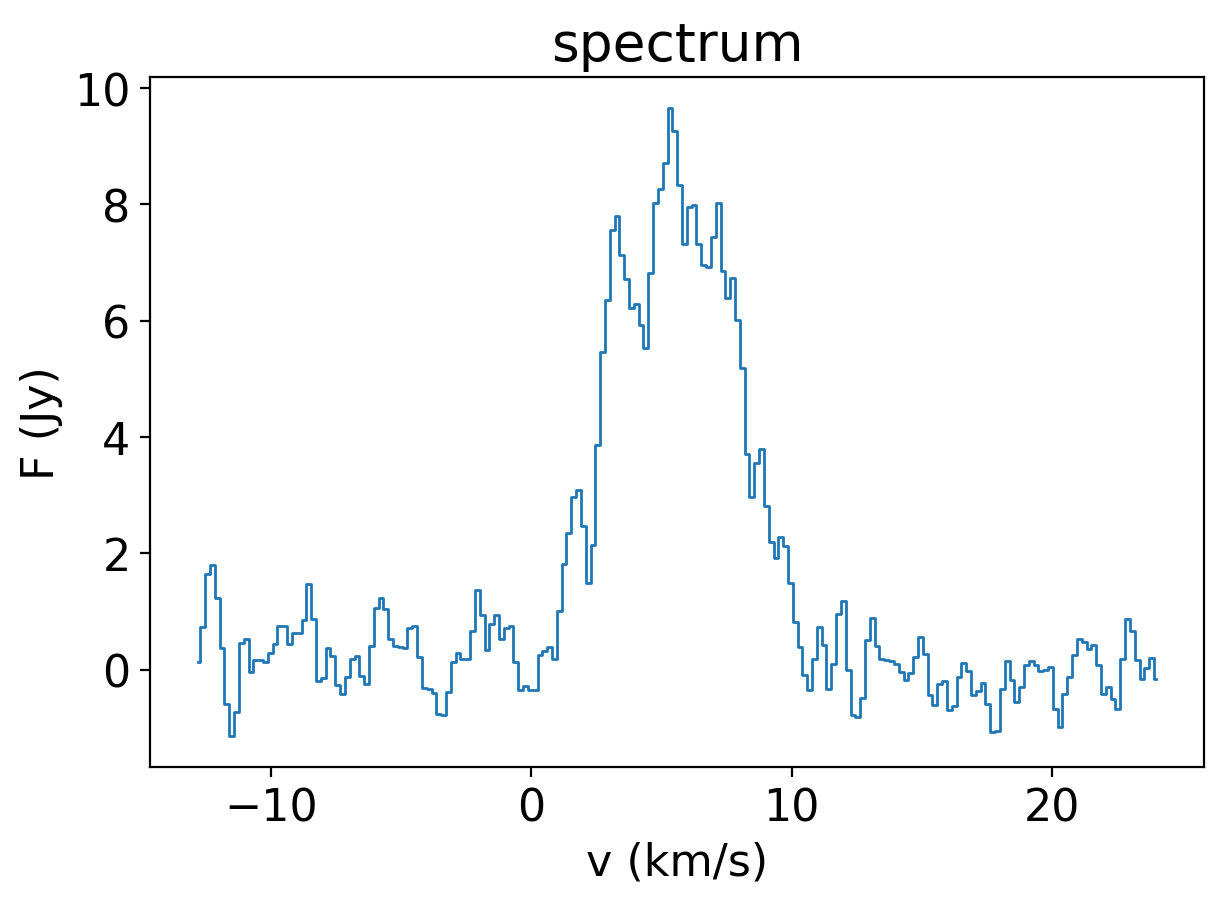}
\end{subfigure}
      \caption{Observations of [C~{\sc I}] in the HD~100546 disk. Left: integrated intensity map. The grey dashed lines indicate the dust rings in the HD~100546 disk. The beam is indicated in the bottom left corner and a 100~au scale bar is shown in the bottom right corner. Right: spectrum centered around the stellar velocity of 5.7~km~s$^{-1}$ \citep{Kama2016}.}
         \label{fig:CI_obs}
\end{figure*}

The integrated intensity map and the spectrum of the [C~{\sc I}] emission in the HD~100546 disk is presented in Fig.~\ref{fig:CI_obs}. 
An overview of the observations used in this work is presented in Table~\ref{tab:trans}. In addition to the observations, the critical density for each line is listed as some molecules are found to emit from elevated disk layers below the critical density.

\begin{table*}[b!]
\caption{\label{tab:trans}Continuum wavelengths and molecular transitions. }
\centering
\begin{tabular}{lcccc}
\hline\hline
Molecule & Transition & beam  & $n_{\rm crit} (50~\text{K})$ & ref. \\
\hline
continuum        & 0.9~mm (high res.)     & $0\farcs05\times 0\farcs03\ (34\degree)$ && 1 \\
continuum        & 0.9~mm (moderate res.) & $0\farcs38\times 0\farcs30\ (31\degree)$ && 2 \\
\ce{^12CO}       & $J=2-1$       & $0\farcs097\times 0\farcs077\ (-16\degree)$ & $2.7\times 10^3$ & 3 \\
\ce{^12CO}       & $J=3-2$       & $0\farcs35\times 0\farcs23\ (60\degree)$ & $ 9.2\times 10^3 $ & 2 \\
\ce{^12CO}       & $J=7-6$       & $0\farcs23\times 0\farcs17\ (30\degree)$ & $1.2\times 10^5 $ & 4 \\
\ce{^13CO}       & $J=2-1$       & $0\farcs098\times 0\farcs077\ (-15\degree)$ & $2.3\times 10^3 $ & 3 \\
\ce{C^18O}       & $J=2-1$       & $0\farcs097\times 0\farcs076\ (-13\degree)$ & $2.3\times 10^3 $ & 3 \\
\ce{C^17O}       & $J=3-2$       & $0\farcs39\times 0\farcs31\ (29\degree)$ & $ 8.5\times 10^3$ & 2 \\\relax  
[C {\sc I}]      & $^3P_2-^3P_1$ & $0\farcs21\times 0\farcs16\ (28\degree)$ & $1.4\times 10^3 $ & 5 \\
HCN              & $J=4-3$       & $0\farcs34\times 0\farcs23\ (60\degree)$ & $1.4\times 10^7$ & 2 \\
CN               & $3_{0, 7/2, 7/2} - 2_{0, 5/2, 5/2}$ & $0\farcs38\times 0\farcs30\ (29\degree)$ & $3.8\times 10^6$ & 2 \\
CN               & $3_{0, 7/2, 9/2} - 2_{0, 5/2, 7/2}$ & $0\farcs38\times 0\farcs30\ (29\degree)$ & $3.8\times 10^6$ & 2 \\
CN               & $3_{0, 7/2, 5/2} - 2_{0, 5/2, 3/2}$ & $0\farcs38\times 0\farcs30\ (29\degree)$ & $3.8\times 10^6$ & 2 \\
\ce{C2H}$^{(a)}$ & $4_{9/2, 5} - 3_{7/2, 4}$ & $0\farcs37\times 0\farcs30\ (27\degree)$ & $ 1.3\times 10^6$ & 2 \\
\ce{C2H}$^{(a)}$ & $4_{9/2, 4} - 3_{7/2, 3}$ & $0\farcs37\times 0\farcs30\ (27\degree)$ & $ 1.2\times 10^6$ & 2 \\
NO$^{(a)}$       & $4_{1, 7/2, 9/2} - 3_{-1, 5/2, 7/2}$ & $0\farcs47\times 0\farcs37\ (31\degree)$ & $2.7\times 10^4 $ & 2 \\
NO$^{(a)}$       & $4_{1, 7/2, 7/2} - 3_{-1, 5/2, 5/2}$ & $0\farcs47\times 0\farcs37\ (31\degree)$ & $2.4\times 10^4 $ & 2 \\
NO$^{(a)}$       & $4_{1, 7/2, 5/2} - 3_{-1, 5/2, 3/2}$ & $0\farcs47\times 0\farcs37\ (31\degree)$ & $2.4\times 10^4 $ & 2 \\
\ce{HCO+}        & $J=4-3$       & $0\farcs34\times 0\farcs23\ (60\degree)$ & $2.6\times 10^6$ & 2 \\ 
\hline
\end{tabular}
\tablefoot{$^{(a)}$ All components of a single molecule are added together assuming the emission is optically thin. The critical densities are calculated assuming that the main collision partner is \ce{H2}. References: 1. \citet{Pineda2019}, 2. \citet{Boothinprep}, 3. \citet{Perez2020} and W\"{o}lfer et al. subm., 4. W\"{o}lfer et al. subm., 5. This work.  }
\end{table*}

\section{DALI} \label{app:dali_method}

\subsection{Model setup}

\subsubsection{Radial and vertical structure} \label{app:dali_structure}
Radially, the gas density follows the solution for a viscously evolving disk model \citep{LyndenBell1974, Hartmann1998, Andrews2011}:
\begin{align}
\Sigma_{\rm gas\ full} &= \Sigma_{\rm c} \left ( \frac{r}{r_{\rm c}} \right )^{-\gamma} e^{-(r/r_{\rm c})^{2-\gamma}},
\end{align}
with $\Sigma_{\rm gas\ full}$ the gas surface density of a full disk, $\Sigma_{\rm c}/e$ the gas surface density at the characteristic radius $r_{\rm c}$, $r$ the radius, and $\gamma$ the power law index of the surface density profile. The surface density is defined from the sublimation radius $r_{\rm subl}$ to the outer radius of the disk $r_{\rm out,\ disk}$. Assuming a typical dust sublimation temperature of 1500~K, the sublimation radius is located at $\sim 0.07~\mathrm{au} \times \sqrt{L/L_{\odot}}$, with $L$ the luminosity of the central star \citep{Dullemond2001}. For the 36~L$_{\odot}$ HD~100546, this corresponds to 0.4~au.

Vertically, we assume that the gas follows a Gaussian distribution with the scale height $h$ given by
\begin{align}
h = h_{\rm c} \left (\frac{r}{r_{\rm c}} \right )^{\psi}, \label{eq:flaring}
\end{align}
with $h_{\rm c}$ the scale height at $r_{\rm c}$ and $\psi$ the flaring index.

   \begin{figure*}[ht!]
   \centering
   \includegraphics[width=\textwidth]{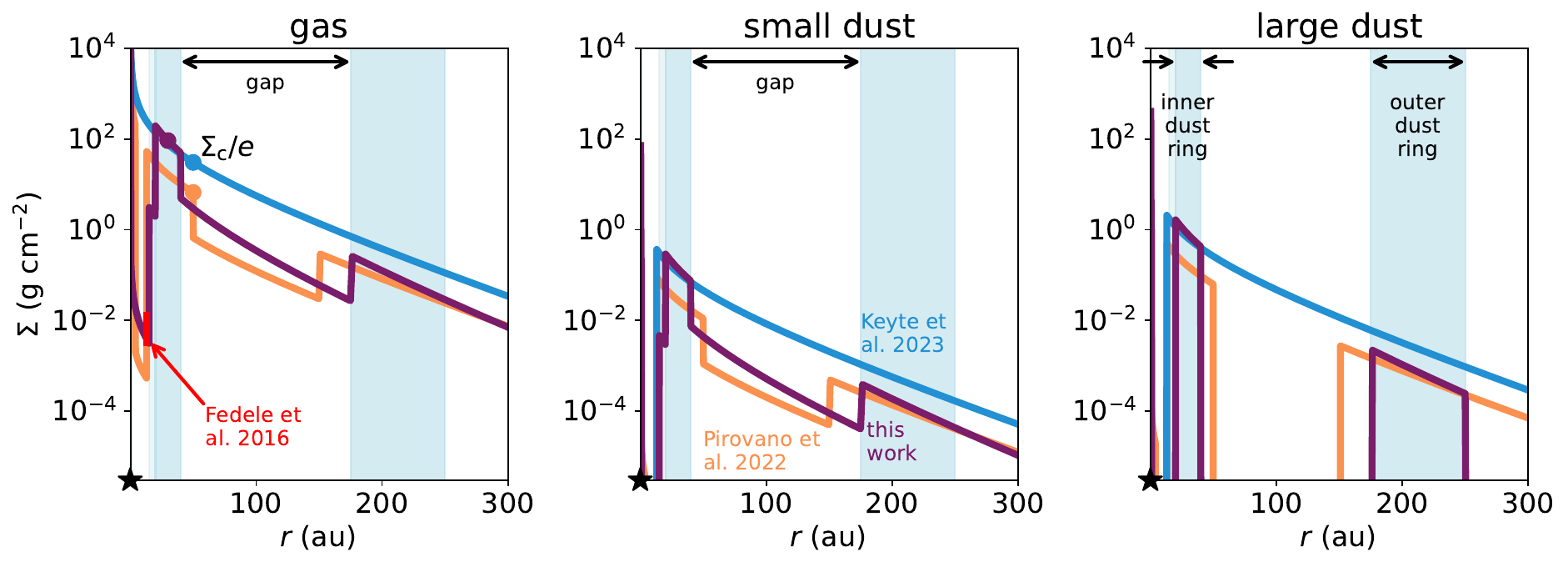}
      \caption{Surface density of the gas (left), small dust (middle), and large dust (right) in three models for the HD~100546 disk. The blue lines show the full gas disk model presented in \citet{Keyte2023}, the orange lines show the model presented in \citet{Pirovano2022}, and the purple lines show the model presented in this work. The shaded blue background indicates the dusty inner disk and the dust rings modelled in this work. The surface density $\Sigma_{\rm c}/e$ at $r_{\rm c}$ is indicated with a scatter point in the left panel. }
         \label{fig:dali_structure_lit}
   \end{figure*}

The dust in DALI is modeled using a small dust population and a large dust population. Both populations have a minimum grain size of 5~nm but the small grains have a maximum size of 1~$\mu$m, whereas the large grains go up to 1~mm. An MRN  distribution with an index of -3.5 is used to set the grain sizes $a$ within these populations \citep{Mathis1977}. Furthermore, the mass ratio of the large grain population compared to that of the small grains is set by $f_{\rm ls}$ such that the small and large dust in a full disk follow:
\begin{align}
\Sigma_{\rm s\ dust\ full} &= \frac{(1-f_{\rm ls}) \Sigma_{\rm gas\ full}}{\Delta_{\rm gd}}\\
\Sigma_{\rm l\ dust\ full} &= \frac{f_{\rm ls} \Sigma_{\rm gas\ full}}{\Delta_{\rm gd}},
\end{align}
with $\Delta_{\rm gd}$ the global gas-to-dust mass ratio. A comparison of the radial structure used in this work to those used by \citet{Pirovano2022} and \citet{Keyte2023} is presented in Fig.~\ref{fig:dali_structure_lit}.

Edge-on disks show that the large dust is settled to the disk midplane \citep{Villenave2020}. Therefore, the vertical scale height of the large dust is reduced by a settling parameter $\chi$. Thus, the small and large dust are not following the same radial and vertical distributions and the average grain size is calculated per cell, which is used for further calculations.

\subsubsection{Initial conditions of the chemical networks} \label{app:dali_chem_init}
The chemical networks are started using mostly molecular initial conditions. Initially, all volatile carbon is locked up in CO and all remaining oxygen is locked up in \ce{H2O} such that the total (gas $+$ ice) C/O ratio is 0.4. In the case of a C/O ratio larger than 1, all oxygen is initialized in the form of CO and the remaining carbon is put in as \ce{CH4}. The precise initial partition of nitrogen over N, \ce{N2}, and \ce{NH3} in protoplanetary disks is unknown, therefore we rescale the predictions of dark cloud models at 0.3~Myr \citep[N: \ce{N2}: \ce{NH3}: total number of N atoms $= 0.48: 0.19: 0.13: 1 $;][]{Walsh2015}. Finally, hydrogen is mainly in \ce{H2} when the model is initialised. An overview of the initial conditions is presented in Table~\ref{tab:dali}. The chemistry is evolved to the age of the HD~100546 disk of 5~Myr \citep{Arun2019} and a cosmic ray ionisation rate consistent with that found in other protoplanetary disks of $10^{-18}$~s$^{-1}$ is adopted \citep{Aikawa2021}.

\begin{table*}
\caption{DALI model parameters. }             
\label{tab:dali}      
\centering          
\begin{tabular}{p{0.6\columnwidth}>{\centering\arraybackslash}p{0.4\columnwidth}p{0.9\columnwidth}}    
    \hline\hline
        \centering\arraybackslash Model parameter & \centering\arraybackslash value & \centering\arraybackslash Description \\ \hline
        \\[-0.7em]
        \textit{Physical structure} &  & \\
		$r_{\mathrm{subl}} - r_{\mathrm{dust\ cav\ in}}$ & $0.4-1$~au & dusty inner disk\\ 
		$\delta_{\rm gas\ cav}, \delta_{\rm s\ dust,\ in}, \delta_{\rm l\ dust,\ in}$ & $10^{-5}, 10^{-2}, 10^{-2}$ & surface density drop for $r_{\mathrm{subl}} < r < r_{\mathrm{cav\ in,\ dust}}$\\
        $r_{\mathrm{dust\ cav\ in}} - r_{\mathrm{gas\ cav\ out}}$ & $1-15$~au & gas cavity\\
        $\delta_{\rm gas\ cav}, \delta_{\rm s\ dust\ cav}, \delta_{\rm l\ dust\ cav}$ & $10^{-5}, 10^{-10}, 10^{-10}$ & surface density drop for $r_{\mathrm{dust\ cav\ in}} < r < r_{\mathrm{gas\ cav\ out}}$\\
		$r_{\mathrm{gas\ cav\ out}} - r_{\mathrm{dust\ cav\ out}}$ & $15-20$~au & region outside gas cavity but inside dust cavity\\
		$\delta_{\rm gas\ cav\ edge}, \delta_{\rm s\ dust\ cav\ edge}, \delta_{\rm l\ dust\ cav\ edge}$ & $10^{-2}, 10^{-2}, 10^{-10}$ & surface density drop for $r_{\mathrm{gas\ cav\ out}} < r < r_{\mathrm{dust\ cav\ out}}$\\
        $r_{\mathrm{gap\ in}}-r_{\mathrm{gap\ out}}$ & $40-175$~au & gap\\
        $\delta_{\mathrm{gas\ gap}}, \delta_{\mathrm{s\ dust\ gap}}, \delta_{\mathrm{l\ dust\ gap}}$ & $10^{-1}, 10^{-1}, 10^{-10}$ & surface density drop for $r_{\mathrm{gap\ in}} < r < r_{\mathrm{gap\ out}}$\\
        $r_{\rm l\ dust\ out}-r_{\rm out}$ & $250-1000$~au & disk region outside the second ring\\
        $\delta_{\rm gas\ out}, \delta_{\rm s\ dust\ out}, \delta_{\rm l\ dust\ out}$ & $1, 1, 10^{-10}$ & surface density drop for $r_{\rm l\ dust\ out} < r < r_{\rm out}$\\                                                   
        $r_{\mathrm{c}}$ & 30~au & Characteristic radius of the surface density profile \\  
        $\Sigma_{\mathrm{c}}$ & 250~g~cm$^{-2}$ & Sets the gas surface density at the characteristic radius $r_{\mathrm{c}}$ \\    
        $M_{\mathrm{disk}}$ & $5\times 10^{-2}$~M$_{\odot}$ & Mass of the disk \\   
        $\gamma$ & 1.1 & Power law index of the surface density profile  \\
        $h_{\mathrm{c}}$ & 0.1 & Scale height angle at the characteristic radius $r_{\mathrm{c}}$ \\
        $\psi$ & 0 & Flaring index of the disk surface density \\
        PAH abundance & 0.1 & Gas-phase abundance of PAHs w.r.t. to ISM value  \\   
        \\[-0.3em]
        \textit{Dust properties} & &  \\
        $\chi$ & 0.2 & Settling of large grains \\
        $f_{\mathrm{ls}}$ & 0.85 & Mass-fraction of grains that is large  \\
        $\Delta_{\mathrm{gas/ dust}}$ & 100 & Gas-to-dust mass ratio   \\
        \\[-0.3em]
        \textit{Stellar properties} & & \\          
        $M_{\star}$ & 2.5~$M_{\odot}$ & Mass of the central star  \\
        $L_{\star}$ & 36~$L_{\odot}$ & Luminosity of the central star   \\
        $L_{\mathrm{X}}$ & $7.9\times 10^{28}$~erg~s$^{-1}$ & X-ray luminosity of the central star  \\
        $T_{\mathrm{X}}$ &  $7\times 10^7$~K & Effective temperature of the X-ray radiation   \\  
        $\zeta_{\mathrm{c.r.}}$ & $1\times 10^{-18}$~s$^{-1}$ & Cosmic ray ionization rate \\ 
        \\[-0.3em]
        \multicolumn{2}{l}{\textit{Observational geometry}}&  \\
        $i$ & 41$\degree$ & Disk inclination ($0\degree$ is face-on)\\
        $d$ & 108.1~pc & Distance to the star \\ 
        \\[-0.3em]
        \textit{Chemistry}$^{(1)}$  &&  \\
        H        & $5.2\times 10^{-5}$ & \\
        He       & $1.4\times 10^{-1}$& \\
        N		 & $3.0\times 10^{-5}$ & \\
        \ce{H2}  & $5.0\times 10^{-1}$ & \\   
        \ce{H2O} & $1.9\times 10^{-4}$ & \\               
        \ce{CO}  & $1.3\times 10^{-4}$ & \\                
        \ce{N2}  & $1.2\times 10^{-5}$ & \\              
        \ce{NH3} & $8.1\times 10^{-6}$ &  \\                      
        \ce{Mg+} & $1.0\times 10^{-11}$ & \\                
        \ce{Si+} & $1.0\times 10^{-11}$ & \\           
        \ce{S+}  & $1.0\times 10^{-11}$ & \\                
        \ce{Fe+} & $1.0\times 10^{-11}$ & \\        
        $t_{\mathrm{end}}$ & 5~Myr & \\ \hline                           
\end{tabular}
\tablefoot{$^{(1)}$~Abundance w.r.t. the total number of hydrogen atoms. All molecules start as gas-phase species. }
    \begin{tablenotes}
      \small
      \item 
    \end{tablenotes}
\end{table*}

\subsection{Model predictions}

\subsubsection{Outer ring in \ce{C^18O} and \ce{C^17O} models}

The fiducial model predicts a ring in the \ce{C^18O} $J=2-1$ transition at $\sim200$~au and a weak ring is also seen in the \ce{C^17O} $J=3-2$ line. This is due to the low column density just inside the dust ring at $150-175$~au. In this region the column density of the CO isotopologues drops below the minimum column density needed to make self-shielding effective. Therefore, \ce{C^18O} and \ce{C^17O} are effectively destroyed in this region. At the outer dust ring, the gas column density increases by an order of magnitude, increasing the \ce{C^18O} and \ce{C^17O} column densities. The \ce{C^18O} and \ce{C^17O} column densities at 175~au are similar to the respective column densities at 150~au, just before self-shielding becomes efficient. Therefore, the \ce{C^18O} and \ce{C^17O} intensity in the outer dust ring is very sensitive to the gas column density in this region.

   \begin{figure}[ht!]
   \centering
   \includegraphics[width=\hsize]{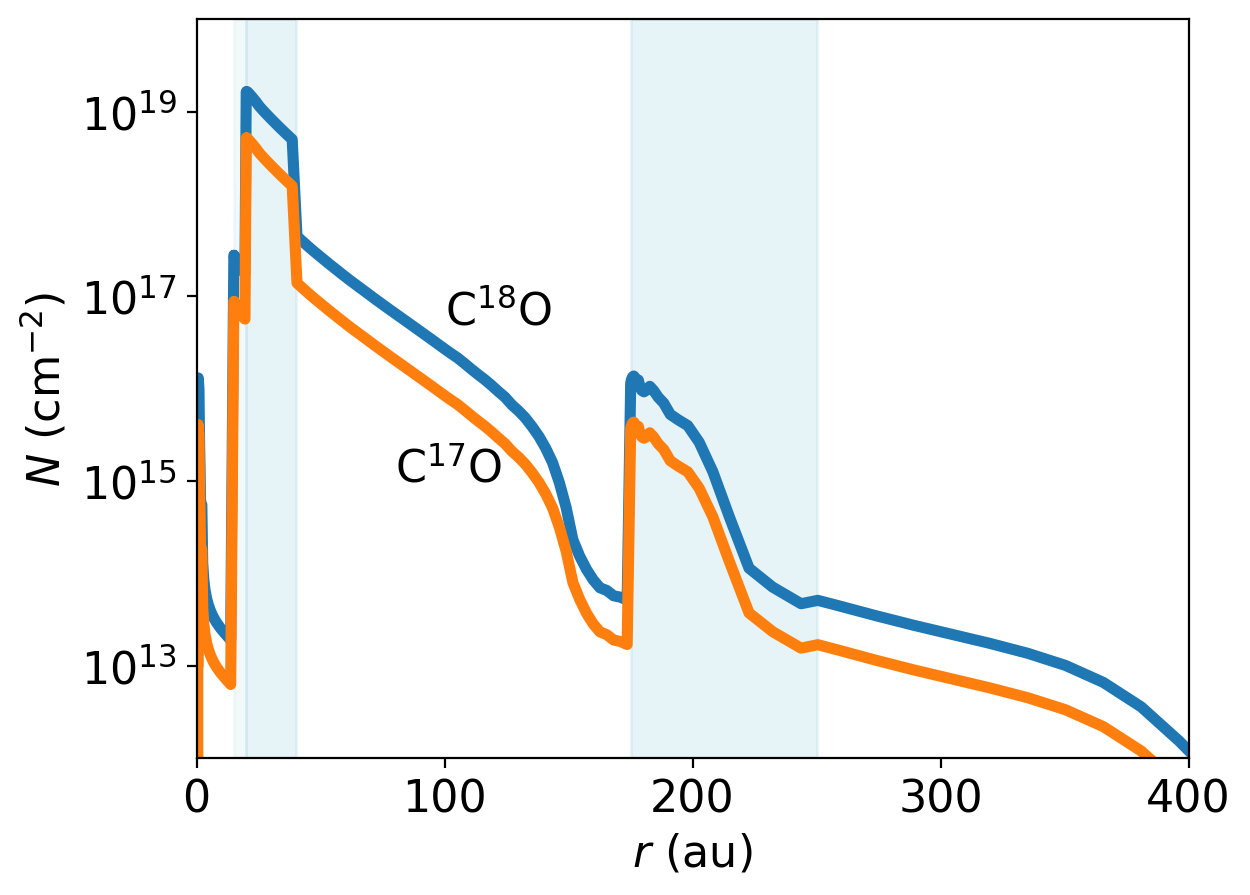}
      \caption{Column density of \ce{C^18O} and \ce{C^17O} in the fiducial disk model. The steep drop in these column densities at $\sim 150-175$~au is due to the lack of self-shielding. Note that the column density of \ce{C^18O} and \ce{C^17O} at 175~au is very similar to the respective column densities in the gap just before self-shielding becomes ineffective. }
         \label{fig:N_COisos}
   \end{figure}

\subsubsection{Emission profiles and emitting heights of CO isotopologues} \label{app:COemission_and_heights}

\begin{figure*}[ht!]
   \centering
   \includegraphics[width=\hsize]{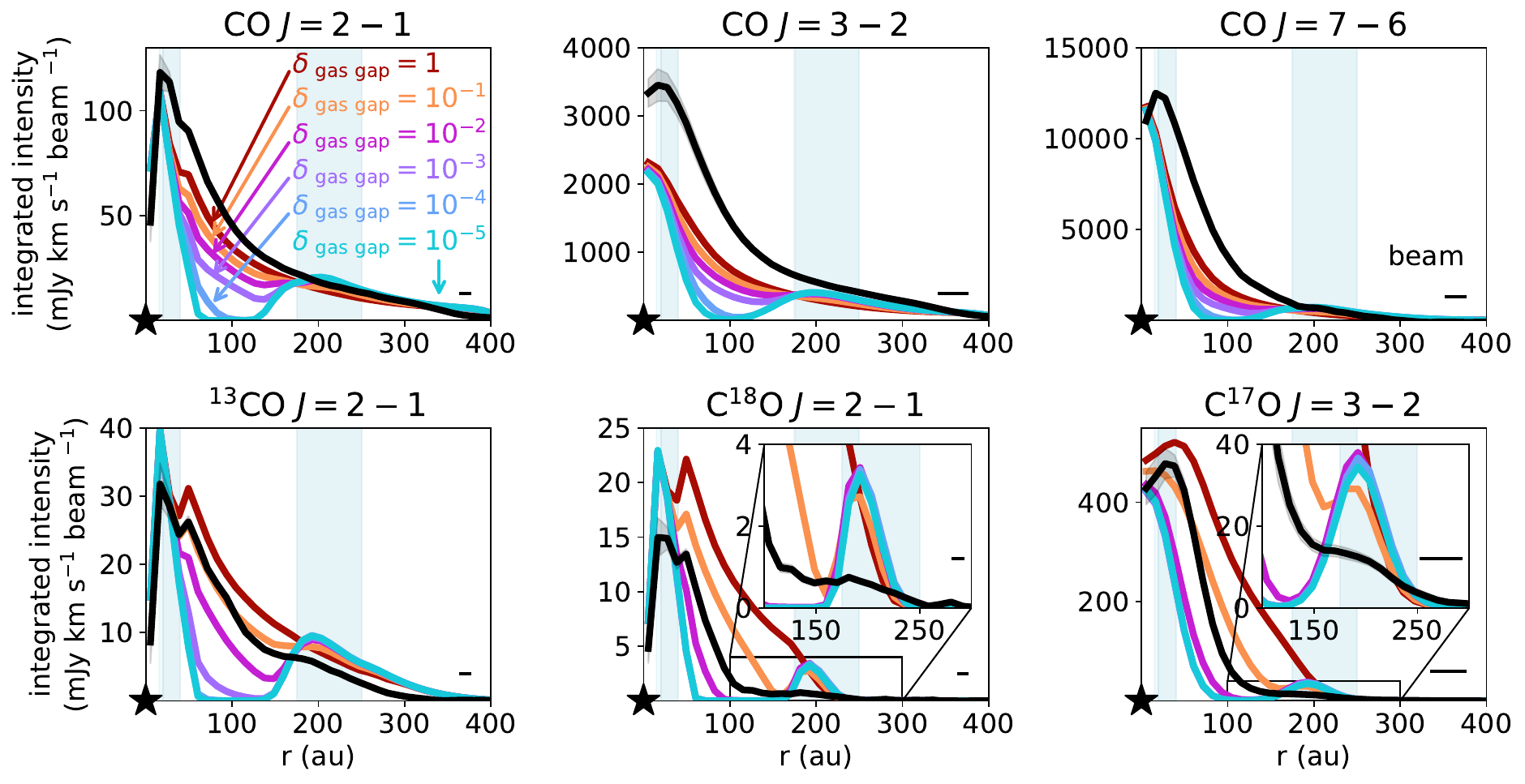}
      \caption{Azimuthally averaged radial profiles of the dust and CO isotopologue emission observed in the HD~100546 disk (black) together with model predictions for different gaps. In the models, the gas density in the gap ($40-175$~au) is reduced by a factor of  1 (no gap; dark red), $10^{-1}$ (orange), $10^{-2}$ (pink), $10^{-3}$ (purple), $10^{-4}$ (blue), and $10^{-5}$ (very deep gap; light blue). For all these models, the drop in the small dust density follows that of the gas. The beam is indicated with the horizontal bar in the bottom right corner of each panel. }
         \label{fig:dali_dgas_gap}
   \end{figure*}

The azimuthally averaged radial profiles of the CO isotopologue emission are presented in Fig.~\ref{fig:dali_dgas_gap}. The models with a drop in gas density and small dust density of 1 (no drop) up to $10^{-5}$ are shown in the same figure with the colored lines. The model without a drop in the gas and small dust density overpredicts the \ce{C^18O} and \ce{C^17O} emission that are the most optically thin and thus the most sensitive to the gas column density. On the other hand, the models with a deeper gap ($10^{-2}$, $10^{-3}$) than the fiducial model ($10^{-1}$), underpredict the \ce{C^18O} and \ce{C^17O} emission in the gap. Additionally, these models show a clear gap in the brighter \ce{^12CO} and \ce{^13CO} lines that is not seen in the observations. This cannot be due to the limited spatial resolution of the observations as the gas gap is fully resolved with 4 beams across the gas gap for \ce{C^17O}, the lowest spatial resolution line in this work, and 14 for \ce{C^18O} which has the highest spatial resolution. Outside the gas gap, at the outer dust ring, the \ce{C^18O} emission is overpredicted. However, a lower gas and thus CO isotopologue column density lowers the \ce{C^17O} column density in the gas gap to a value below its minimum value to self-shield. Therefore, a model that fits the \ce{C^18O} in that region will not reproduce the shelf in \ce{C^17O}. Therefore, the CO isotopologue observations show that the gas density is depleted by at most a factor of $10^{-1}$ between the two dust rings under the assumption that the gas and dust gap have the same widths.

   \begin{figure*}[ht!]
   \centering
   \includegraphics[width=\hsize]{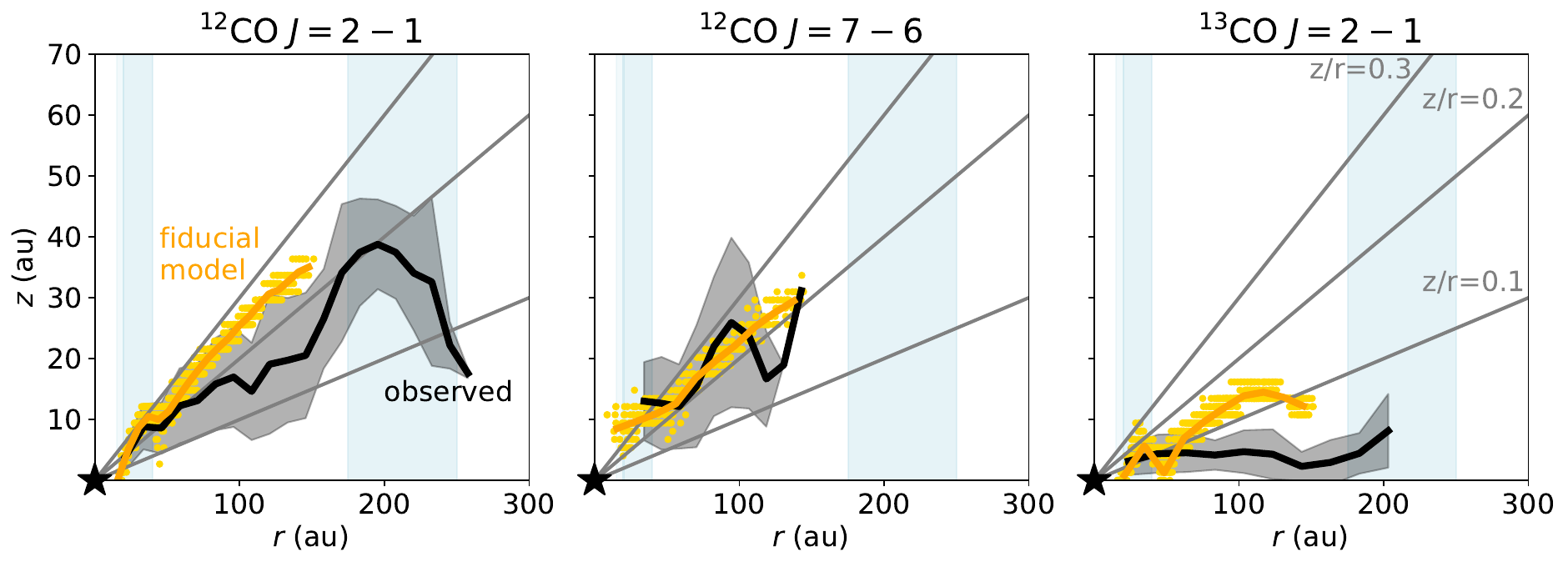}
      \caption{ Emitting heights of the \ce{^12CO} $J=2-1$, and $J=7-6$, and the \ce{^13CO} $J=2-1$ transition. The observed emitting heights are presented in black and those of the fiducial model in orange.}
         \label{fig:emitting_heights}
   \end{figure*}

The emitting heights of the \ce{^12CO} $J=2-1$ and $J=7-6$ transitions together with that of the \ce{^13CO} $J=2-1$ transition are presented in Fig.~\ref{fig:emitting_heights}. These emitting heights are derived using the ALFAHOR package by masking the individual channels after visually inspecting the data \citep{PanequeCarreno2023}. W\"{o}lfer et al. subm. found evidence for an asymmetry in the vertical structure between the blue and red shifted sides of this disk. As the DALI models used in this work are azimuthally symmetric, the observed emitting heights are presented as an average of both disk sides. 

The fiducial model roughly reproduces the observed emitting height of the \ce{^12CO} $J=2-1$ and $J=7-6$ transitions within the scatter for a radius up to 150~au, as expected with the $\sim 10$~au scatter in the \ce{^12CO} $J=2-1$ emitting surface. Still, emitting surfaces of \ce{^12CO} and \ce{^13CO} in the IM~Lup, HD~163296, and the MWC~480 disks generally follow the same vertical modulations in the derived \ce{^13CO} emitting layer as are seen in the \ce{^12CO} layer despite the large scatter. Therefore, the derived \ce{^12CO } $J=2-1$ surface in the HD~100546 disk is robust. The lack of vertical segregation between different transitions of one molecule is seen in other DALI models as well (Paneque-Carre{\~n}o in prep.) and the emitting height of the \ce{^12CO} $J=2-1$ transition observed in the HD~100546 disk is similar to that observed in other disks \citep{Law2022, Law2023, Stapper2023}. Finally, the \ce{^13CO} emission is seen very close to the disk midplane at a height of only a few au in the observations. The model reproduces the trend that the \ce{^13CO} emission comes from deeper disk layers than the \ce{^12CO}, and is consistent within $\sim 2\sigma$ from the observations up to 150~au.

\subsubsection{Emission profiles and column densities of [C~{\sc I}], HCN, CN, \ce{C2H}, NO, and \ce{HCO+} for different gas gaps}\label{app:molecules_radial}

\begin{figure*}
   \centering
   \includegraphics[width=\hsize]{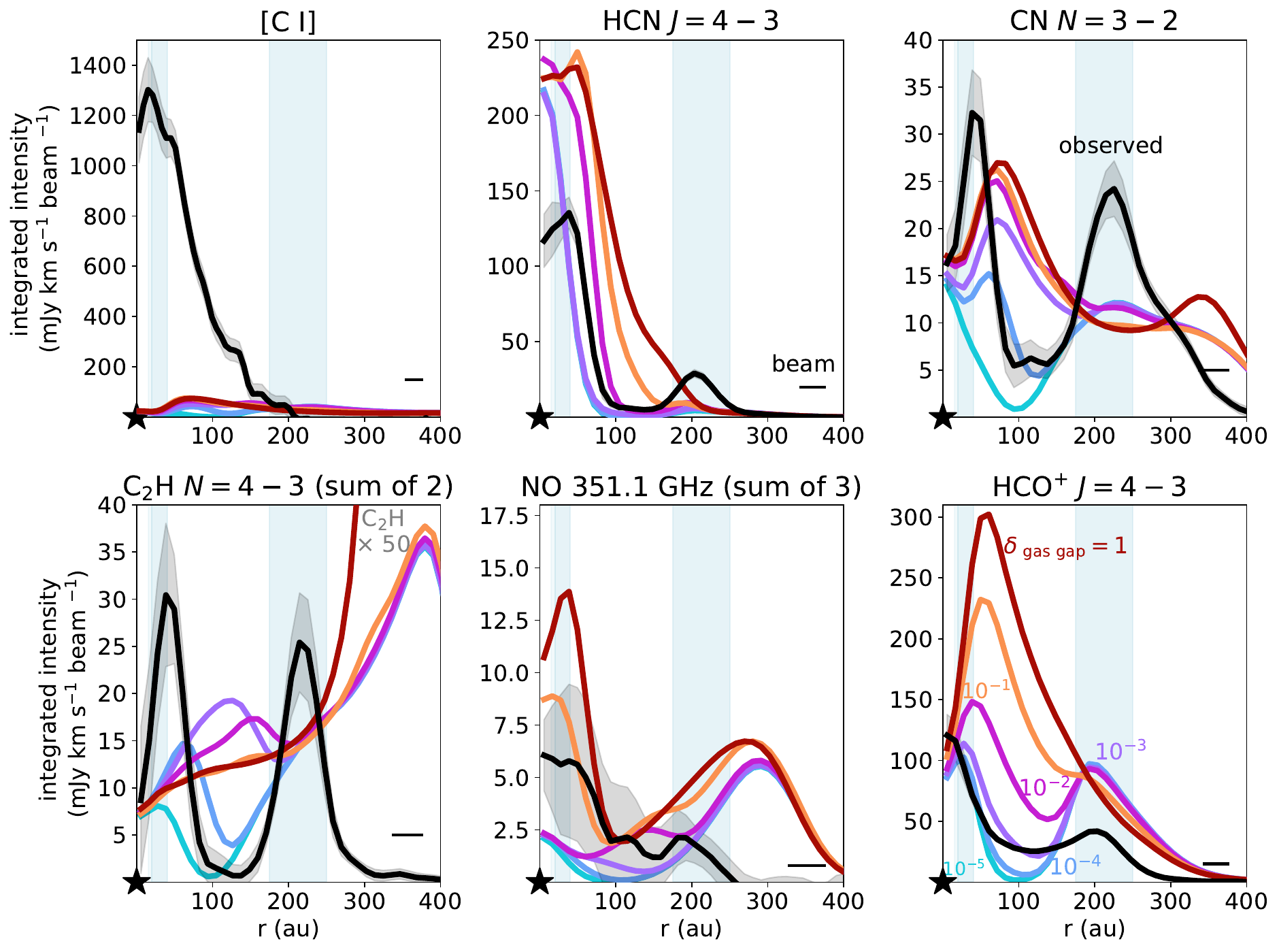}
      \caption{Azimuthally averaged radial profiles of the [C~I], HCN, CN, \ce{C2H}, NO, and \ce{HCO+} emission predicted by models with a gas gap depth of 1 (no gap; dark red), $10^{-1}$ (orange), $10^{-2}$ (pink), $10^{-3}$ (purple), $10^{-4}$ (blue), and $10^{-5}$ (very deep gap; light blue). The \ce{C2H} emission is increased by a factor of 50 for all models to show it on the same scale as the observations. The peak \ce{C2H} intensity of the full disk model is 2~mJy~km~s$^{-1}$~beam$^{-1}$. The small dust density is dropped by the same factor in the gap. The observed intensities are indicated with the black lines. }
         \label{fig:dali_azi_avg_dgap}
   \end{figure*}

\begin{figure*}
   \centering
   \includegraphics[width=\hsize]{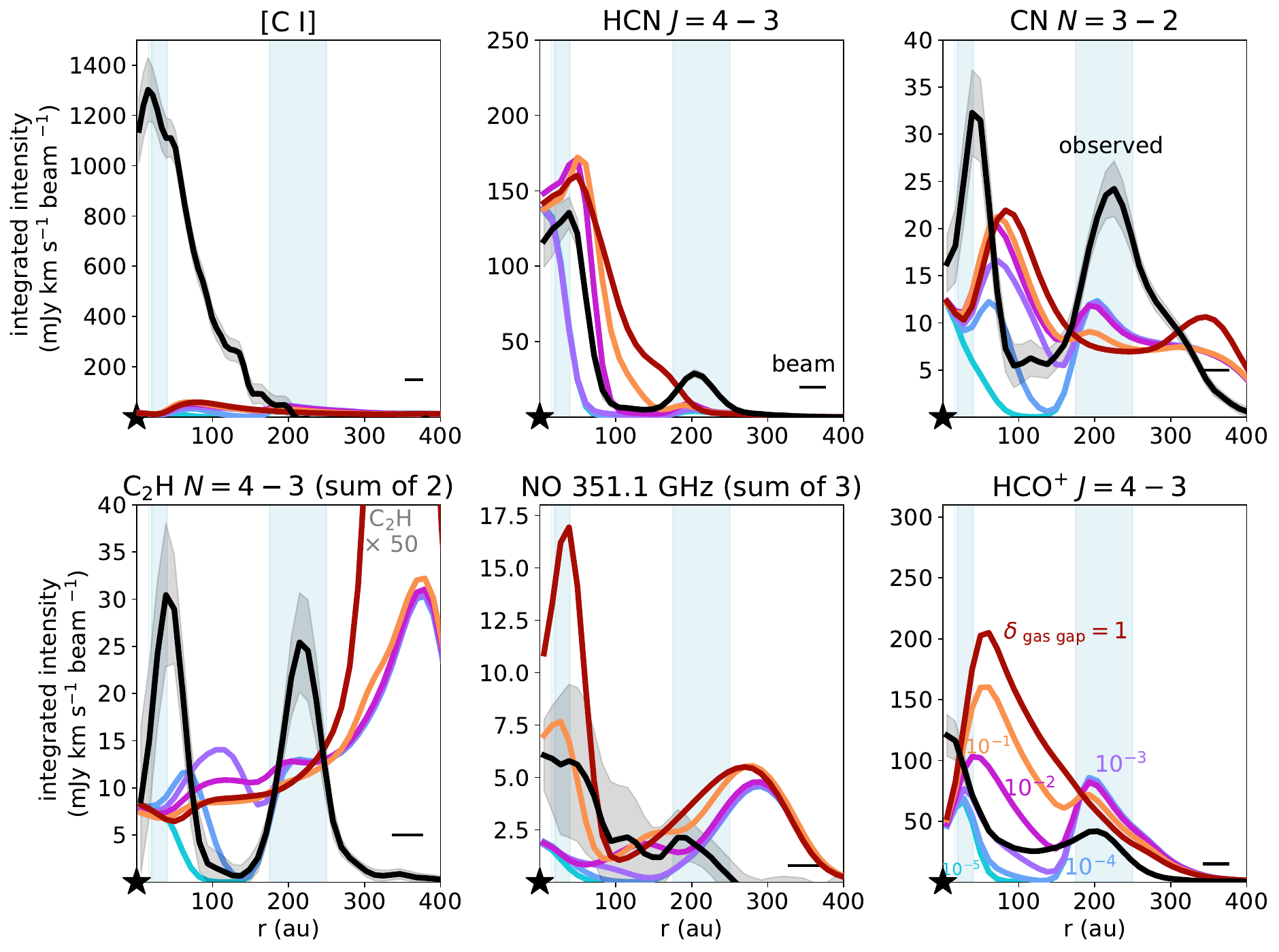}
      \caption{Same as Fig.~\ref{fig:dali_azi_avg_dgap} but then for face-on disk. The modelled \ce{C2H} emission is increased by a factor of 50 to show it on the same scale as the observations.}
         \label{fig:dali_azi_avg_dgap_face-on}
   \end{figure*}

The azimuthally averaged radial profiles for the HCN, CN, \ce{C2H}, NO, and \ce{HCO+} moment 0 maps presented in Fig.~\ref{fig:mom0_fid} are presented in Fig.~\ref{fig:dali_azi_avg_dgap}. Instead of the \ce{C^18O} emission, the profile for the atomic carbon line is presented in the top left corner. The colored lines indicate the models for different gas gap depths between the two dust rings, identical to those in e.g. Fig.~\ref{fig:dali_dgas_gap}. The rings seen in the observations (black) do not coindice with those seen in the models in general. Additionally, the modelled [C~{\sc I}] emission  inside 200~au is up to more than one order of magnitude weaker than what is observed. At larger radii, the model slightly overpredicts the observations, but emission at these scales may be missing due to the small maximum recoverable scale (2" diameter, 108~au radius) of the observation.  As the models predict that the molecules are abundant in an elevated layer at $z/r \sim 0.2-0.3$, the disk inclination of 41$\degree$ causes deprojection effects that wash out some of the rings in the CN and move those in \ce{C2H} to the location of the outer dust ring. Therefore, the azimuthally averaged radial profiles predicted by a face-on disk model are presented in Fig.~\ref{fig:dali_azi_avg_dgap_face-on}.

   \begin{figure*}[]
   \centering
   \includegraphics[width=\hsize]{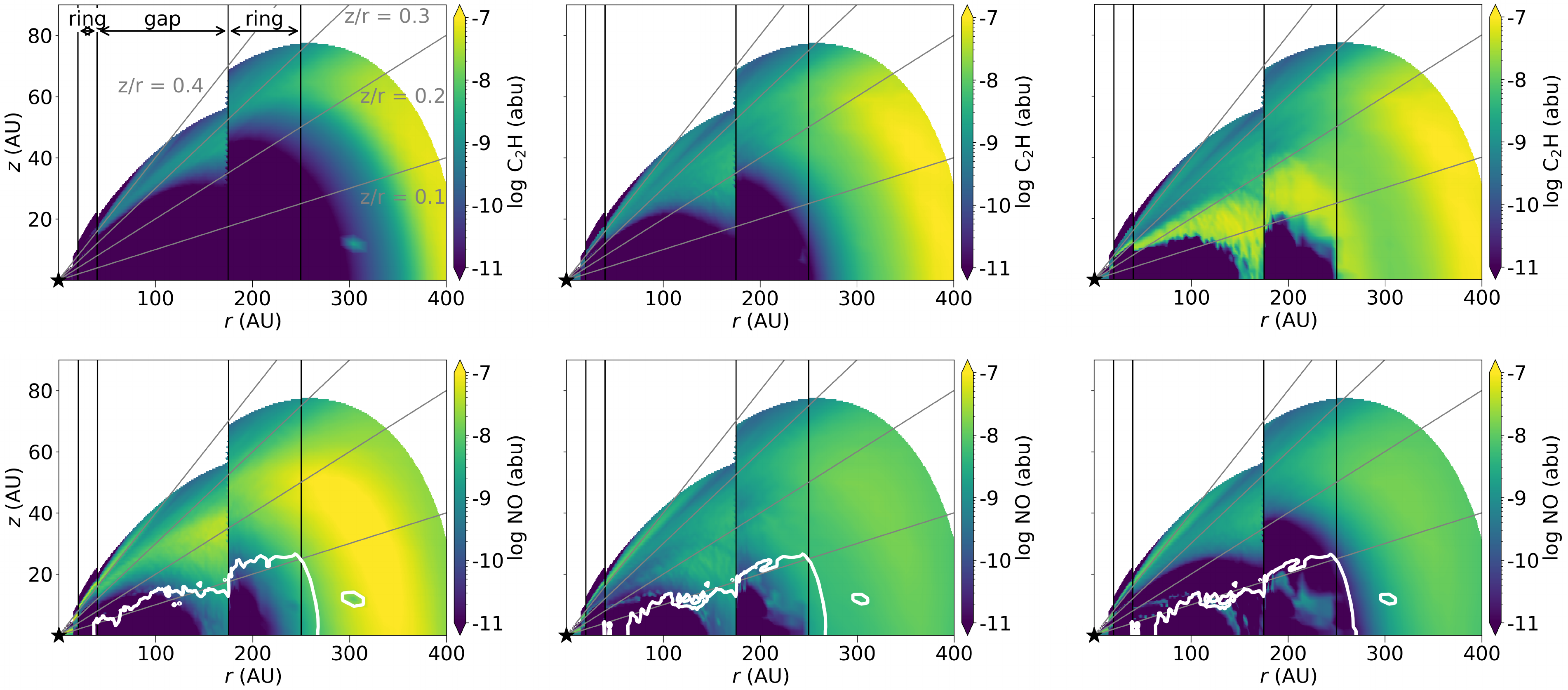}
      \caption{The \ce{C2H} (top row) and NO (bottom row) abundance for different C/O ratios. From left to right: C/O = 0.4 (fiducial value), 0.99, and 1.01. Only the regions with a gas number density above $10^5$~cm$^{-3}$ are shown. The white contour indicates the NO snowline.}
         \label{fig:dali_2D_dC_O}
   \end{figure*}

   \begin{figure*}
   \centering
   \includegraphics[width=\hsize]{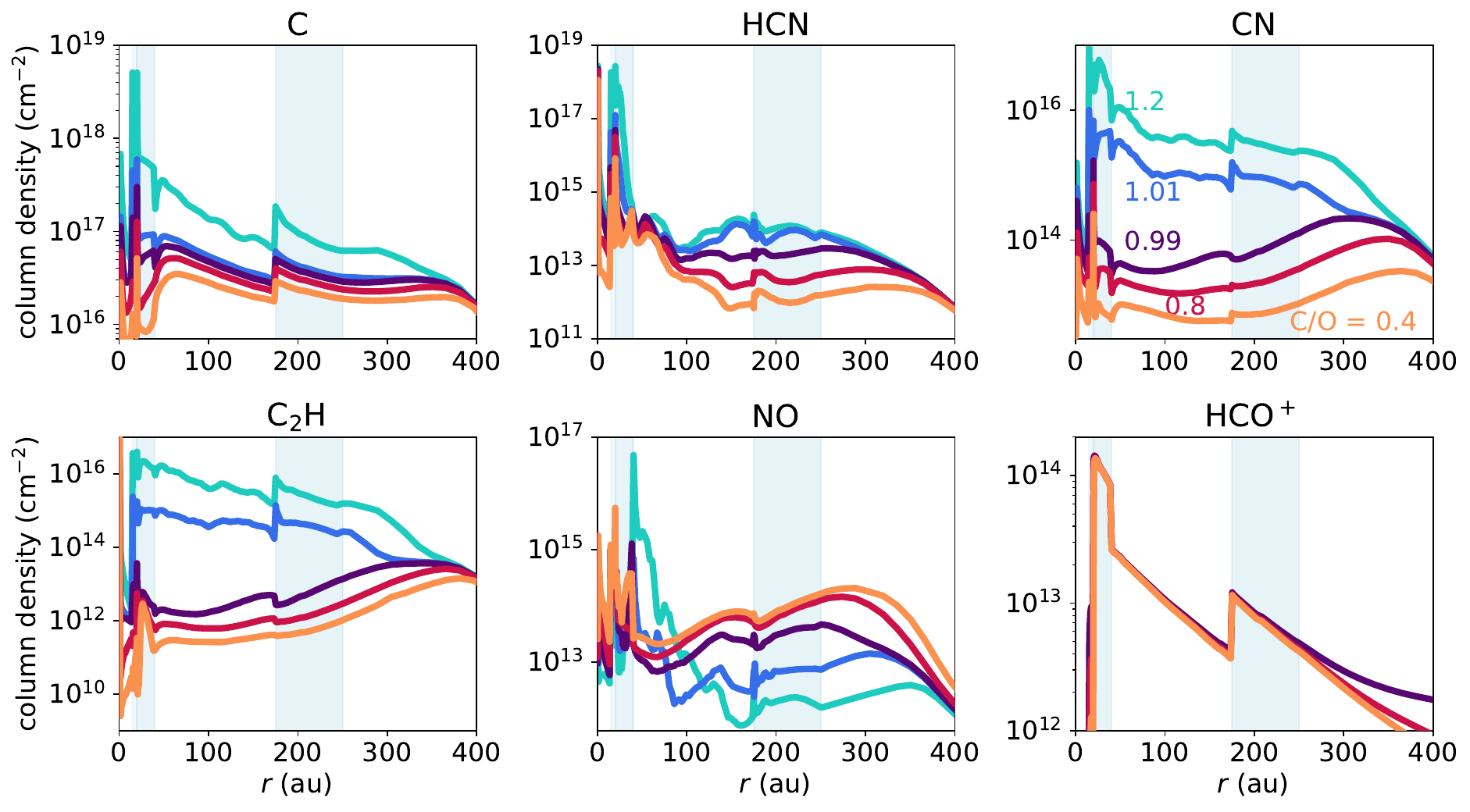}
      \caption{Column densities of the molecular lines studied in this work for different C/O ratios. }
         \label{fig:N_dC_O}
   \end{figure*}

Similar to the trend seen for the \ce{HCO+} column density, the \ce{HCO+} emission in the gap decreases for deeper gap depths. Even for the shallowest gas gap ($\delta_{\rm gas\ gap} = 10^{-1}$), a shallow gap and a ring are seen in the \ce{HCO+} emission. For deeper gas gaps, the double ring nature of the \ce{HCO+} emission becomes more pronounced (see Fig.~\ref{fig:dali_azi_avg_dgap} for a more quantitative figure). The modelled intensity in the outer ring lays within a factor of 2.5 of the data for all models. To reproduce the diffuse emission observed in the gas gap, a model with a moderately deep gap of $10^{-2}-10^{-3}$ is needed. These models as well as those, with even deeper gap of $10^{-4}-10^{-5}$ match the intensity of the \ce{HCO+} in the gas cavity within $2.5\sigma$, though the models predict a ring at $30-40$~au instead of centrally peaked emission.

The CN emission shows a bright inner ring just outside the inner dust ring and diffuse emission outwards. For gaps up to $\delta_{\rm gas\ gap} = 10^{-3}$, the CN intensity stays more or less constant. For deeper gaps, the CN intensity follows the trends seen in the column density and the inner ring becomes weaker. The models with gaps deeper than $10^{-2}$ all show emission from an elevated layer as the outer ring consists of two ellipses that are offset along the disk minor axis. This is not seen in the data where the outer CN ring is centred at the position of the star. Apart from this, the model with a deep gap of 4 orders of magnitude less gas resembles the observations quite closely as two distinct rings with an intensity within a factor of $\sim3$ of the observations are seen. 

\ce{C2H} emits as a bright outer ring seen at 375~au due to the high column density in this region. This is much further out than the second ring seen in the data and not affected by the depth of the gas gap. Inside the gap, a ring at low intensity is seen. This ring follows the trend seen in the column density, moves inwards, and brightens as the gas gap becomes deeper. The bright ring seen in the model with a $10^{-4}$ deep gas gap is only peaks one beam further out than the ring seen in the observations but the modelled intensity is a factor of 40 too low. 
The inward travelling ring is also seen in the NO emission causing a very bright inner ring for the two models with the deepest gas gaps. The outer ring seen just outside the outer dust ring in the models is not detected in the observations.

Finally, the HCN shows a bright inner ring with an intensity within a factor of two of the observations for the models with a shallow gas gap. For the models with a gap deeper than $10^{-2}$ a gap and then a very weak ($\lesssim 5$~\%) ring at the location of the outer dust ring are seen as highlighted by the white contours in Fig.~\ref{fig:dali_mom0_dgap}. This ring has an intensity of only $3-5$ times weaker than what is observed.

\subsubsection{Abundances and column densities for different C/O ratios}
Figure~\ref{fig:dali_2D_dC_O} presents the 2D abundance structure of \ce{C2H} (top row) and NO (bottom row) for the model with the fiducial C/O ratio of 0.4 and models with a higher C/O ratio of 0.99 and 1.01. The \ce{C2H} and NO molecules show opposite trends where the \ce{C2H} abundance increases with an increasing C/O ratio and the NO abundances decreases.
The column densities of C, HCN, CN, \ce{C2H}, NO, and \ce{HCO+} for different C/O ratios are presented in Fig.~\ref{fig:N_dC_O}. The column densities of the molecules that contain a carbon atom but no oxygen, C, HCN, CN, and \ce{C2H}, all increase with increasing C/O. In particular, the CN and \ce{C2H} are sensitive to a C/O ratio increasing from 0.99 to 1.01 as their column densities increase by one and two orders of magnitude, respectively. The NO molecule, that only carries an oxygen atom and no carbon, shows the opposite trend in general. The only exception to this is the high NO column density just outside the inner dust ring in the models with C/O $>1$. Finally, the \ce{HCO+} is not sensitive to the C/O ratio in most disk regions as it carries both oxygen and carbon atoms.

\subsubsection{The effect of grain growth on \ce{C2H}}

\begin{figure*}
   \centering
   \includegraphics[width=\hsize]{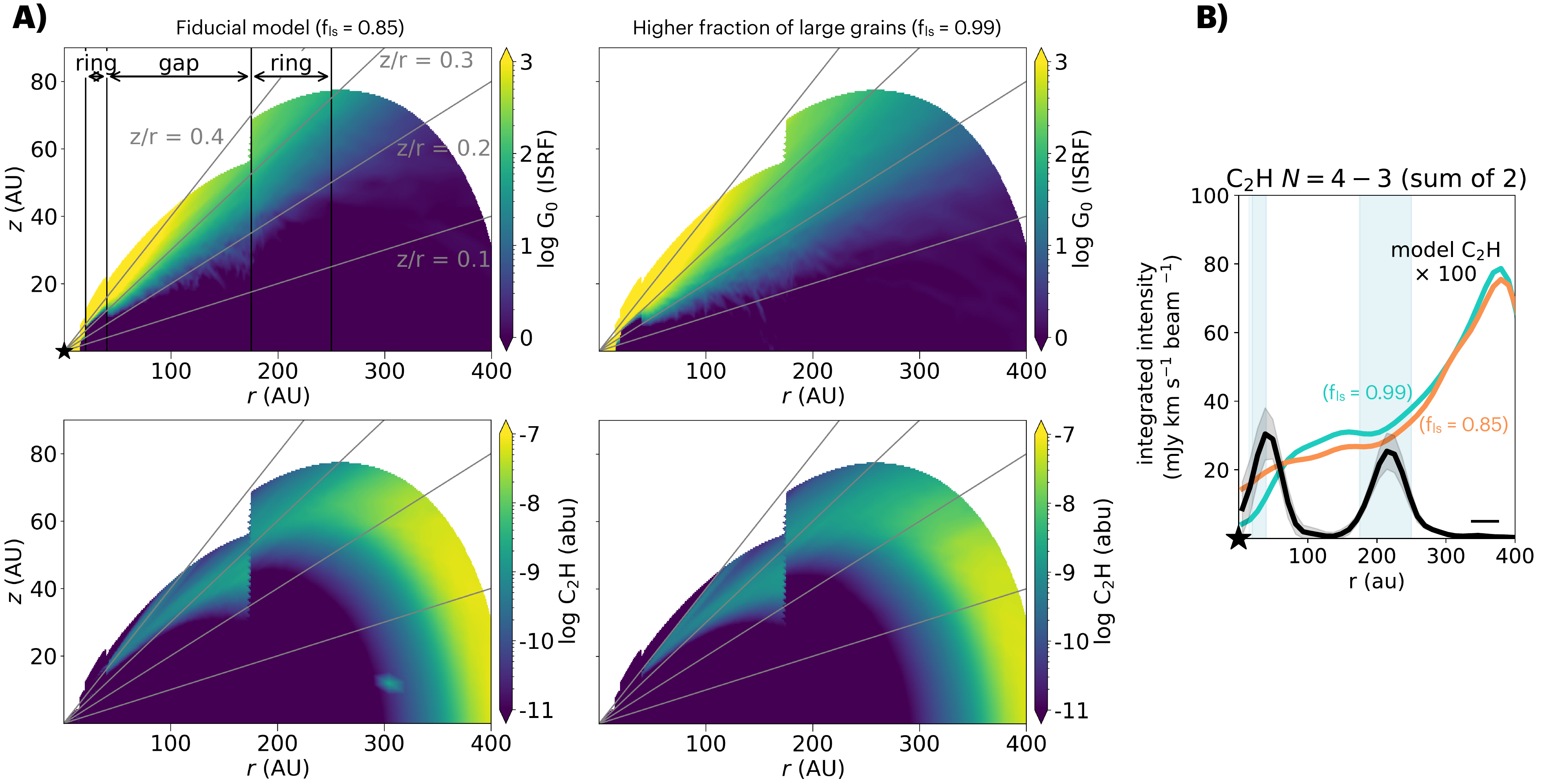}
      \caption{UV field and \ce{C2H} abundance map for the fiducial model and a model with grain growth (panel A) and the predicted \ce{C2H} emission (panel B). }
         \label{fig:fls}
   \end{figure*}

The \ce{C2H} molecule is particularly sensitive to the C/O ratio and the UV field. In this subsection, the effect of grain growth, modelled as an increase in the fraction of large grains from 0.85 to 0.99, is presented. An increased fraction of large grains effectively removes small grains from the upper layers of the disk. This effect is shown in the top row of Fig.~\ref{fig:fls}A. The \ce{C2H} abundance map is mainly affected in the surface layer around 300~au. This is also reflected in the emission that is similar between the two models (Fig.~\ref{fig:fls}B).

\subsection{Molecular ratios} \label{app:dC_O_ratios}

\subsubsection{Column density ratios for different gap depths}

   \begin{figure*}[ht!]
   \centering
   \includegraphics[width=14cm]{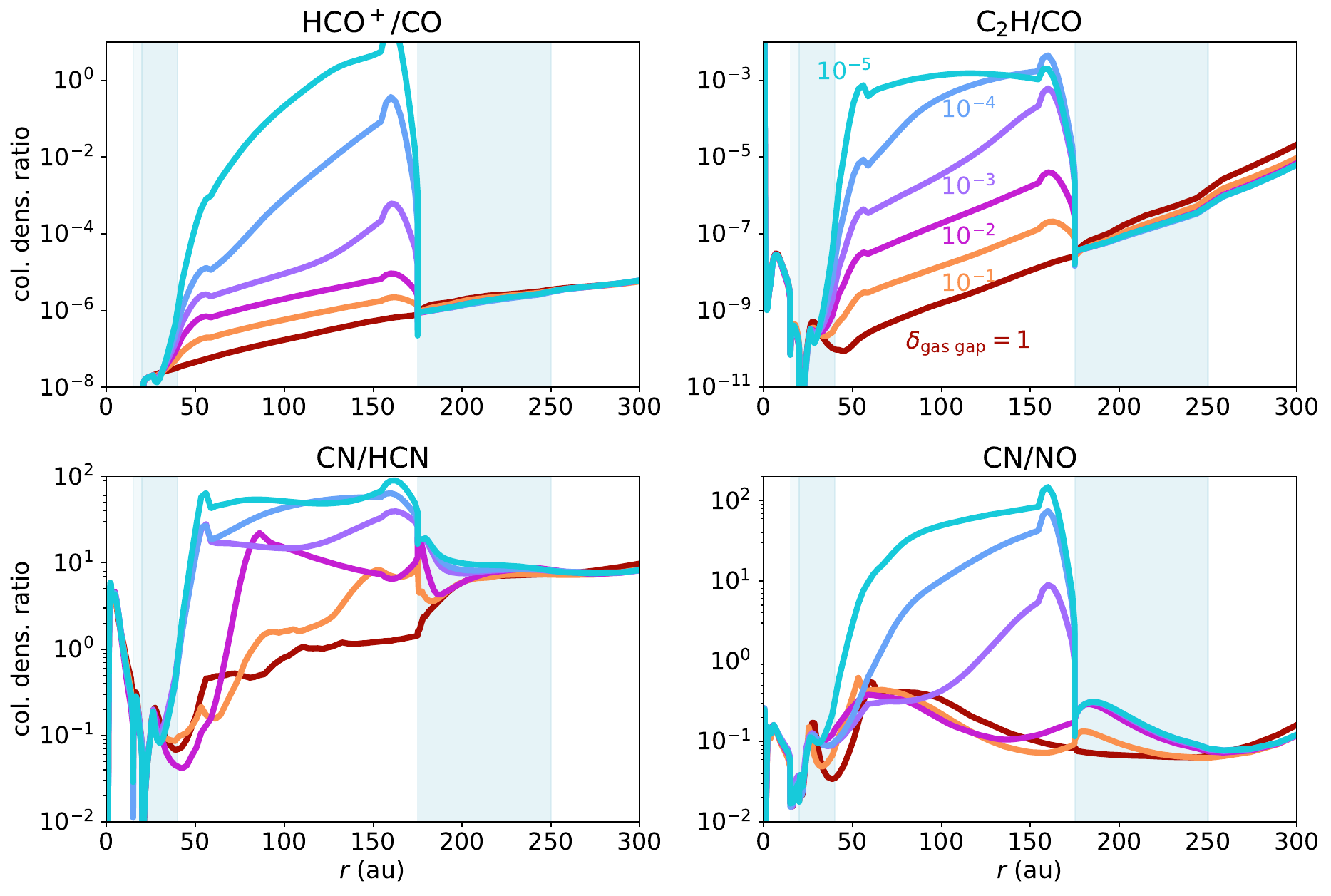}
      \caption{Column density ratios of \ce{HCO+}/CO (top left), \ce{C2H}/CO (top right), CN/HCN (bottom left), and CN/NO (bottom right) for different gas gap depths. }
         \label{fig:Nratios_diff_gaps}
   \end{figure*}

The \ce{HCO+}/CO ratio for gaps up to two orders of magnitude deep is mainly driven by the CO density that is directly proportional to the gas density and the \ce{HCO+} density that goes with the square root of the gas density (see Figs.~\ref{fig:dali_N_dgap} and \ref{fig:Nratios_diff_gaps}). For deeper gaps, CO photodissociation drives the \ce{HCO+}/CO ratio up. Even though no \ce{HCO+} photodissociation is included in the small network, its photodissociation rate is 44 times lower than that of CO \citep{Heays2017}. Therefore, the \ce{HCO+}/CO ratio is most sensitive to the gas density.

The \ce{C2H}/CO ratio is primarily driven by the drop in the CO column density for gaps up to $10^{-3}$ as the \ce{C2H} column density only starts to drop steeply for gaps deeper than that. This causes the \ce{C2H}/CO ratio to saturate at a few $10^{-3}$ in the outer regions of the $10^{-5}$ deep gas gap. 
The CN/HCN ratio rapidly increases when the gap depth is lowered from $10^{-1}$ to $10^{-2}$ as the midplane HCN component is removed inside of $\sim 80$~au. For deeper gaps, the ratio settles between a value of $10-100$ inside the gap. Outside the gap, CN is a factor of ten more abundant than HCN. 

In contrast to the former ratios, the CN/NO ratio is almost independent of gap depths for a disk with no gas gaps and gaps with up to two orders of magnitude less gas. This is due to the CN and NO column density that are both insensitive to the gap depths for these models. For deeper gaps, both the CN and the NO column density decrease as the gas gap becomes deeper but the CN column density does so at a slower rate than that of NO. Therefore, the CN/NO column density ratio is higher inside the gap than outside.

   \begin{figure*}[ht!]
   \centering
   \includegraphics[width=14cm]{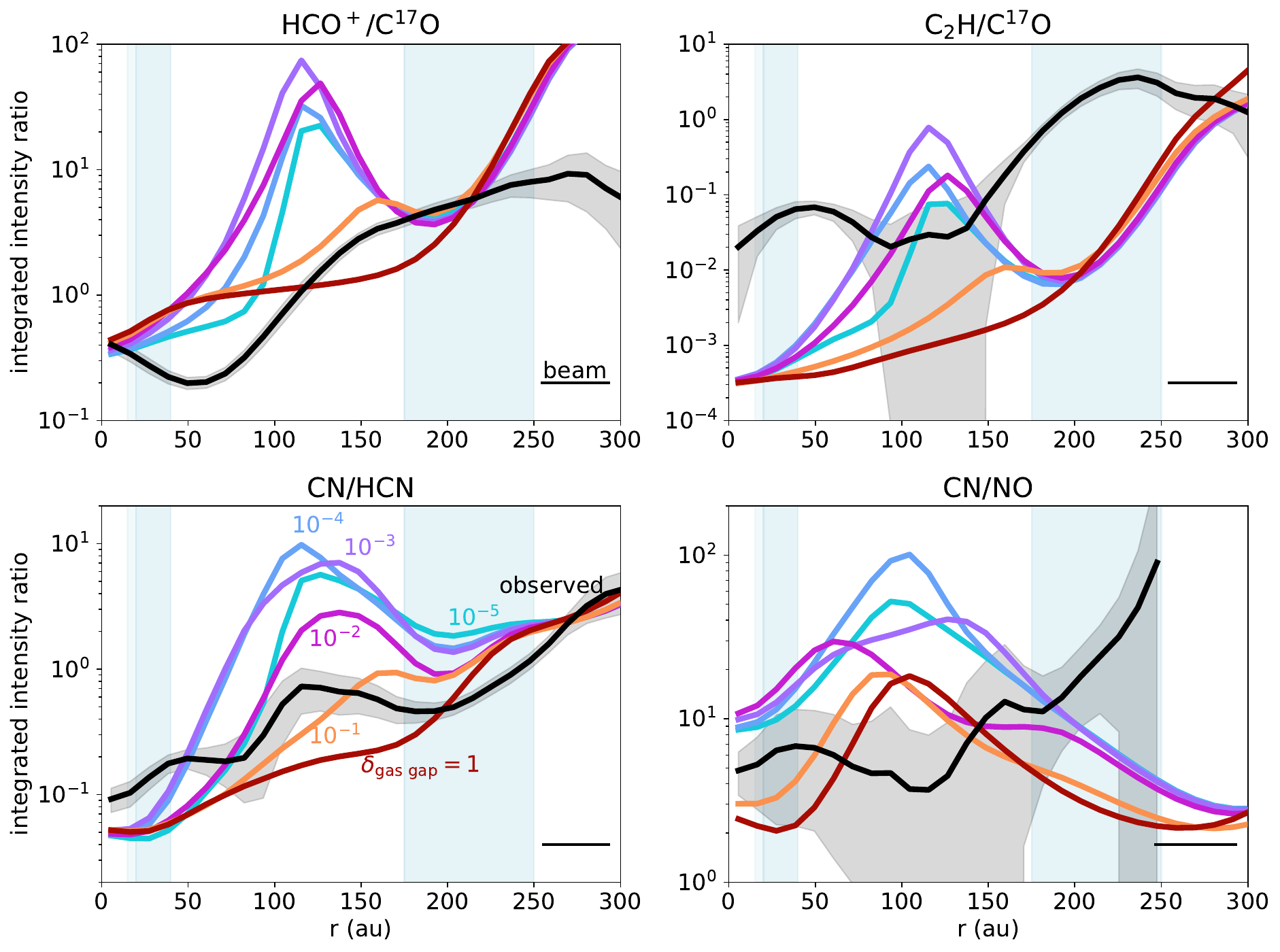}
      \caption{Emission line ratios of \ce{HCO+}/CO (top left), \ce{C2H}/CO (top right), CN/HCN (bottom left), and CN/NO (bottom right) for different gas gap depths.     }
         \label{fig:Iratios_diff_gaps}
   \end{figure*}

   \begin{figure*}[ht!]
   \centering
   \includegraphics[width=0.7\hsize]{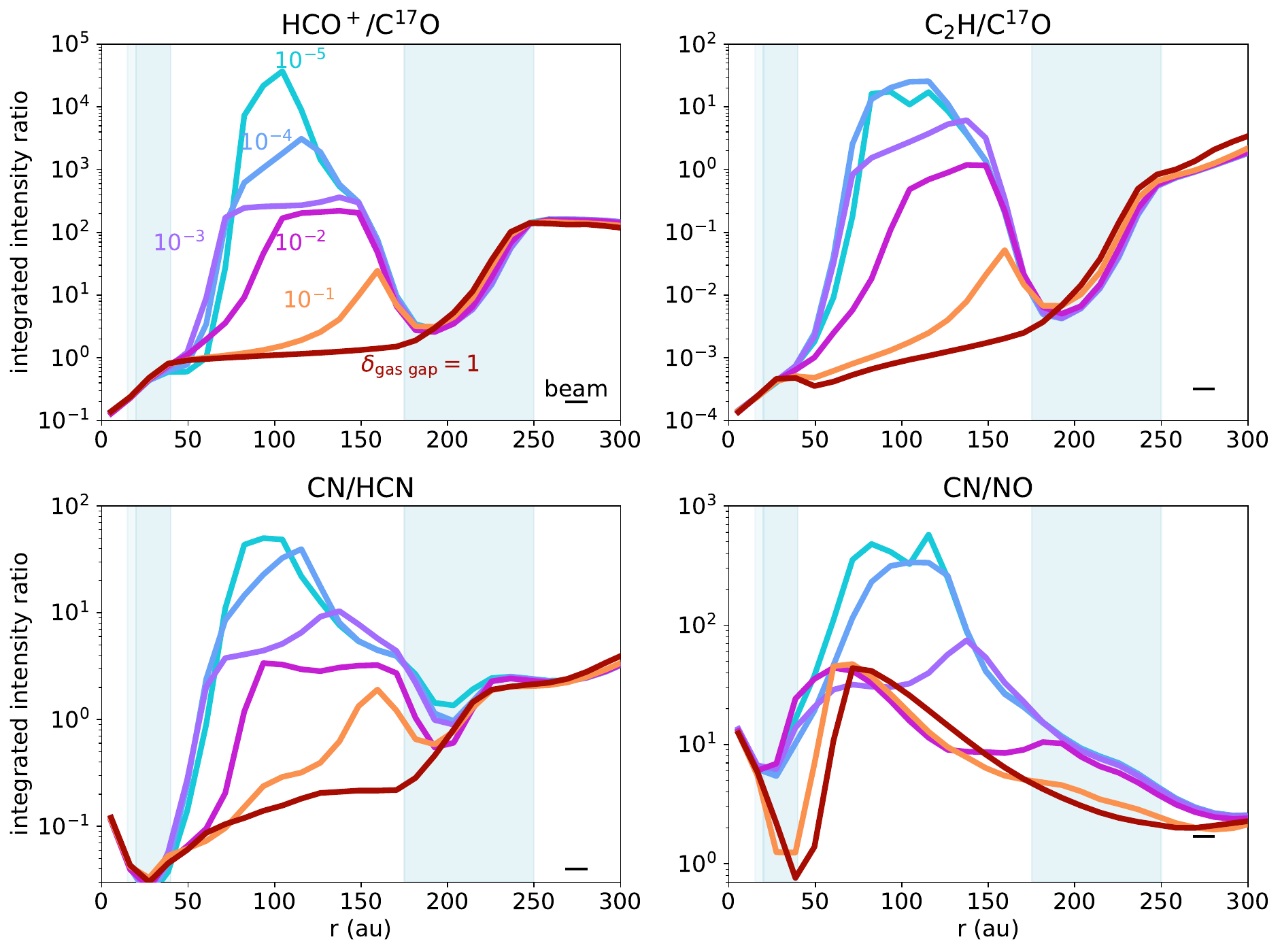}
      \caption{Same as Fig.~\ref{fig:Iratios_diff_gaps} but then for a small beam of $0\farcs1$.  } 
         \label{fig:Iratios_diff_gaps_small_beam}
   \end{figure*}

\subsubsection{Emission line ratios for different gap depths}
The corresponding figure for the emission lines is presented in Fig.~\ref{fig:Iratios_diff_gaps}. For each ratio, the profile of the molecule at the highest spatial resolution is smoothed to the resolution of the image at lower resolution. The modelled ratios of \ce{HCO+}/CO, \ce{C2H}/CO, CN/HCN, and CN/NO all peak inside the gas gap if the gas gap is at least $10^{-1}$ deep. The column density ratio increases with gap depth for most regions inside the gas gap, however, the ratio of the integrated intensity does not. This is due to the beam size of the observations. The size and orientation of the beam is such that the dust gap is sampled by only $\sim2$ beams across the disk minor axis. Therefore, the line ratio in the gap is dominated by the bright emission at the gap edges. If the models are convolved with a $3-4$ times smaller beam of $0\farcs1$, the gap depth can be traced up to a very deep gas gap of $10^{-4}$ times less gas, see Fig.~\ref{fig:Iratios_diff_gaps_small_beam}. 

All models reproduce the \ce{HCO+}/\ce{C^17O} ratio seen inside the inner dust ring ($r<20$~au) and at 200~au where the outer dust ring is located. The observations show that this ratio first decreases in the gap and then increases, whereas the models with a gas gap show a peak at 125~au ($\delta_{\rm gas\ gap} = 10^{-2}-10^{-5}$) or 150~au ($\delta_{\rm gas\ gap} = 10^{-1}$). The fiducial model with a gap of only one order of magnitude best reproduces the observed ratio outside 100~au. The other models show a ratio of \ce{HCO+}/\ce{C^17O} that is too high, indicating that the ionisation is too high or the abundance of gas-phase water as traced by \ce{HCO+} is too low in the gap. The latter is consistent with the detection of cold, gas-phase water lines originating from $\sim40-300$~au in this disk \citep{vanDishoeck2021, Pirovano2022}. This is because the modelled \ce{HCO+} abundance does not account for photodesorbed water. 

The observed \ce{C2H}/\ce{C^17O} ratio peaks just outside the inner dust ring and at the outer edge of the outer dust ring. The models with a gas gap show the opposite behaviour where a peak at $\sim 100$~au in the \ce{C2H}/\ce{C^17O} ratio is predicted. Only the model without a drop in the gas density reproduces the overall trend of the observations with an offset of two orders of magnitude. This offset is driven by the modelled \ce{C2H} emission being too weak by that same factor. 

The ratio of CN/HCN increases with radius, with a peak seen inside the gas gap and a decrease in the outer dust ring. The models with a gas gap predict the same trend across the disk, with the ratio agreeing within a factor of $\sim4$ for the models with gaps up to $10^{-2}$. The observations are most consistent with a gap depth of one to two orders of magnitude. The UV field in the models with deeper gas gaps is too intense to reproduce the CN/HCN ratio.

Finally, the CN/NO ratio is presented in the bottom right panel of \ref{fig:Iratios_diff_gaps}. The CN/NO ratio of $\sim5$ at the inner dust ring is reproduced by the models that do not have a very deep gas gap of $10^{-4}-10^{-5}$. For the latter models, the NO intensity is much higher, lowering the CN/NO ratio. In the outer dust ring the observed ratio is $\sim 10$, whereas the models underpredict this by an order of magnitude. 

In summary, the line ratios of \ce{HCO+}/CO, \ce{C2H}/CO, and CN/HCN at the spatial resolution of the observations are only sensitive to the gas gap depths up to a gap with two orders of magnitude less gas. The latter ratio is well reproduced by the models, whereas the \ce{C2H}/\ce{C^17O} is not due to the weak \ce{C2H} emission. The \ce{HCO+}/\ce{C^17O} ratio can be used to trace gas gaps that are up to two orders of magnitude depleted in gas.

      \begin{figure*}
   \centering
   \includegraphics[width=0.7\hsize]{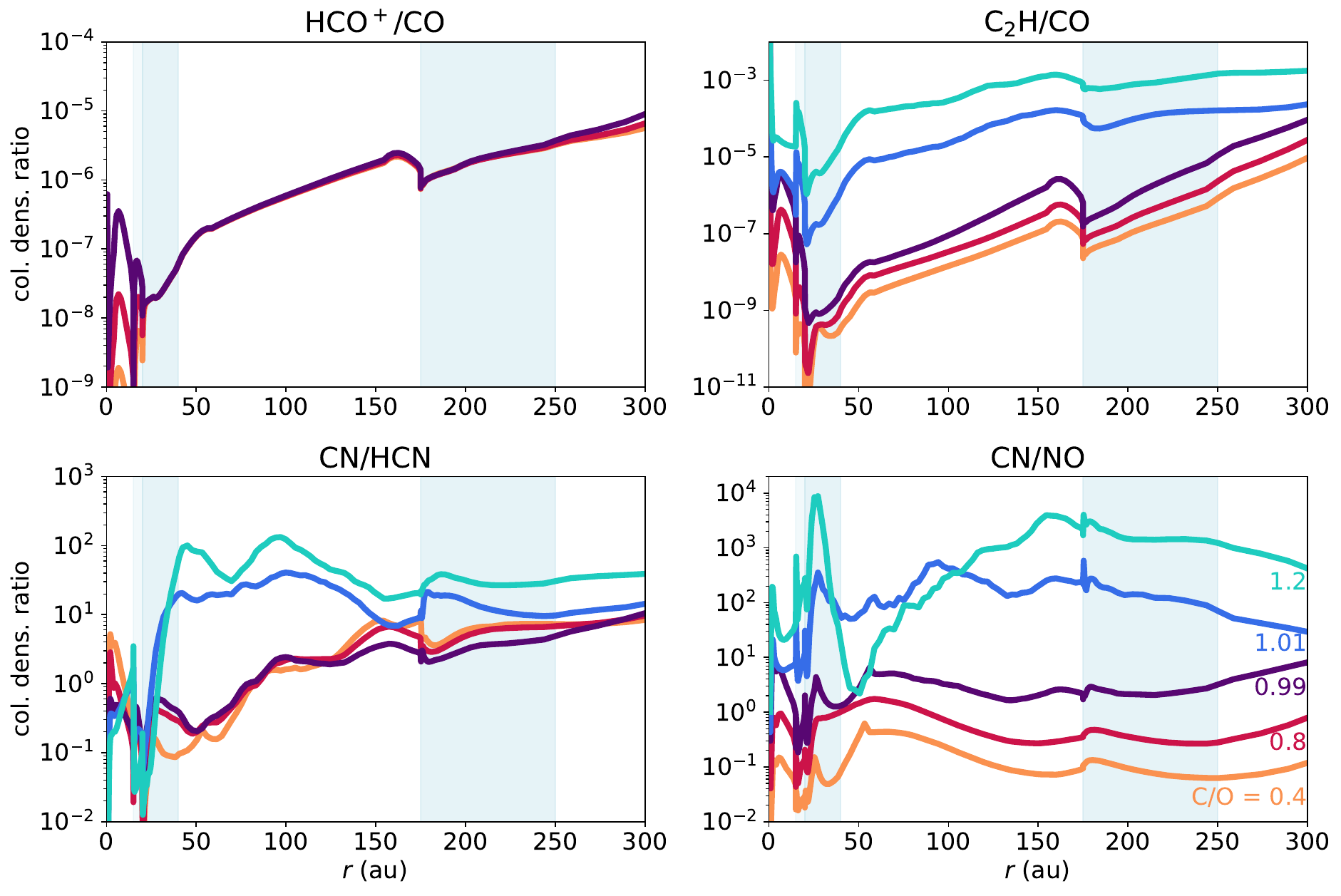}
      \caption{Column density ratios for different molecules and different C/O ratios. }
         \label{fig:N_dC_O_ratios}
   \end{figure*}

   \begin{figure*}
   \centering
   \includegraphics[width=0.7\hsize]{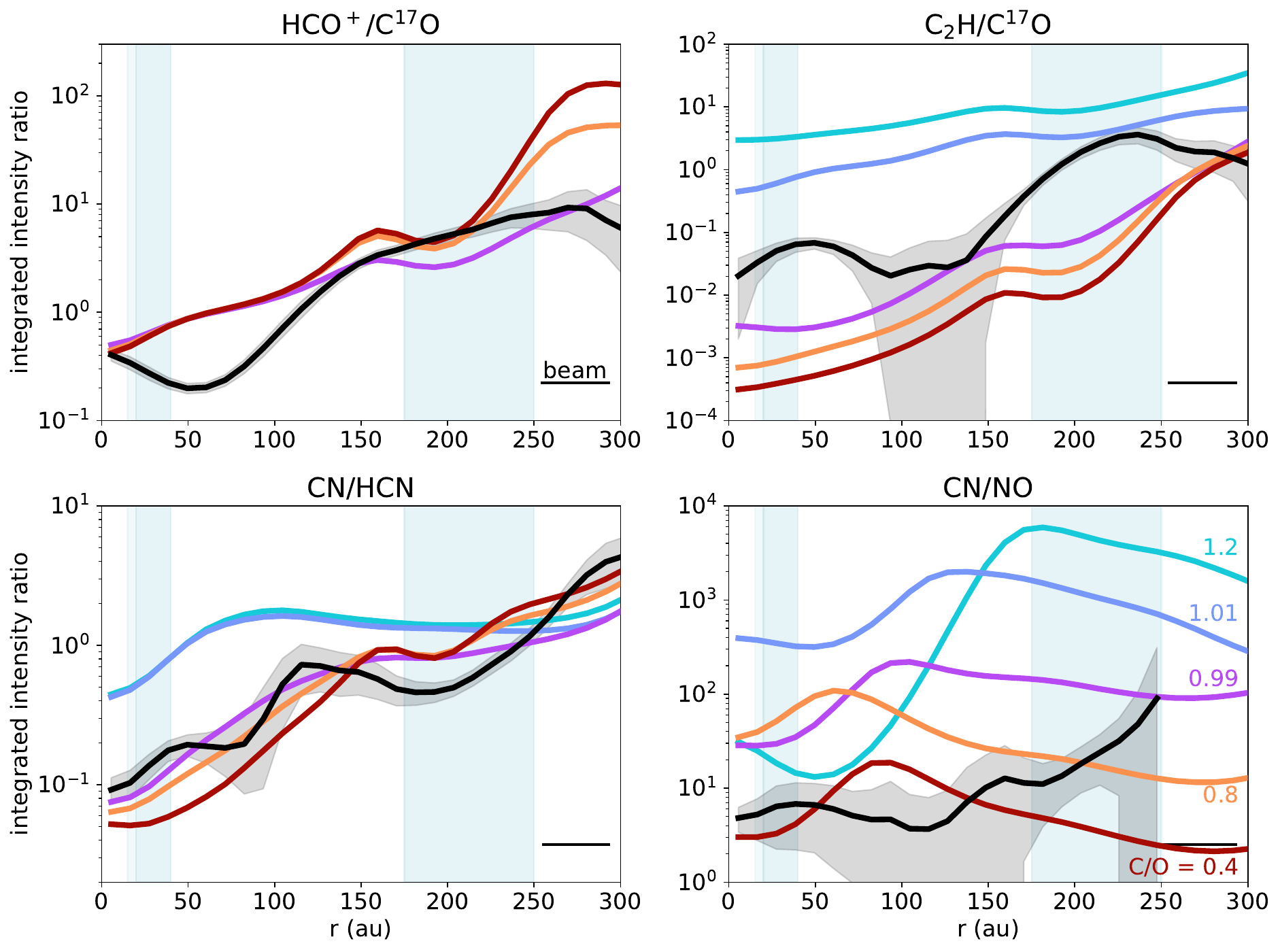}
      \caption{Emission line ratios of \ce{HCO+}/CO (top left), \ce{C2H}/CO (top right), CN/HCN (bottom left), and CN/NO (bottom right) for different gas gap depths. The black line indicated the observed emission line ratio for each pair.  }
         \label{fig:I_dC_O_ratios}
   \end{figure*}

\subsubsection{Column density and line ratios for different C/O ratios}   \label{app:ratios_C_O}
   
The column density ratios for various molecules as function of the C/O ratio are presented in Fig.~\ref{fig:N_dC_O_ratios}. The corresponding emission line ratios are presented in Fig.~\ref{fig:I_dC_O_ratios}. The \ce{HCO+}/\ce{C^17O} ratio presented in the top left corner does not depend on the C/O ratio of the gas for most disk regions. Only outside the outermost dust ring, the \ce{HCO+}/\ce{C^17O} ratio decreases by one order of magnitude down to the observed value for a C/O of 0.99 instead of 0.4. 

The \ce{C2H}/\ce{C^17O} ratio is driven by the \ce{C2H} emission that increases steeply with an increasing C/O ratio. Between the two dust rings, the observed \ce{C2H}/\ce{C^17O} is best explained by a model with a C/O just below 1 at 0.99, whereas in the outer dust ring the model with a C/O of 1.01 is close to the observed ratio. The C/O ratio greatly affects the value of the \ce{C2H}/\ce{C^17O} ratio, but it does not change the location of the peaks and dips in the profile. Therefore, a global change in the C/O ratio alone cannot explain the observations.

Increasing the C/O ratio increases the CN intensity more than it does so for the HCN. Therefore, the CN/HCN ratio increases with increasing C/O ratio. The morphology of the observed ratio is best reproduced by a C/O of $0.4-0.8$, whereas the value inside the gap is reproduced within $\sim 1\sigma$ for C/O = 0.99. The CN/NO ratio on the other hand is best reproduced by a low C/O of 0.4. In the outer disk, the CN/NO ratio increases by one order of magnitude for a C/O of 0.4, 0.8, 0.99, 1.01, and 1.2. Even though the S/N on the CN/NO ratio is low, the value in the outer dust ring is indicative of a slightly elevated C/O ratio of 0.8.

\subsection{Background UV}

\begin{figure*}
   \centering
   \includegraphics[width=\hsize]{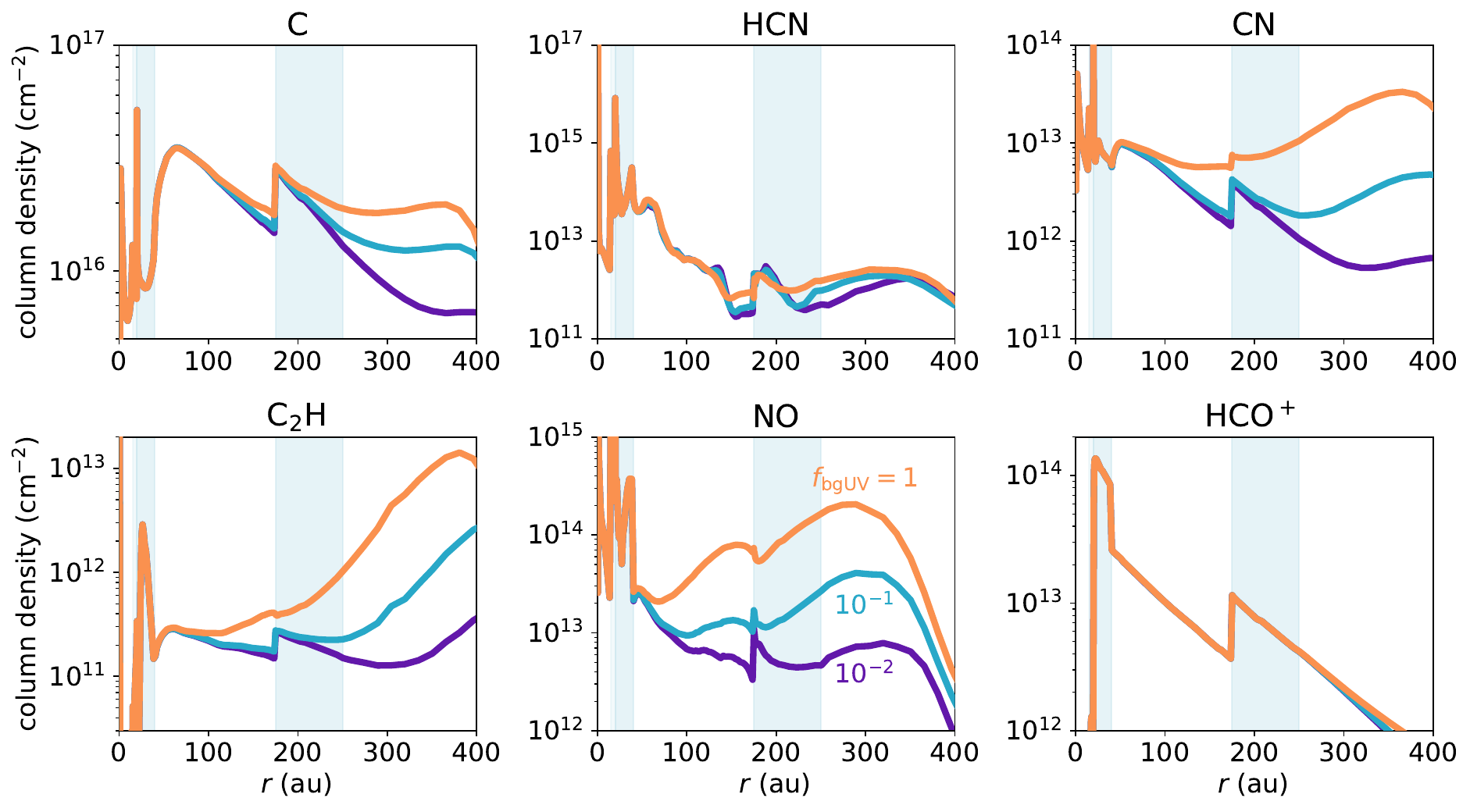}
      \caption{Column densities of different molecules studied in this work for different background UV radiation fields.}
         \label{fig:N_dbgUV}
   \end{figure*}

In Fig.~\ref{fig:N_dbgUV}, the column densities for models with a lower background UV radiation field are presented. This parameter mainly affects the column densities in the outer disk where a lower UV field lowers the column density of C, CN, \ce{C2H}, and NO, whereas it increases that of HCN somewhat due to less efficient photodissociation. 

\subsection{Flaring} \label{app:flaring}

\begin{figure*}
   \centering
   \includegraphics[width=0.95\hsize]{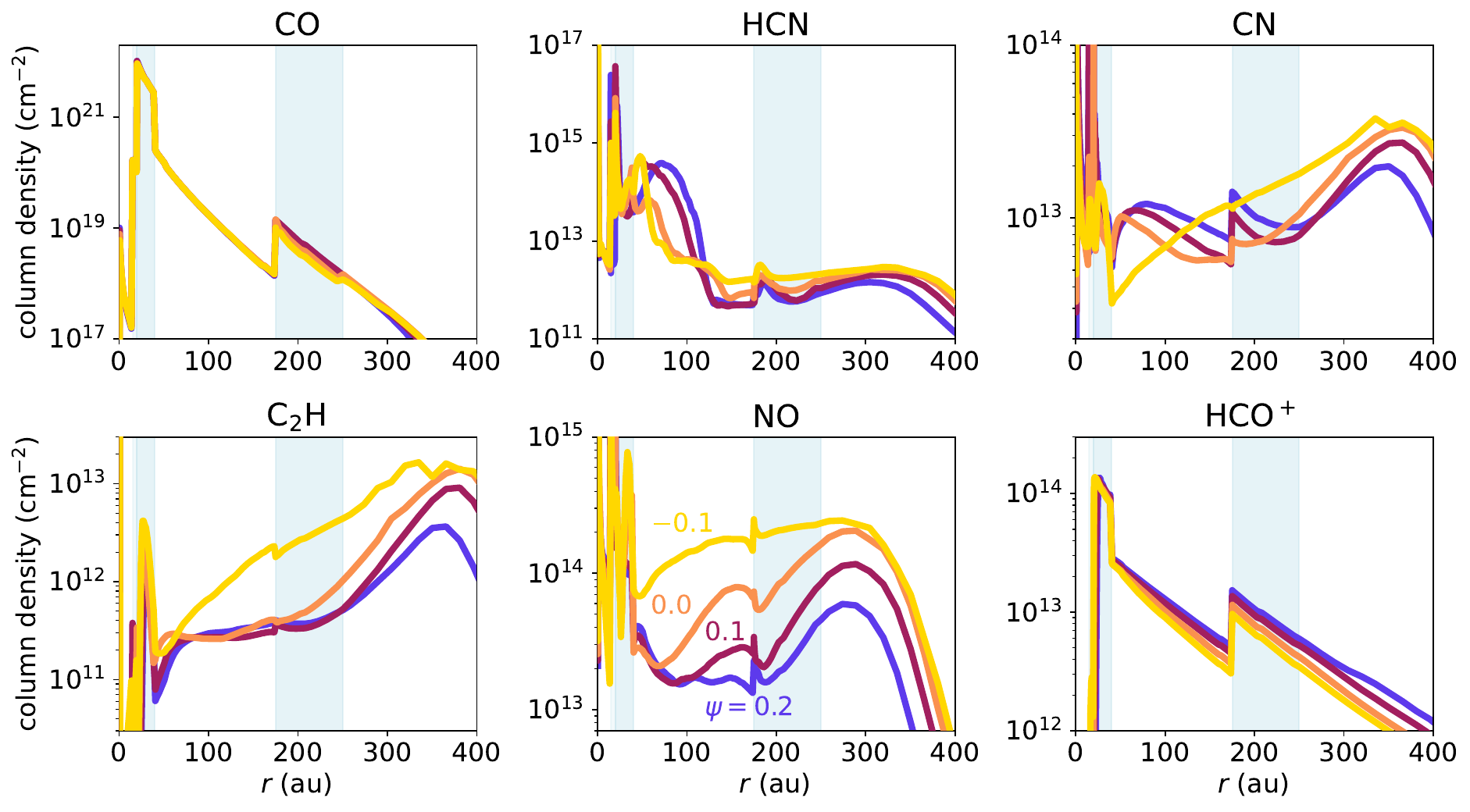}
      \caption{Column densities of different molecules studied in this work for different flaring indices.  }
         \label{fig:N_psi}
   \end{figure*}

\begin{figure*}
   \centering
   \includegraphics[width=0.95\hsize]{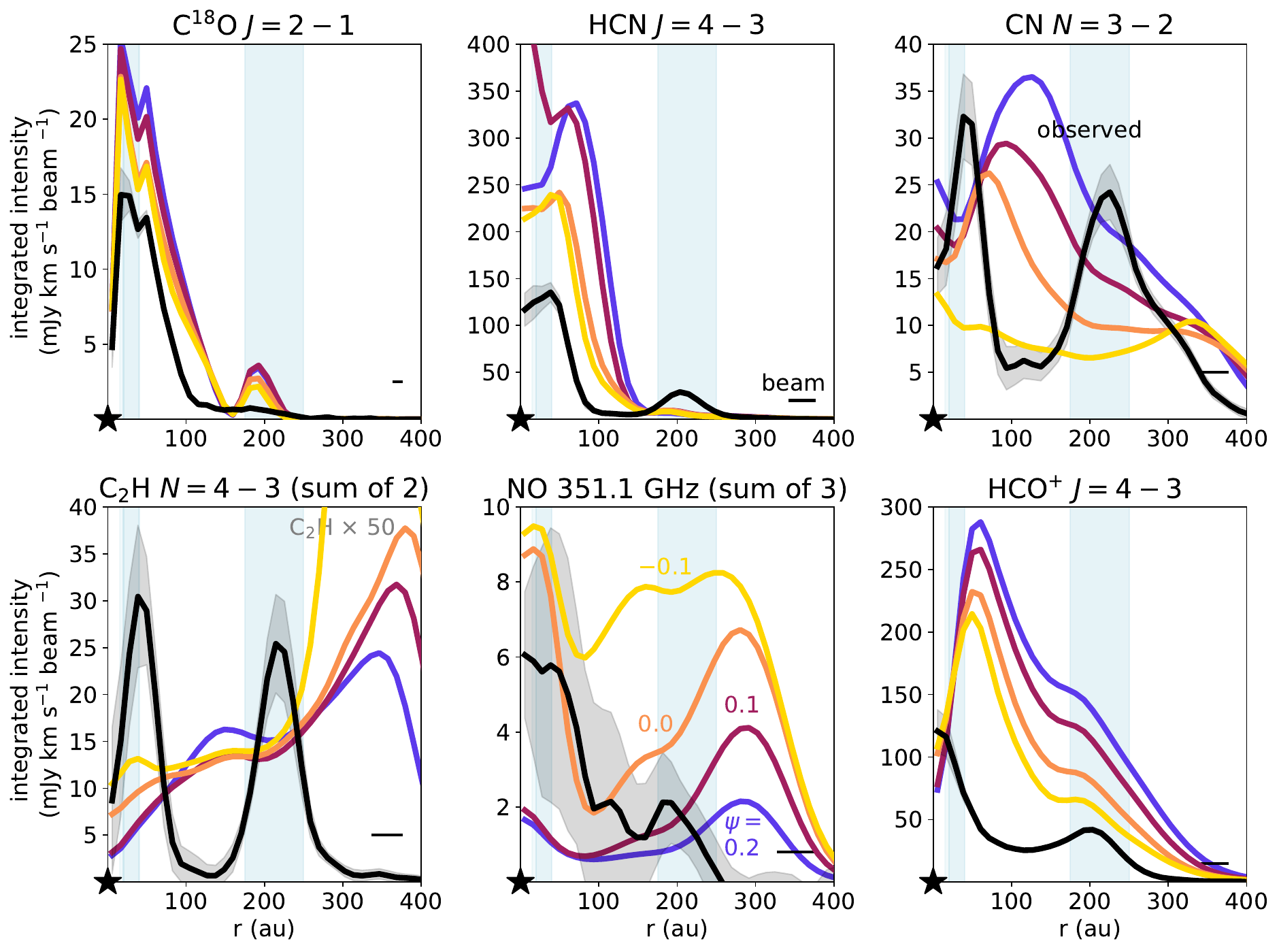}
      \caption{Molecular line emission of different molecules studied in this work for different flaring indices.  }
         \label{fig:I_psi}
   \end{figure*}

The column density and azimuthally averaged radial profiles of CO, HCN, CN, \ce{C2H}, NO, and \ce{HCO+} are presented in Fig.~\ref{fig:N_psi} and \ref{fig:I_psi}, respectively. Flaring alone has a minor to no effect on the CO and \ce{HCO+} column densities, whereas those of HCN, CN, \ce{C2H}, and NO do generally increase with decreasing flaring indices. These differences in the column densities are partially countered by the changing temperature structure as less flared disks intercept less UV. 

\subsection{X-ray luminosity} \label{app:Lx}

   \begin{figure*}
   \centering
   \includegraphics[width=0.95\hsize]{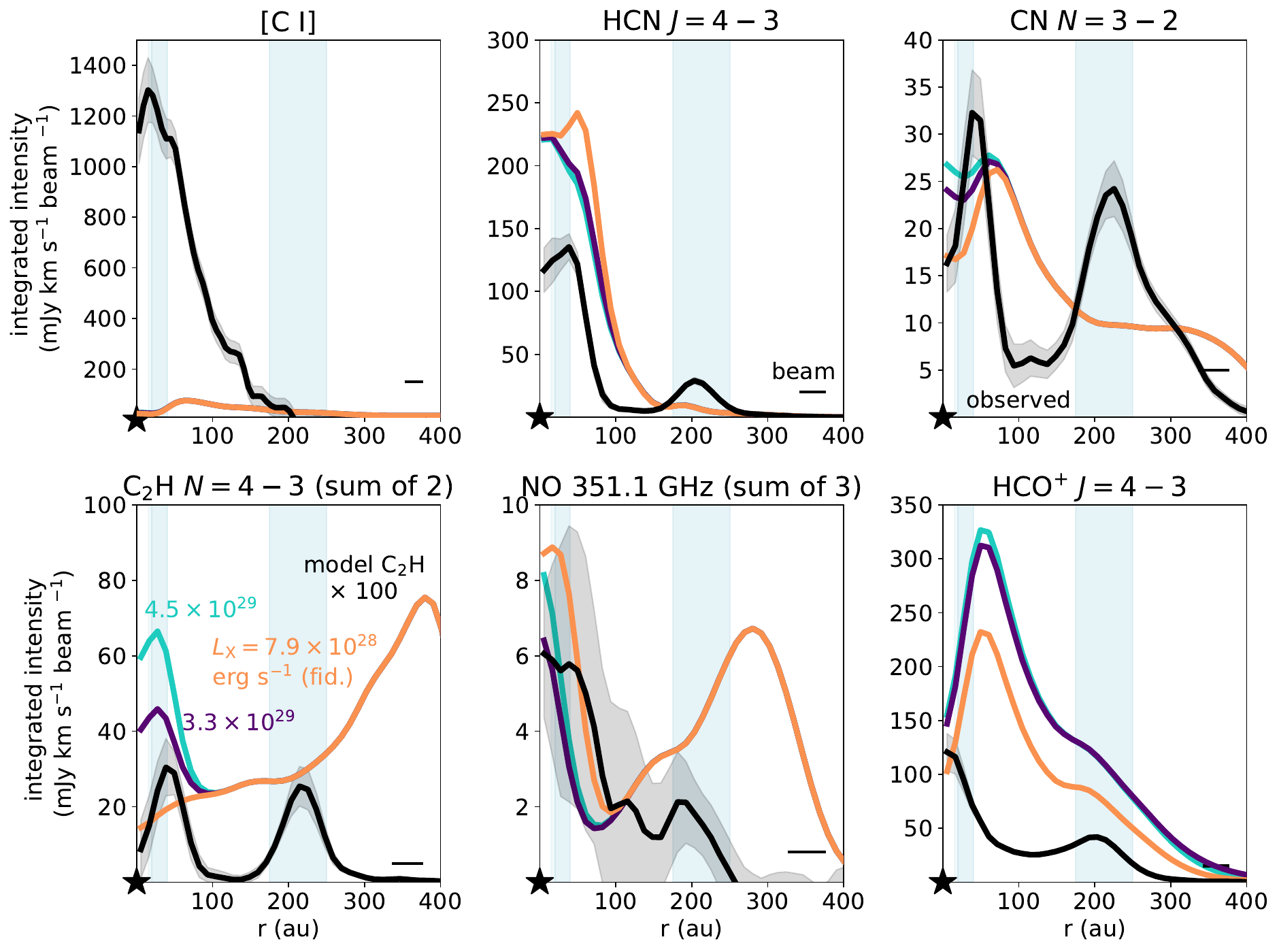}
      \caption{Molecular line emission of different molecules studied in this work for different X-ray luminosities. The effect of an increased X-ray luminosity on the \ce{HCO+} is modelled as an increase in the effective cosmic ray ionisation rate to $2\times 10^{-18}$~s$^{-1}$ (purple) and $2\times 10^{-18}$~s$^{-1}$ (cyan), respectively. }  
         \label{fig:I_Lx}
   \end{figure*}

Fig.~\ref{fig:I_Lx} presents the intensity of [C~{\sc I}], HCN, CN, \ce{C2H}, NO, and \ce{HCO+} for the mean and maximum X-ray luminosities reported in \citet{Skinner2020} that are a factor of 4 and 6 higher than that in the fiducial model. The higher X-ray luminosity mainly affects the emission in the inner 100~au of the disk. The morphology of the \ce{C2H} emission is better reproduced by the model with a higher X-ray luminosities due to an inner ring at 30~au in the \ce{C2H} in these models. However, these models predict CN emission in the inner tens of au that is not seen in the observations.  
The effect of the increased X-ray luminosity on the \ce{HCO+} emission is modelled as in increase in the effective cosmic ray ionisation rate by a factor of 2 and 5 respectively. This increases the modelled \ce{HCO+} intensity by less than a factor of 2 in all disk regions.

\end{appendix}

\end{document}